\newif\ifcomment
\newif\ifdraft
\newif\iflatexdiff
\latexdifftrue

\newif\ifmassfullsub
\newif\ifmasspartsub
\newif\ifmasssepbg
\masssepbgtrue

\def\dvers{v1.0}
\def\dtitle{Production of $\pi^0$ and $\eta$ mesons up to high transverse momentum in pp collisions at 2.76 TeV} 
\def\stitle{Neutral meson production up to high $\pt$ at 2.76 TeV} 
\documentclass[ALICE,manyauthors,12pt]{cernphprep}
\usepackage[comma,square,numbers,sort&compress]{natbib}
\usepackage{hyperref}
\usepackage{graphicx}  
\usepackage{dcolumn}   
\usepackage{bm}        
\usepackage{amssymb}   
\usepackage{amsfonts}
\usepackage{graphics}
\usepackage{grffile}   
\usepackage{epsfig}
\usepackage[loose]{units}
\usepackage[usenames,dvipsnames]{xcolor}
\usepackage[normalem]{ulem} 
\usepackage{color}
\usepackage[utf8]{inputenc}
\usepackage[T1]{fontenc}
\usepackage{subfigure}
\iflatexdiff
\RequirePackage{color}\definecolor{RED}{rgb}{1,0,0}\definecolor{BLUE}{rgb}{0,0,1}

\fi

\newcommand{\pz}           {\ensuremath{\pi^{0}}}
\newcommand{\pzs}          {\ensuremath{\pi^{0}}s}
\newcommand{\lzt}  	   {\ensuremath{\sigma^{2}_{\rm long}}}
\newcommand{\Vz}  	   {\ensuremath{V^{0}}}
\newcommand{\dEdx}         {\ensuremath{{\rm d}E/{\rm d}x}}

\newcommand{\kzs}          {K$^{0}_{\rm S}$}
\newcommand{\kzl}          {K$^{0}_{\rm L}$}

\newcommand{\pp}           {pp}

\newcommand{\PCM}          {\mbox{PCM}}
\newcommand{\PHOS}         {\mbox{PHOS}}
\newcommand{\EMC}          {\mbox{EMC}}
\newcommand{\mEMC}         {\mbox{mEMC}}
\newcommand{\PCMEMC}       {\mbox{PCM-EMC}}
\newcommand{\PE}           {\mbox{P-E}}
\newcommand{\Pythia}       {\mbox{PYTHIA}}

\newcommand{\dNdeta}       {\mathrm{d}N_\mathrm{ch}/\mathrm{d}\eta}

\newcommand{\degree}       {\ensuremath^{\rm o}}
\newcommand{\gevc}         {GeV/$c$}
\newcommand{\s}            {\ensuremath{\sqrt{s}}}

\newcommand{\pt}           {\ensuremath{p_{\mathrm{T}}}}
\newcommand{\ptHard}       {\ensuremath{p_{\mathrm{T, hard}}}}

\newcommand{\abs}[1]       {\ensuremath{\left|#1\right|}}
\newcommand{\avg}[1]       {\ensuremath{\left\langle#1\right\rangle}}

\newcommand{\Rt}           {\ensuremath{R_{\rm Trig}}}
\newcommand{\kapt}         {\ensuremath{\kappa_{\rm Trig}}}
\newcommand{\dd}           {\ensuremath{\mathrm{d}}}

\newcommand{\RConv}        {\ensuremath{R_{\rm conv}}}
\newcommand{\ZConv}        {\ensuremath{Z_{\rm conv}}}

\newcommand{\EtaV}         {\ensuremath{\eta_{\rm \Vz}}}
\newcommand{\psip}         {\ensuremath{\psi_{\rm pair}}}
\newcommand{\chit}         {\ensuremath{\chi_{\rm red}^2}}
\newcommand{\psipm}        {\ensuremath{\psi_{\rm pair, max}}}
\newcommand{\chitm}        {\ensuremath{\chi^2_{\rm red, max}}}
\newcommand{\qt}           {\ensuremath{q_{\mathrm{T}}}}
\newcommand{\qtm}          {\ensuremath{q_{\mathrm{T, max}}}}
\newcommand{\alpham}       {\ensuremath{\alpha_{\mathrm{max}}}}
\newcommand{\PhiConv}      {\ensuremath{\varphi_{\rm conv}}}
\newcommand{\etatopi}      {\ensuremath{\eta/\pi^0}}

\newcommand{\lsim}         {\stackrel{<}{\sim}}

\newcommand{\Fig}[1]       {Fig.~\ref{#1}}
\newcommand{\Figure}[1]    {Figure~\ref{#1}}
\newcommand{\Sect}[1]      {Sect.~\ref{#1}}
\newcommand{\Section}[1]   {Section~\ref{#1}}

\newcommand{\Eq}[1]        {Eq.~\ref{#1}}

\newcommand{\Tab}[1]       {Tab.~\ref{#1}}

\newcommand{\Ref}[1]       {Ref.~\cite{#1}}
\newcommand{\Refs}[1]      {Refs.~\cite{#1}}

\newcommand{\com}[1]       {}
\iflatexdiff

\renewcommand{\xout}[1]    {\textcolor{red}{\sout{#1}}}
 
\newcommand{\old}[1]       {{\textcolor{red}{\sout{#1}}}}

\else

\renewcommand{\xout}[1]    {}
\newcommand{\old}[1]       {\relax}

\fi

\graphicspath{{./img/}}
\ifdraft
\usepackage{lineno}
\linenumbers
\modulolinenumbers[1]
\usepackage{fancyhdr}
\fi
\begin{document}
\begin{titlepage}
\PHyear{2017}
\PHnumber{019} 
\PHdate{30 Jan}   
\title{\dtitle}
\ShortTitle{\stitle}
\Collaboration{ALICE Collaboration%
         \thanks{See Appendix~\ref{app:collab} for the list of collaboration members}}
\ShortAuthor{ALICE Collaboration} 
\begin{center}
\ifdraft
\today\\ \color{red}DRAFT \dvers\ \hspace{0.3cm} \$Revision: 3268 $\color{white}:$\$\color{black}\vspace{0.3cm}
\else
\today
\fi
\end{center}
\begin{abstract}
The invariant differential cross sections for inclusive $\pz$ and $\eta$ mesons at midrapidity were measured in pp collisions at $\sqrt{s}=2.76$~TeV for transverse momenta $0.4<\pt<40$~GeV/$c$ and $0.6<\pt<20$~GeV/$c$, respectively, using the ALICE detector. 
This large range in $\pt$ was achieved by combining various analysis techniques and different triggers involving the electromagnetic calorimeter~(EMCal).
In particular, a new  single-cluster, shower-shape based method was developed for the identification of high-$\pt$ neutral pions, which exploits that the showers originating from their decay photons overlap in the EMCal.
Above $4$~\gevc, the measured cross sections are found to exhibit a similar power-law behavior with an exponent of about $6.3$.
Next-to-leading-order perturbative QCD calculations differ from the measured cross sections by about $30$\% for the $\pz$, and between $30$--$50$\% for the $\eta$ meson, while generator-level simulations with \Pythia\ 8.2 describe the data to better than $10$--$30$\%, except at $\pt<1$~\gevc.
The new data can therefore be used to further improve the theoretical description of $\pz$ and $\eta$ meson production.
\end{abstract}
\end{titlepage}
\newpage
\setcounter{page}{2}
\section{Introduction}
\label{sec:intro}
Measurements of identified hadron spectra in proton-proton~(pp) collisions are well suited to constrain predictions from Quantum Chromodynamics~(QCD)~\cite{Gross:1973ju}.
Such predictions are typically calculated in the pertubative approximation of QCD~(pQCD) based on the factorization of the elementary short-range scattering processes (such as quark--quark, quark--gluon and gluon--gluon scatterings) involving large momentum transfer~($Q^2$) and long-range universal properties of QCD that need to be experimentally constrained.
The universal properties are typically modeled by parton distribution functions~(PDFs), which describe the kinematic distributions of quarks and gluons within the proton in the collinear approximation, and fragmentation functions (FFs), which describe the probability for a quark or gluon to fragment into hadrons of a certain type. 
The cross section for the production of a given hadron of type H can be written as a sum over parton types 
\begin{equation}
E\frac{d^3\sigma^{\rm H}}{d\vec{p}} = \sum_{a,b,c} f_a(x_1,Q^2) \otimes f_b(x_2,Q^2) \otimes D_c^H(z_c,Q^2) \otimes d\hat{\sigma}_{ab\rightarrow cX}(Q^2,x_1,x_2)\,, 
\end{equation} 
where $f_i(x)$ denotes the proton PDF of parton $i$ carrying a fraction $x$ of the proton's longitudinal momentum, $D^H_i(z_i)$ the FF of parton $i$ into hadron H carrying a fraction $z_i$ of the parton's momentum, and $d\hat{\sigma}_{ij\rightarrow kX}$ the inclusive short-distance scattering cross section of partons $i$ and $j$ into $k$~(see e.g.~\cite{Brock:1994er}).

Measurements of hadron production provide constraints on the PDFs and FFs, which are crucial for pQCD predictions, and at LHC energies probe rather low values of $x\sim0.001$ and $z\sim0.1$. 
The neutral pion~($\pz)$ is of special interest because as the lightest hadron it is abundantly produced, and at LHC collision energies below a transverse momentum~($\pt$) of $20$ GeV/$c$ dominantly originates from gluon fragmentation. 
While the collision energy ($\sqrt{s}$) dependence of $\pz$ cross sections has been useful for guiding the parametrization of the FFs~\cite{deFlorian:2014xna}, experimental data for neutral pions~\cite{Abelev:2012cn,Abelev:2014ypa} at the LHC are not available above $20$~GeV/$c$, where quark fragmentation starts to play a role.
The new $\pz$ data presented in this paper extend our previous measurement~\cite{Abelev:2014ypa} in pp collisions at $\sqrt{s} = 2.76$~TeV to $\pt$ values of $40$~GeV/$c$ allowing one to investigate the $\pt$ dependence of the $\pz$ cross section at high transverse momentum.
In addition, we present the cross section of the $\eta$ meson, which due to its strange quark content provides access to the study of possible differences of fragmentation functions with and without strange quarks~\cite{Aidala:2010bn}.
Furthermore, the $\eta$ meson constitutes the second most important source of decay photons and electrons after the $\pz$.
Hence, $\pz$ and $\eta$ meson spectra over a large $\pt$ range are needed for a precise characterization of the decay photon~(electron) background for direct photon~(semileptonic open charm and beauty) measurements. 

The new measurement of the $\pz$ cross section is a result of five analyses using data from various ALICE detector systems and different identification techniques.
The decay photons are either measured directly in the Electromagnetic Calorimeter~(EMCal), the Photon Spectrometer~(PHOS) or via the photon conversion method~(\PCM).
In the \PCM\ measurement, the photons are reconstructed via their conversions into $e^{+}e^{-}$ pairs within the detector material, where the $e^{+}e^{-}$ pairs are reconstructed with the charged-particle tracking systems.
The $\pz$ is reconstructed statistically using the invariant mass technique.
At high $\pt$, where the decay photons are too close together to be resolved individually, the $\pz$ can still be measured via the characteristic shape of their energy deposition in the EMCal.  
We combine statistically independent analyses where {\it(i)} both photons are individually resolved in the EMCal~(\EMC), {\it(ii)} one photon is identified in the EMCal and one is reconstructed via its conversion to $e^{+}e^{-}$ (\PCMEMC), and {\it(iii)} the photon pair's energy is merged in the EMCal~(\mEMC). 
Finally, the previously published measurements based on methods where both photons are reconstructed with (iv)~\PHOS\ or (v)~\PCM\ are included as well~\cite{Abelev:2014ypa}. 
The addition of the EMCal based measurements extends the $\pt$ reach from $12$~GeV/$c$ to $40$~GeV/$c$, the highest $\pt$ for identified hadrons achieved so far.  
The $\eta$ meson cross section that was previously not available at $\s=2.76$ TeV is measured in the range from $0.6$ to $20$~GeV/$c$ using the \PCM, \PCM-EMC\ and \EMC\ methods. 
Consequently, the $\etatopi$ ratio is measured in the same $\pt$ range.

The article is organized as follows: 
\Section{sec:aliceDet} briefly describes the experimental setup.
\Section{sec:datasamples} describes the data samples and event selection.
\Section{sec:reco} describes the neutral meson reconstruction techniques and corresponding corrections for the cross section measurements.
\Section{sec:corrsys} discusses the systematic uncertainties of the various measurements.
\Section{sec:results} presents the data and comparison with calculations and
\Section{sec:summary} provides a summary.

\section{ALICE detector}
\label{sec:aliceDet}
A detailed description of the ALICE detector systems and their performance can be found in~\Refs{Aamodt:2008zz,Abelev:2014ffa}. 
The new measurements primarily use the Electromagnetic Calorimeter (EMCal), the Inner Tracking System (ITS), and the Time Projection Chamber (TPC) at mid-rapidity, which are positioned within a $0.5$~T solenoidal magnetic field.  
Two forward scintillator arrays (V0A and V0C) subtending a pseudorapidity~($\eta$) range of $2.8 < \eta < 5.1$ and $-3.7 < \eta < -1.7$, respectively, provided the minimum bias trigger, which will be further discussed in the next section.

The ITS~\cite{Aamodt:2008zz} consists of two layers of Silicon Pixel Detectors (SPD) positioned at a radial distance of \unit[3.9]{cm} and \unit[7.6]{cm}, two layers of Silicon Drift Detectors (SDD) at \unit[15.0]{cm} and \unit[23.9]{cm}, and two layers of Silicon Strip Detectors (SSD) at \unit[38.0]{cm} and \unit[43.0]{cm} from the beamline.
The two SPD layers cover a pseudorapidity range of $|\eta|<2$ and $|\eta| < 1.4$, respectively. 
The SDD and the SSD subtend $|\eta|<0.9$ and $|\eta|<1.0$, respectively.
The primary vertex can be reconstructed with a precision of $\sigma_{z(xy)} = A/\sqrt{(\dNdeta)^\beta} \oplus B$, where $A \approx600$~($300$)~$\mu$m, for the longitudinal~($z$) and transverse~($xy$) directions, respectively, $B\approx40$~$\mu$m and $\beta\approx1.4$. 

The TPC~\cite{Alme:2010ke} is a large (90~m$^3$) cylindrical drift detector filled with a Ne/CO$_2$ gas mixture. 
It covers a pseudorapidity range of $|\eta|<0.9$ over the full azimuthal angle for the maximum track length of 159 reconstructed space points.
The ITS and the TPC were aligned with respect to each other to a precision better than $\unit[100]{\mu m}$ using tracks from cosmic rays and proton-proton collisions~\cite{Aamodt:2010aa}.
The combined information of the ITS and TPC allows one to determine the momenta of charged particles in the range of $0.05$ to $100$ GeV/$c$ with a resolution between $1$\% at low $\pt$ and $10$\% at high $\pt$.
In addition, the TPC provides particle identification via the measurement of the specific energy loss (d$E$/d$x$) with a resolution of $\approx$5\%. 
The tracking detectors are complemented by the Transition Radiation Detector~(TRD) and a large time-of-flight~(TOF) detector.
These detectors were used to estimate the systematic uncertainty resulting from the non-perfect knowledge of the material in front of the EMCal.

The EMCal~\cite{Cortese:2008zza} is a layered lead-scintillator sampling calorimeter with wavelength shifting fibers for light collection. 
The overall EMCal covers $107\degree$ in azimuth and $-0.7 \le \eta \le 0.7$ in pseudorapidity.
The detector consists of $12288$ cells~(also called towers) with a size of $\Delta \eta \times \Delta \varphi = 0.0143\times 0.0143$ corresponding to about twice the effective Moli\`{e}re radius; the cells are read out individually.  
With a depth of $24.6$~cm, or $\approx20$ radiation lengths, $2\times2$ cells comprise a physical module. 
The $3072$ modules are arranged in $10$ full-sized and $2$ one-third-sized supermodules, consisting of $12\times24$ and $4\times24$ modules, respectively, of which only the full-sized modules, corresponding to an azimuthal coverage of $100\degree$, were readout for the data recorded in $2011$--$2013$.~\footnote{The detector was installed in its complete configuration by early 2012, while $4$ and $10$ full-sized supermodules were present in 2010 and 2011, respectively.} 
The modules are installed with a radial distance to the nominal collision vertex of $4.28$~m at the closest point, and assembled to be approximately projective in $\eta$. 
The scintillation light from each cell is collected with wavelength shifting fibers that are connected to a $5\times 5$~mm$^2$ active-area avalanche photodiode. 
The relative energy and position resolutions improve with rising incident energy of the particle\com{, and was studied at the CERN PS and SPS using a well-controlled test beam of electron}~\cite{Abeysekara:2010ze}. 
The energy resolution can be described by a constant and two energy dependent terms parametrized as $\frac{\sigma_E}{E} = A^2 \oplus \frac{B^2}{E} \oplus \frac{C^2}{E^2}$\% with $A=1.7\pm0.3$, $B=11.3\pm0.5$, $C=4.8\pm0.8$ and $E$ in GeV.
The position resolution is linear as a function of $1/\sqrt{E}$ and parametrized as $1.5 {\rm mm} + \frac{5.3 {\rm mm}}{\sqrt{E}}$ with $E$ in GeV.
Starting with the highest cell $E_{\rm seed}>0.5$~GeV, the energy depositions from directly adjacent \mbox{EMCal} cells with $E_{\rm cell}>0.1$~GeV are combined to form clusters representing the total energy and physical position of incident particles~\cite{Abelev:2014ffa}. 
The clustering algorithm allows only one local energy maximum in a cluster; if a second is found a new cluster is initiated. 
Each cell is restricted to only be part of one cluster. 
Individual cells were calibrated using the $\pz$ mass peak position evaluated cell-by-cell, achieving a relative variation of below $1$\%.
\section{Data samples and event selection}
\label{sec:datasamples}
The data presented in this paper were recorded during the 2011 and 2013 periods with pp collisions at $\s = \unit[2.76]{TeV}$.
Various EMCal triggers were employed and, while the majority of the minimum bias data were recorded in 2011, the 2013 running period took advantage of higher threshold EMCal triggers to collect a notable high-$\pt$ data sample.
For the pp data collected in 2011, the minimum bias trigger~(MB$_\mathrm{OR}$) required a hit in either V0 detector or a hit in the SPD, while it required hits in both V0 detectors for the data collected in 2013~(MB$_\mathrm{AND}$).
The respective cross sections were determined based on van-der-Meer scans, and found to be $\sigma_{\rm MB_{AND}} =47.7\pm0.9$~mb with $\sigma_{\rm MB_{AND}}/\sigma_{\rm MB_{OR}}=0.8613\pm0.0006$ and  $\sigma_{\rm MB_{AND}}/\sigma_{\rm inel} = 0.760^{+0.052}_{-0.028}$~\cite{Abelev:2012sea}.
For the normalisation of the 2013 data, for which there was no vdM scan, the uncertainty $\sigma_{\rm MB_{AND}}$ was conservatively increased to $4$\%, to account for possible variations of the MB$_{\rm AND}$ trigger efficiency between 2011 and 2013.
The resulting uncertainty due to the luminosity determination is $2.5$\% for both datasets together. 

The EMCal issues triggers at two different levels, Level~0~(L0) and Level~1~(L1).
The events accepted at L0 are further processed at L1.
The L0 decision, issued latest $1.2$~$\mu$s after the collision, is based on the analog charge sum of $2\times2$ adjacent cells evaluated with a sliding window algorithm within each physical Trigger Region Unit~(TRU) spanning $4\times24$ cells in coincidence with a minimum bias trigger.
The L1 trigger decision, which must be taken within $6.2$~$\mu$s after the collision, can incorporate additional information from different TRUs, as well as other triggers or detectors. 
The data presented in this paper used the photon~(EG) trigger at L1, which extends the $2\times2$ sliding window search across neighboring TRUs, resulting in a $\approx30$\% larger trigger area than the L0 trigger.

\begin{figure}[t!]
  \centering
  \includegraphics[width=0.7\textwidth]{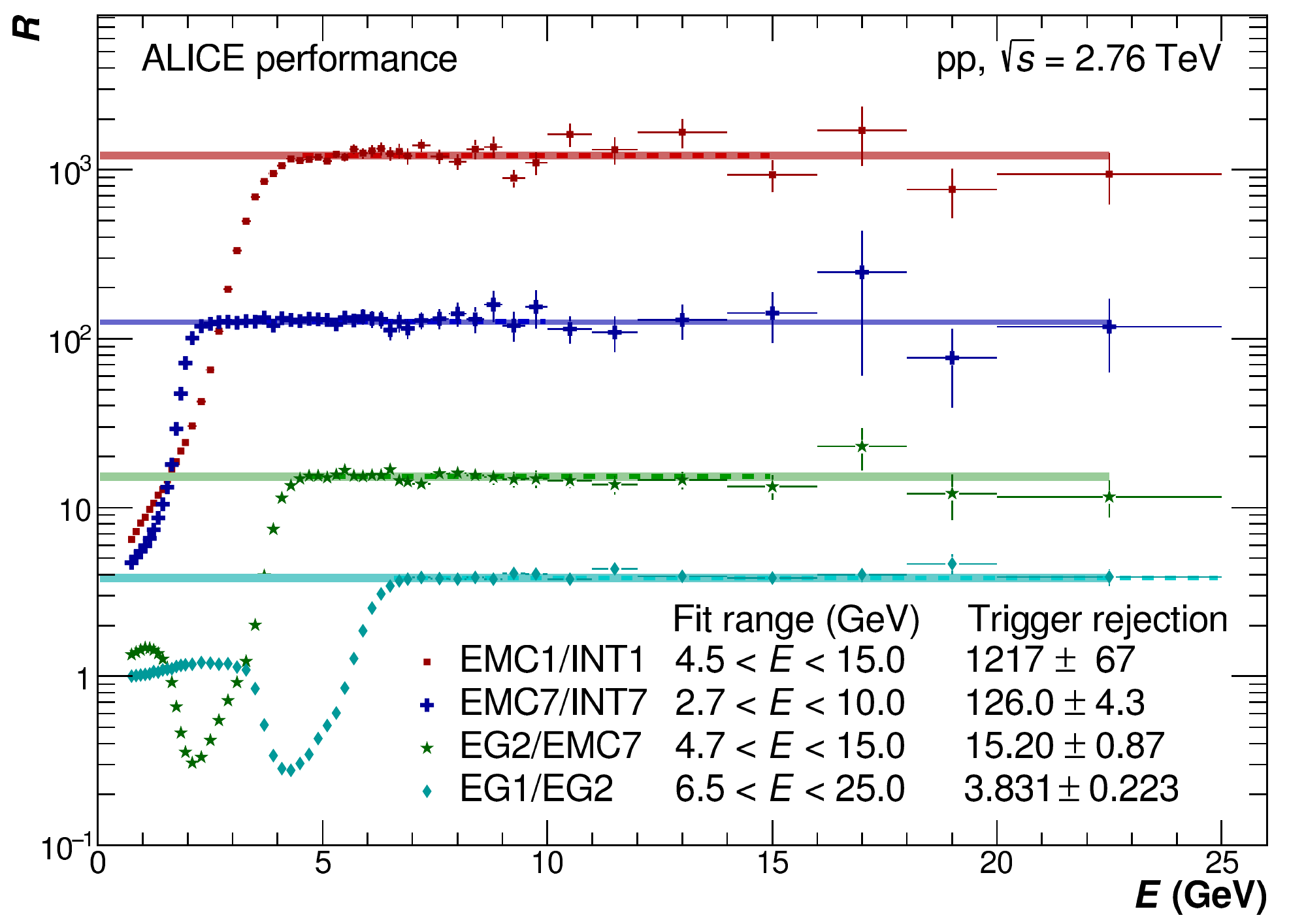}
  \caption{Energy dependence of ratios between cluster spectra for EMC1/INT1, EMC7/INT7, EG2/EMC7 and EG1/EG2. The trigger names INT1 and INT7 denote the minimum bias triggers MB$_{\rm OR}$ and MB$_{\rm AND}$ respectively. 
           The trigger names EMC1, EMC7, EG2 and EG1 denote the EMCal triggers at L0 in 2011 and 2013, and the EMCal triggers at L1 in 2013 with increasing threshold respectively. 
           The individual trigger rejection factors and their respective fit ranges in the plateau region are indicated as well. 
           The final rejection factors with respect to the minimum bias trigger are given in \Tab{tab:EventStat}.} 
  \label{fig:trignorm}
\end{figure}

In 2011, only the L0 trigger was used with one threshold~(EMC1), while in 2013, one L0~(EMC7) and two L1 triggers~(EG1, EG2) with different thresholds were used, as summarized in \Tab{tab:EventStat}.
The lower L1 trigger threshold in 2013 was set to approximately match the L0 threshold in 2011 for consistency. 
In case an event was associated with several triggers, the trigger with the lowest threshold was retained.

However, the thresholds are configured in the hardware via analog values, not actual units of energy. 
Their transformation into energy values directly depends on the energy calibration of the detector. 
For a reliable normalization of each trigger, the Trigger Rejection Factor ($\Rt$) is used.
The $\Rt$ takes into account a combination of the efficiency, acceptance and the downscaling of the respective triggers.
It can be obtained from the ratio~$R$ of the number of clusters reconstructed in EMCal triggered events to those in minimum bias events at high cluster energy $E$ where $R$ should be approximately constant~(plateau region), assuming the trigger does not affect the cluster reconstruction efficiency, but only the overall rate of clusters.
To reduce the statistical uncertainties on the normalization for the higher threshold triggers, $\Rt$ was always estimated with respect to the trigger with the next lower threshold in the EMCal or the respective minimum bias trigger if no lower EMCal trigger was available.
By consecutively multiplying the individual rejection factors up to the minimum bias trigger, the final $\Rt$ was obtained with respect to the minimum bias trigger.
The energy dependence of the ratios between cluster spectra of the relevant trigger combinations~(EMC1/INT1, EMC7/INT7, EG2/EMC7 and EG1/EG2) are shown in \Fig{fig:trignorm}.
At low $E$, there is a minimum at roughly the threshold of the lower-level trigger for EG2/EMC7 and EG1/EG2, while at high $E$ there is a pronounced plateau for every trigger combination.
The averages above the threshold in the plateau region, which represent $\Rt$ for the respective trigger combinations, are indicated by a line whose width represents the respective statistical uncertainty.
The corresponding systematic uncertainties were obtained by varying the range for the fit of the plateau region.
Finally, the values for the average trigger rejection factors above the threshold with respect to the corresponding minimum bias triggers are given in \Tab{tab:EventStat}. 
For the \PCMEMC\ and \EMC\ analyses, all available triggers were used, while for \mEMC\ only the EMC1, EG2 and EG1 triggers were included.
The collected integrated luminosities for minimum bias and EMCal triggers
\begin{equation}
L_{\rm int} = \frac{N_{\rm trig}}{\sigma_{\rm MB}}\,R_{\rm trig}\,,
\label{eq:Li}
\end{equation}
where $\sigma_{\rm MB}$ refers to $\sigma_{\rm MB_{OR}}$ for 2011 and $\sigma_{\rm MB_{AND}}$ for 2013, are summarized in \Tab{tab:EventStat}.
The statistical uncertainties on $\Rt$ are treated as systematic uncertainties on the integrated luminosity. 

\begin{table}[t!]
\centering  
 \begin{tabular}{rlllll}
  Year &  Trigger             & Trigger   & Approx.\      & Trigger rejection   & $L_{\text{int}}$\\ 
       &                      & name      & threshold     & factor (\Rt)        & $(\text{nb}^{-1})$\\ \hline
  2011 &  MB$_\mathrm{OR}$     & INT1      & $0$           & $1$                 & $0.524 \pm 0.010$\\ 
       &  EMCal L0            & EMC1      & $3.4$~GeV     & $1217 \pm 67$       & $13.8 \pm 0.806$\\ \hline
  2013 &  MB$_\mathrm{AND}$    & INT7      & $0$           & $1$                 & $0.335 \pm 0.013$\\ 
       &  EMCal L0            & EMC7      & $2.0$~GeV     & $126.0 \pm 4.3$     & $1.19 \pm 0.062$\\ 
       &  EMCal L1 (G2)       & EG2       & $3.5$~GeV     & $1959 \pm 131$      & $6.98 \pm 0.542$\\ 
       &  EMCal L1 (G1)       & EG1       & $5.5$~GeV     & $7743 \pm 685$      & $47.1 \pm 4.57$\\ \hline
 \end{tabular}
  \caption{Approximate trigger threshold and corresponding trigger rejection factor for EMCal triggers, as well as integrated luminosity for minimum bias and various EMCal triggers.} 
  \label{tab:EventStat}
\end{table}

Monte Carlo~(MC) samples were generated using PYTHIA8~\cite{Sjostrand:2007gs} and PHOJET~\cite{Engel:1995sb}.
The correction factors obtained independently from the two MC samples were found to be consistent, and hence combined.
For mesons with $\pt>5$~GeV/$c$, as in the triggered or merged cluster analyses, PYTHIA6~\cite{Sjostrand:2006za} simulations enriched with jets generated in bins of the hard scattering~(\ptHard) were used.
All MC simulations were obtained for a full ALICE detector description using the GEANT3~\cite{Brun:1987ma} framework and reconstructed with the same algorithms as for the data processing.  

\begin{figure}[t!]
  \centering
  \includegraphics[width=0.49\textwidth]{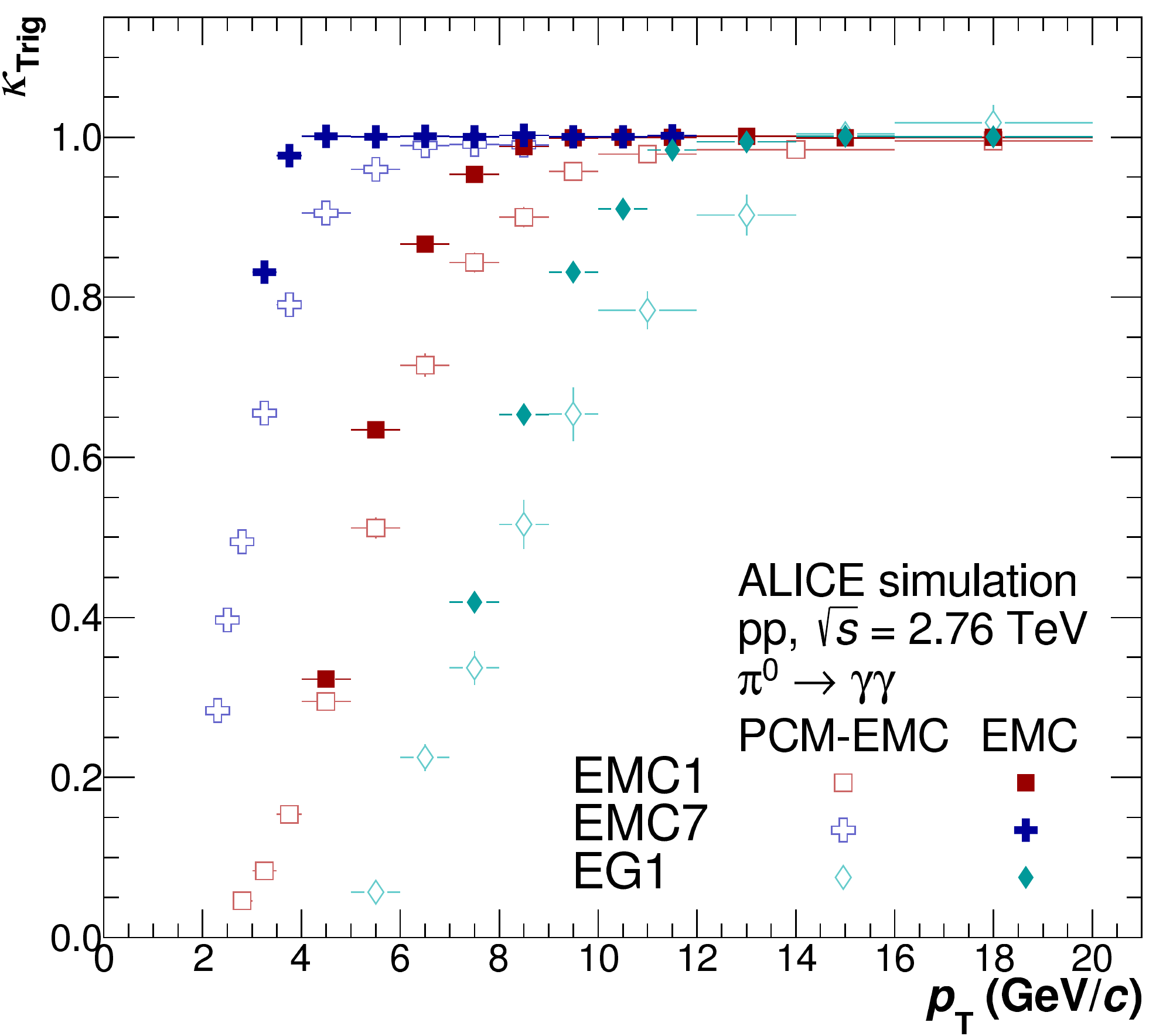}
  \includegraphics[width=0.49\textwidth]{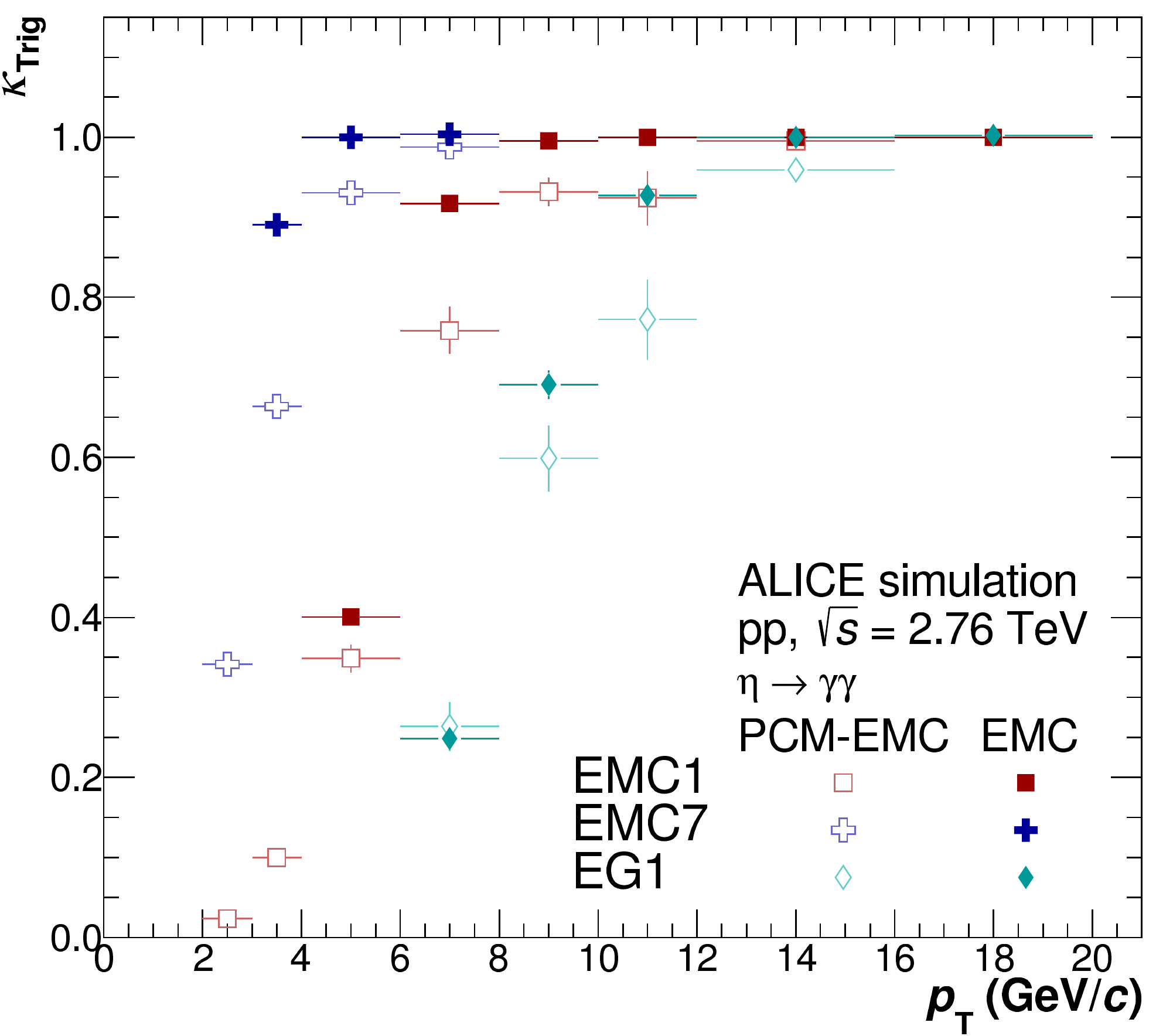}
  \caption{Efficiency bias $\kapt$ induced by different triggers (EMC1, EMC7 and EG1) for neutral pions~(left panel) and $\eta$ mesons (right panel) for \PCMEMC~(open symbols) and \EMC~(closed symbols).} 
  \label{fig:trigeff}
\end{figure}

The different triggers of the EMCal\com{, which significantly increase the sampled luminosity, } affect the properties of the reconstructible mesons, like the energy asymmetry~($\alpha=\frac{E_1-E_2}{E_1+E_2}$) of the decay photons, and hence significantly alter the reconstruction efficiency above the trigger threshold in the trigger turn-on region.
The efficiency biases $\kapt$ induced by the triggers were simulated using the approximate thresholds and their spread for different TRUs.
The bias was defined as the ratio of the $\pz$ or $\eta$ reconstruction efficiency in triggered events over that in minimum bias events.
\Figure{fig:trigeff} shows the $\pt$ dependence of $\kapt$ for different triggers and reconstruction methods for the $\pz$ and $\eta$ meson. 
While $\kapt$ is unity for the \mEMC\ analysis in the considered kinematic range, it is significantly below one for the \PCMEMC\ and \EMC\ neutral meson reconstruction, and reaches $\approx1$ only at about twice the trigger threshold. 
The corresponding correction factors are found to be larger for the \PCMEMC\ compared to the \EMC\ method, and larger for the $\eta$ than the $\pz$ meson.
This is a consequence of the much lower energy threshold imposed on the photons reconstructed with \PCM, which leads to wider opening angle and asymmetry distributions of the reconstructible mesons.
At low $\pt$, $\kapt$ also exhibits the effect of the trigger on subleading particles, for which the efficiency in triggered events is strongly reduced. 
However, the various triggers are only used if the meson momentum is at least 1.5 times the trigger threshold, thus the effect on the subleading particles is neglible.

In the offline analysis, only events with a reconstructed vertex with $|z_\mathrm{vtx}| < \unit[10]{cm}$ with respect to the nominal interaction vertex position along the beam direction were used. 
The finite primary vertex reconstruction efficiency for the MB$_\mathrm{OR}$(MB$_\mathrm{AND}$) trigger of about $0.92$ ($0.98$) is taken into account in the normalization of the respective minimum bias triggers.
Furthermore, only events with exactly one reconstructed vertex were accepted to remove pileup from in- and out-of-bunch collisions. 
While the in-bunch pileup is negligible after the vertex selection, the out-of-bunch pileup accumulating in the TPC due to its readout time of $\unit[90]{ms}$, needs to be subtracted statistically for the mesons measured with \PCM, as described in \Ref{Abelev:2014ypa}.
For the $\pz$~($\eta$) mesons reconstructed with \PCM\ the out-of-bunch pileup correction ranges from $20\%$~($9\%$) at low $\pt$ to about $3\%$ above $4$~\gevc.
Analyses involving the EMCal are not affected because contributions of clusters from different bunch crossings are suppressed by a suitable selection of clusters within a certain time window around the main bunch crossing.

\section{Neutral meson reconstruction}
\label{sec:reco}
Neutral mesons decaying into two photons fulfill
\begin{equation}
 M = \sqrt{2E_{1}E_{2}(1-\cos\theta_{12})}
 \label{eq:invariantmass}
\end{equation} 
where $M$ is the reconstructed mass of the meson, $E_{1}$ and $E_{2}$ are the measured energies of two photons, and $\theta_{12}$ is the opening angle between the photons measured in the laboratory frame. 
Photon candidates are measured either by a calorimeter or by \PCM.
Neutral meson candidates are then obtained by correlating photon candidates measured either by \EMC, \PHOS\ or \PCM\ exclusively, or by a combination of them (\PCMEMC).
The corresponding $\pz$ and $\eta$ meson measurements are described in \Sect{sec:diclusterPi0}.
The typical opening angle $\theta_{12}$ decreases with increasing $\pt$ of the meson due to the larger Lorentz boost. 
For $\pz$ mesons with $\pt$ above $5$--$6$ GeV/$c$, the decay photons become close enough so that their electromagnetic showers overlap in neighboring calorimeter cells of the EMCal.
At $\pt$ above $15$ GeV/$c$, the clustering algorithm can no longer efficiently distinguish the individual showers in the EMCal, and $\pz$ mesons can be measured by inspecting the shower shape of single clusters, referred to as ``merged'' clusters and explained in \Sect{sec:singlePi0}.

To be able to directly compare the reconstruction performances of the various measurement techniques and triggers, the invariant differential neutral meson cross sections were expressed as
\begin{equation}
E \frac{\dd^3 \sigma}{\dd p^3} = \frac{N_{\rm rec}}{\pt\,\Delta\pt\,\kapt\,\varepsilon}\,\frac{1}{L_{\rm int}}\,\frac{1}{\rm BR}
\label{eq:InvXsection}
\end{equation}
with the inverse of the normalized efficiency
\begin{equation}
\frac{1}{\varepsilon} = \frac{1}{2\pi\,A\,\Delta y\,}\frac{P}{\varepsilon_{\rm rec}}
\label{eq:effi}
\end{equation}
and integrated luminosity~(see \Eq{eq:Li}).
The measured cross sections were obtaind by correcting the reconstructed meson yield~$N_{\rm rec}$ for reconstruction efficiency~$\varepsilon_{\rm rec}$, purity~$P$ and acceptance~$A$, efficiency bias~$\kapt$, integrated luminosity~$L_{\rm int}$, as well as for the $\pt$ and $y$ interval ranges, $\Delta\pt$ and $\Delta y$, respectively, and the $\gamma\gamma$ decay branching ratio $BR$.
For invariant mass methods, the effect of reconstructed photon impurities on the meson purity are significantly reduced due to the subtraction of the combinatorial background, and hence the resulting meson impurities were neglected.
For the \mEMC\ method, the $\pz$ purity correction was obtained from MC simulations tuned to data. 
In the case of neutral pions, the contribution from secondary $\pzs$ was subtracted from $N_{\rm rec}$ before applying the corrections. 
The contribution from weak decays was estimated for the different methods by simulating the decays of the \kzs\ and $\Lambda$ using their measured spectra~\cite{Abelev:2014laa}, taking into account the reconstruction efficiencies, as well as resolution and acceptance effects for the respective daughter particles
The contribution from neutral pions produced by hadronic interactions in the detector material was estimated based on the full detector simulations using GEANT3.
Finally, the results were not reported at the center of the $\pt$ intervals used for the measurements, but following the prescription in \Ref{Lafferty:1994cj} at slightly lower $\pt$ values, in order to take into account the effect of the finite bin width $\Delta \pt$. 
The correction was found to be less than $1$\% in every $\pt$ interval for the $\pz$, and between $1$--$4$\% for the $\eta$ meson.

\subsection{Invariant mass analyses}
\label{sec:diclusterPi0}
Applying \Eq{eq:invariantmass}, the invariant mass distribution is obtained by correlating all pairs of photon candidates per event.
The neutral meson yield is then statistically extracted using the distinct mass line shape for identification of the signal and a model of the background.
In the following, only the new measurements are described. 
Details of the \PCM\ and PHOS $\pz$ measurements can be found in Refs.~\cite{Abelev:2012cn,Abelev:2014ypa}.

\begin{table}[t!]
\centering  
 \begin{tabular}{ll}
  \textbf{Track selection} \\\hline
  Track quality selection & $\pt > 0.05$ GeV/$c$ \\ 
                          & $N_{\rm TPC\ cluster}/N_{\rm reconstructible\ clusters}>0.6$ \\ 
                          & $|\eta|< 0.9$ \\
  Electron selection      & $-4<n\sigma_{\rm e}<5$\\
  Pion rejection          & $n\sigma_{\rm \pi}<1$ for $0.4< p <3.5$ GeV/$c$,\\
                          & $n\sigma_{\rm \pi}<0.5$ for $p >3.5$ GeV/$c$ (\PCM)\\
                          & $n\sigma_{\rm \pi}<1$ for $p>0.4$ GeV/$c$ (\PCMEMC)\\
  \textbf{Photon criteria} \\\hline
  Conversion point        & $|\EtaV|<0.9$ \\
                          & $5$~cm $<\RConv<180$~cm \\ 
                          & $|\ZConv|<240$~cm \\ 
                          & $0\leq|\PhiConv|\leq 2\pi$ \\ 
                          &$\text{cos}(\theta_{\text{point}})>0.85$ \\
  Photon quality          & $|\psip| < \psipm -\frac{\psipm}{\chitm} \chit$,\\
                          & with \psipm$~=~0.1$ and \chitm$~=~30$\\
  Armenteros-Podolanski   & $\qt < \qtm \sqrt{1 - \frac{\alpha^2}{\alpham^2}}$, \\
                          &  with $\qtm=0.05$ \gevc\ and $\alpham=0.95$ \\
 \hline
 \end{tabular}
  \caption{Criteria for photon candidate selection for \PCM.} 
  \label{tab:pcmcuts}
\end{table}

For the reconstruction of photons with \PCM, only tracks from secondary vertices without kinks with a minimum momentum of $0.05$~\gevc\ were taken into account. 
The tracks had to be reconstructed within the fiducial acceptance of the TPC and ITS and with at least $60$\% of the reconstructible track points in the TPC.
The photon momentum resolution is better than $1.5$\% at low $\pt$, resulting from the precise determination of the track momenta by the TPC.
Furthermore, the associated energy loss measured in the TPC was required to be within $-4 < n\sigma_{e} < 5$ of the electron expectation, where $n\sigma_{X} = (\dEdx - \avg{\dEdx_{X}})/\sigma_{X}$ with $\avg{\dEdx_{X}}$ and $\sigma_{X}$ the average energy loss and resolution for particle $X$, respectively.
The contamination from charged pions was suppressed by excluding all track candidates within $n\sigma_{\pi} < 1$ of the pion expectation.
The charged pion rejection was applied for track momenta between $0.4 < p < 3.5$~\gevc\ for \PCM\ and $p > 0.4$~\gevc\ for \PCMEMC, while for \PCM\ it was released to $n\sigma_{\pi} < 0.5$ above $p = 3.5$~\gevc.
Only conversions which were pointing to the primary vertex and could be reconstructed with a conversion point with $5<\RConv<180$~cm within the acceptance of the ITS and TPC were considered.
Compared to previous \PCM\ standalone measurements~\cite{Abelev:2014ypa}, the photon candidate selection criteria were optimized in order to reduce the combinatorial background. 
In particular, a two dimensional selection on the reduced $\chi^2$ of the photon conversion fit and the angle between the plane defined by the conversion pair and the magnetic field $|\psip|$ was introduced to suppress random $e^+e^-$ pairs.
Furthermore, the selection in the Armenteros-Podolanski variables~\cite{podolanski:1954} was tightened to reduce the contamination from \kzs\ and $\Lambda$ decays.
A summary of the conversion photon selection criteria is given in \Tab{tab:pcmcuts}.

Clusters in the EMCal were reconstructed by aggregating cells with $E_{\rm cell}>0.1$~GeV to a leading cell energy with at least $E_{\rm seed}>0.5$~GeV, and were required to have only one local maximum.
Photon candidates were obtained from reconstructed clusters by requiring a cluster energy of $0.7$~GeV to ensure acceptable timing and energy resolution and to remove contamination from minimum-ionizing~($\lsim 300$~MeV) and low-energy hadrons.
Furthermore, a cluster had to contain at least two cells to ensure a minimum cluster size and to remove single cell electronic noise fluctuations.
Clusters which could be matched to a track propagated to the average shower depth in the EMCal~(at $440$~cm) within $|\Delta\eta|$ and $|\Delta\varphi|$ criteria that depend on track $\pt$ as given in \Tab{tab:emccuts}, were rejected to further reduce contamination by charged particles. 
The track-to-cluster matching efficiency amounts to about $97$\% for primary charged hadrons at cluster energies of $E_{\rm clus} > 0.7$ GeV, decreasing slowly to $92$\% for clusters of $50$ GeV. 
The removal of matched tracks is particularly important for the \PCMEMC\ method as otherwise a severe auto-correlation between the clusters originating from one of the conversion electrons and the conversion photon would be introduced.
Such auto-correlated pairs strongly distort the shape of the invariant mass distribution between the $\pz$ and $\eta$ mass peak region. 
The standard track matching applied to each conversion leg allowed for the removal of these auto-correlation pairs with an efficiency of more than $99\%$ since the corresponding track was already found.
An additional distinction between clusters from mainly photons, electrons and neutrons is based on their shower shape.
The shower shape can be characterized by the larger eigenvalue squared of the cluster's energy decomposition in the EMCal $\eta$--$\varphi$ plane.
It is expressed as 
\begin{equation}
 \lzt = 0.5 \left( \sigma_{\varphi\varphi}^2+\sigma_{\eta\eta}^2+\sqrt{(\sigma_{\varphi\varphi}^2-\sigma_{\eta\eta}^2)^2 + 4\sigma^4_{\varphi\eta}} \right)
 \label{eq:lambda02}
\end{equation}
where $\sigma^2_{xz}=\left<x\,z\right> - \left<x\right>\left<z\right>$ and $\left<x\right>=\frac{1}{w_{\rm tot}}\sum w_i x_i$ are weighted over all cells associated with the cluster in the $\varphi$ or $\eta$ direction.
The weights $w_i$ logarithmically depend on the ratio of the energy of a given cell to the cluster energy, as $w_i=\max \left(0,4.5+\log E_i/E\right)$, and $w_{\rm tot}=\sum w_i$~\cite{AWES1992130}. 
Nuclear interactions, in particular for neutrons, create an abnormal signal when hitting the corresponding avalanche photodiodes for the readout of the scintillation light\com{ and typically generate a response}.
Such a signal is mainly localized in one high-energy cell with a few surrounding low-energy cells, and can be removed by requiring $\lzt>0.1$.
While the showers from electrons and photons tend to be similar, they can be distinguished based on their elongation, as most of the low-$\pt$ electrons will hit the EMCal surface at an angle due to the bending in the magnetic field. 
Most of the pure photons are reconstructed with a $\lzt\approx0.25$; only late conversions elongate the showers beyond this. 
Thus, rejecting clusters with $\lzt > 0.7$~$(0.5)$ for \EMC~(\PCMEMC) rejects the contamination from late conversion electrons significantly. 
At very high transverse momenta~($> 10$~GeV/$c$), it also rejects part of the contamination from neutral pions for which both photons have been reconstructed in a single cluster. 
Contributions of clusters from different bunch crossings were suppressed by a suitable selection of clusters within a certain time window around the main bunch crossing.
A summary of the selection criteria for EMCal photon candidates is given in \Tab{tab:emccuts}.

\begin{table}[t!]
\centering  
 \begin{tabular}{ll}
  \textbf{Cluster reconstruction} & \\ \hline
  Minimum cell energy         & $E_{\rm cell}>0.1$ GeV\\
  Minimum leading cell energy & $E_{\rm seed}>0.5$ GeV\\
  \textbf{Cluster selection} \\ \hline
  Selection in $\eta$     & $\abs{\eta}<0.67$, $\unit[1.40]{rad}< \varphi < \unit[3.15]{rad}$ \\
  Minimum cluster energy  & $E_{\rm clus} > 0.7$ GeV \\ 
  Minimum number of cells & $N_{\rm cells}\geq2$ \\ 
  Cluster-shape parameter & $0.1<$ $\lzt < 0.5$ (\PCMEMC) \\ 
                          & $0.1<$ $\lzt < 0.7$ (\EMC) \\ 
                          & $\lzt > 0.27$ (\mEMC) \\ 
  Cluster time            & $|t_{\rm clus}|\leq 50$~ns (2011) \\
                          & $-35$~ns~$<t_{\rm clus}<30$~ns (2013) \\
  Cluster--track matching & $|\Delta\eta| \leq 0.010 + \left({\pt + 4.07}\right)^{-2.5}$ \\ 
                          & $|\Delta\varphi| \leq 0.015 + \left({\pt + 3.65}\right)^{-2}$\\
  \hline  
 \end{tabular}
  \caption{Criteria for photon candidate selection for EMCal-based methods.}
  \label{tab:emccuts}
\end{table} 

The good momentum resolution for the \PCM\ photon was exploited to derive an improved correction for the relative energy scale, as well as for the residual misalignment of the EMCal between data and simulation.
The neutral pion mass was evaluated for the \PCMEMC\ method as a function of the EMCal photon energy for data and simulation. 
A correction for the cluster energy was deduced which for a given simulation adjusts the neutral pion mass peak position to the measured position in the data as a function of the cluster energy.
Above $1$~GeV, the corrections for the various MC datasets are typically about 3\%. 

\ifmassfullsub
  \begin{figure}[t!]
    \includegraphics[width=0.45\textwidth]{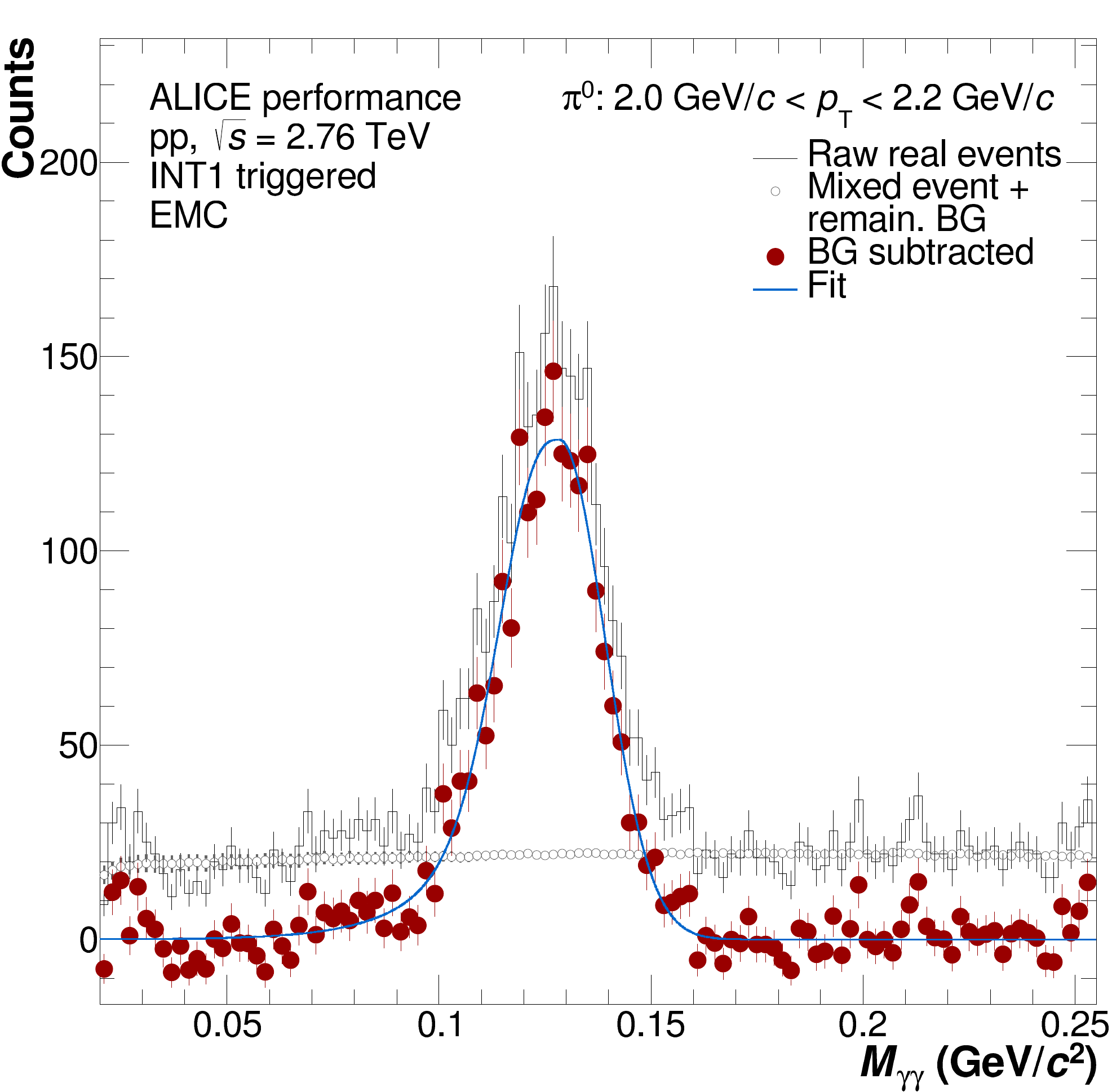}\hspace{0.3cm}
    \includegraphics[width=0.45\textwidth]{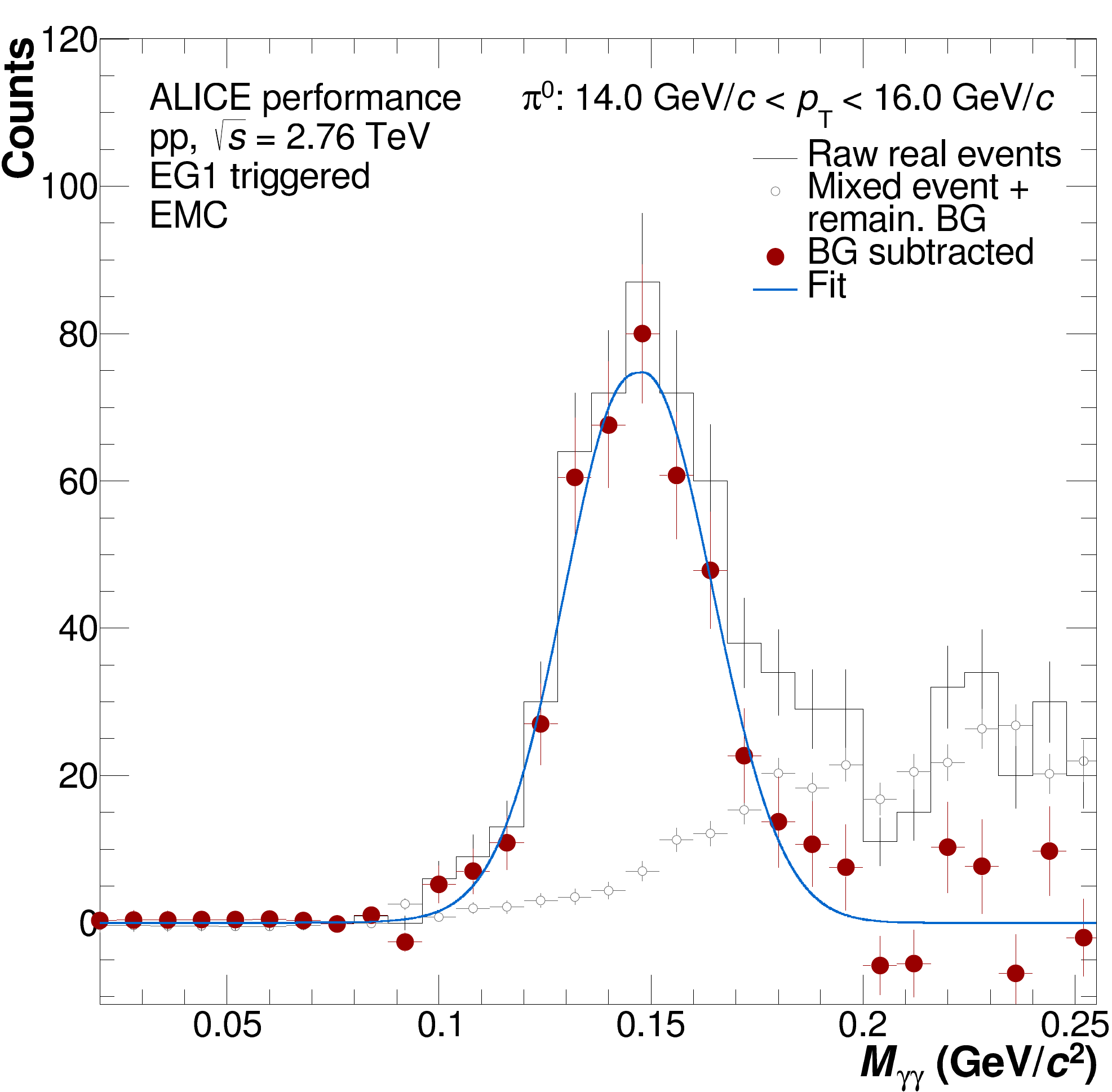}\\
    \includegraphics[width=0.45\textwidth]{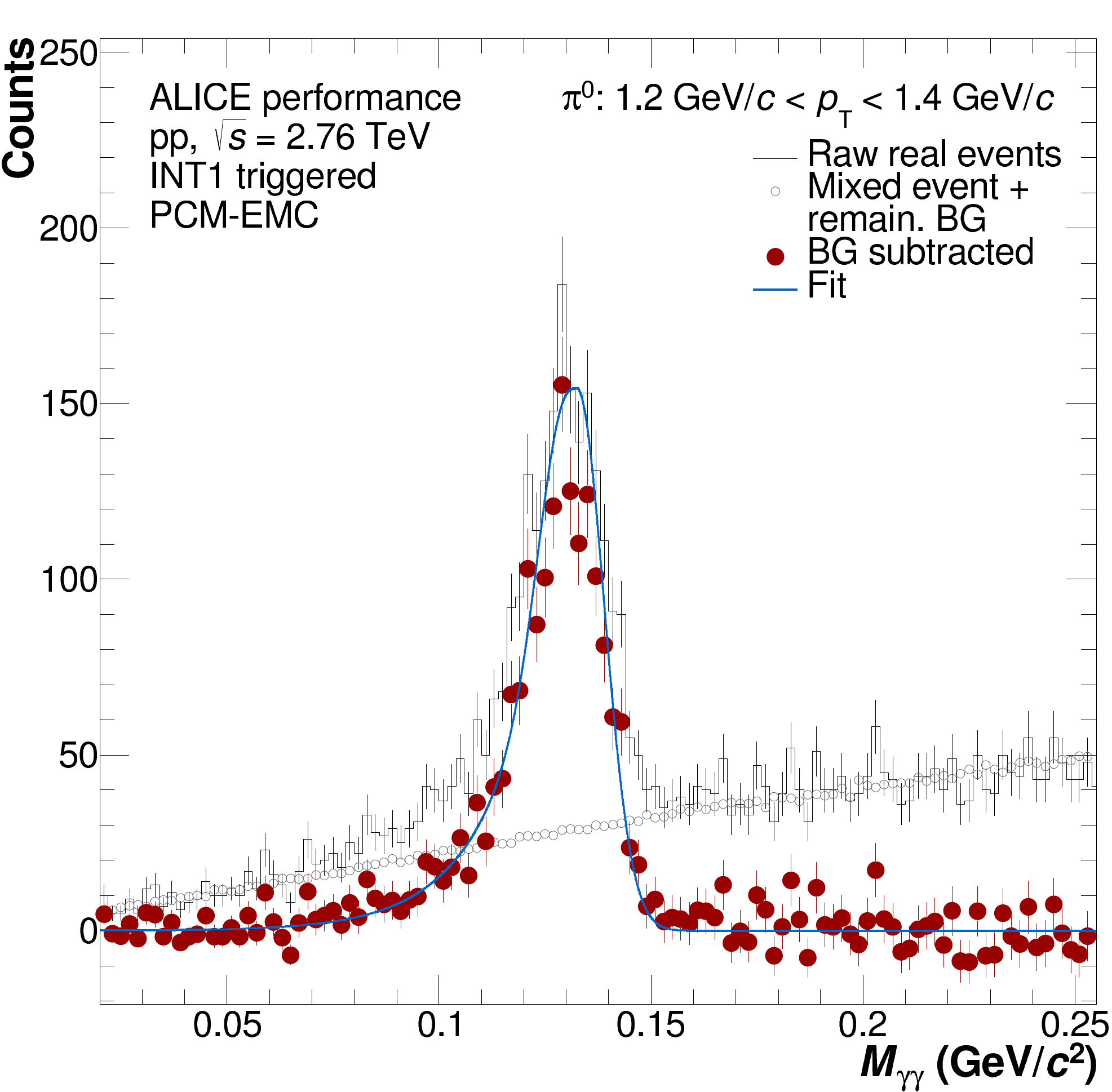}\hspace{0.3cm}
    \includegraphics[width=0.45\textwidth]{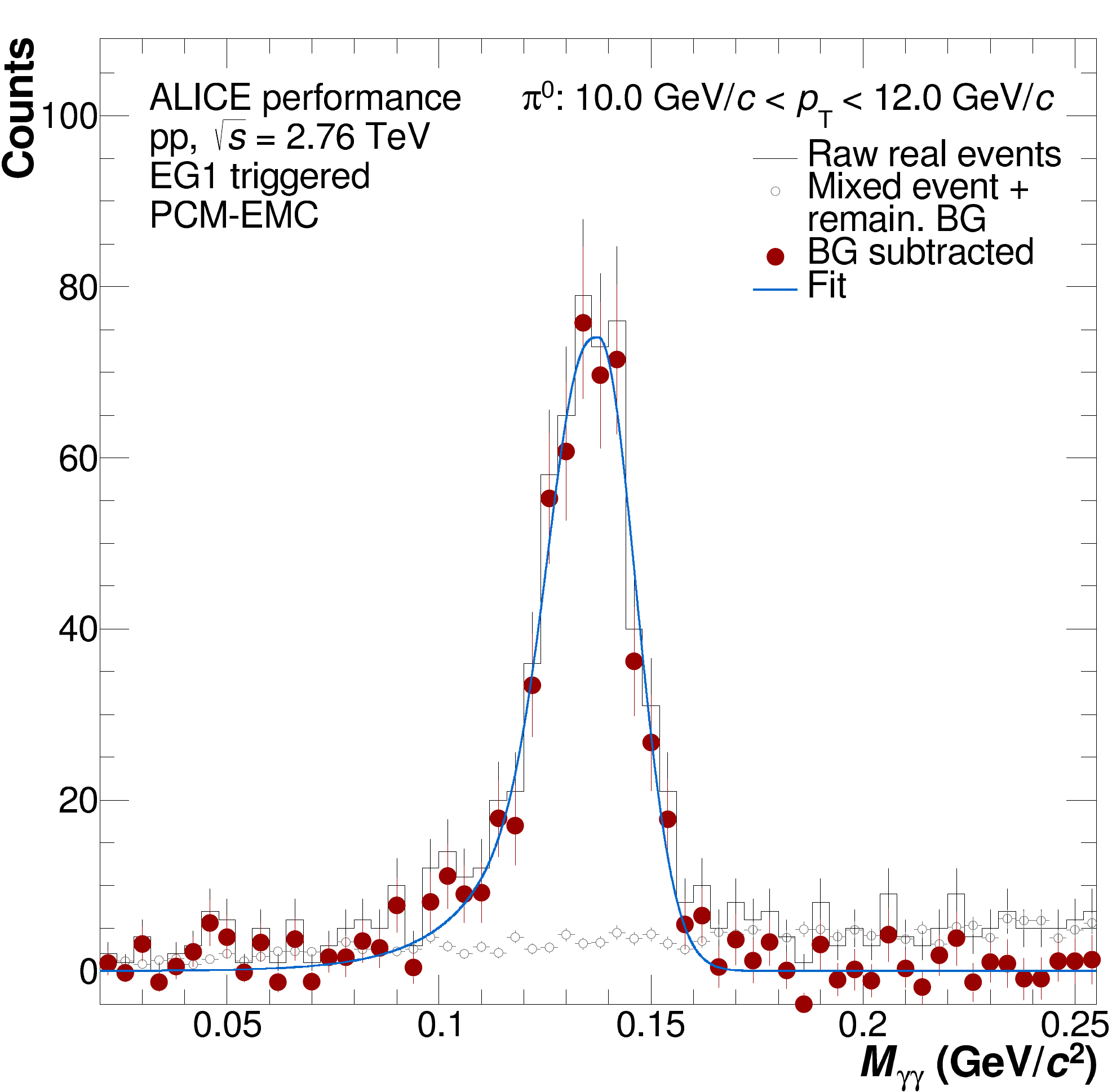}\\
    \caption{Invariant mass distributions in the $\pz$ peak region for INT1~(left panels) and EG1~(right panels) triggers and \EMC~(top panels) and \PCMEMC~(bottom panels) methods.} 
    \label{fig:pi0masspeaks}
  \end{figure}
  \begin{figure}[t!]
    \includegraphics[width=0.45\textwidth]{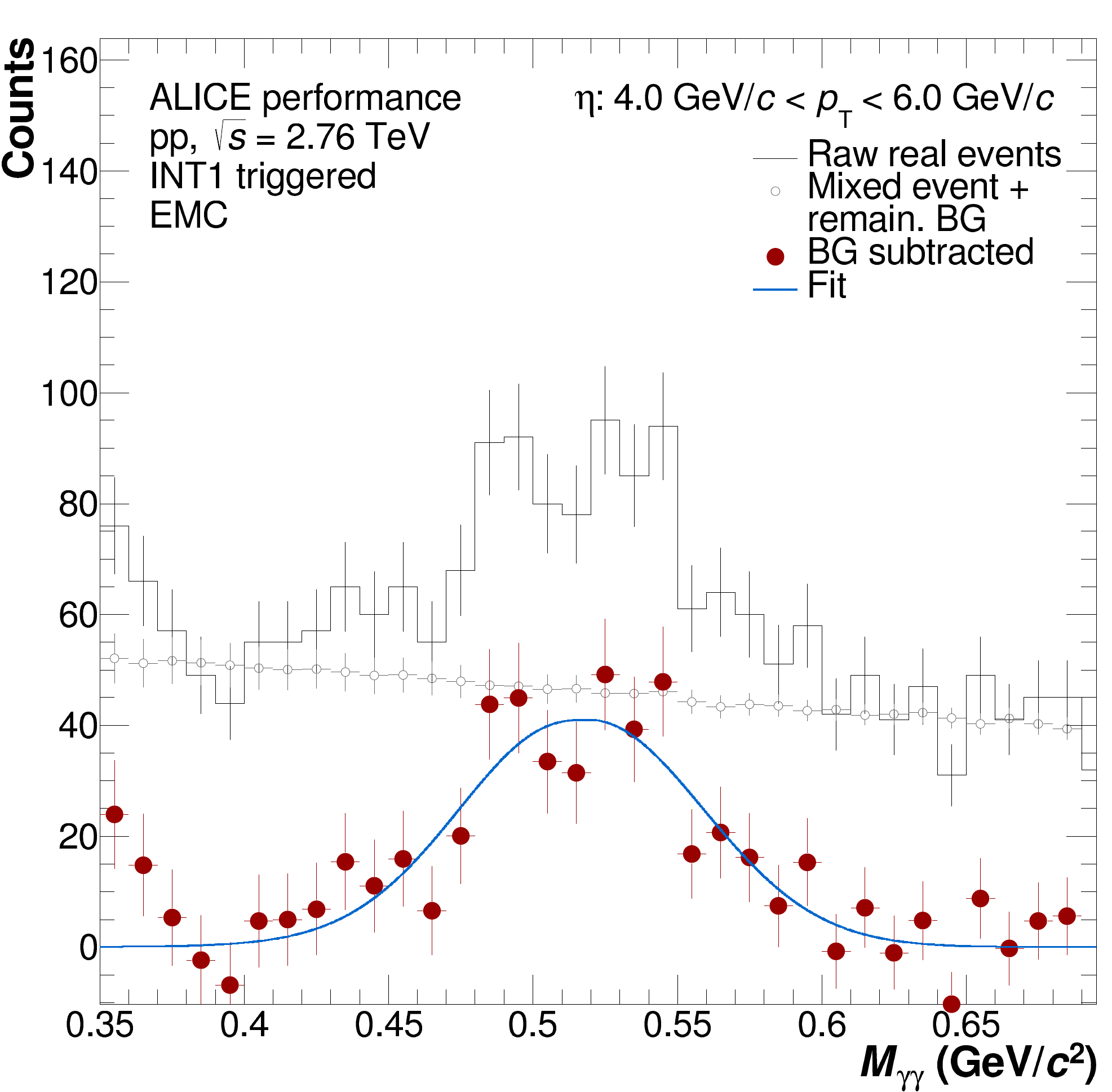}\hspace{0.3cm}
    \includegraphics[width=0.45\textwidth]{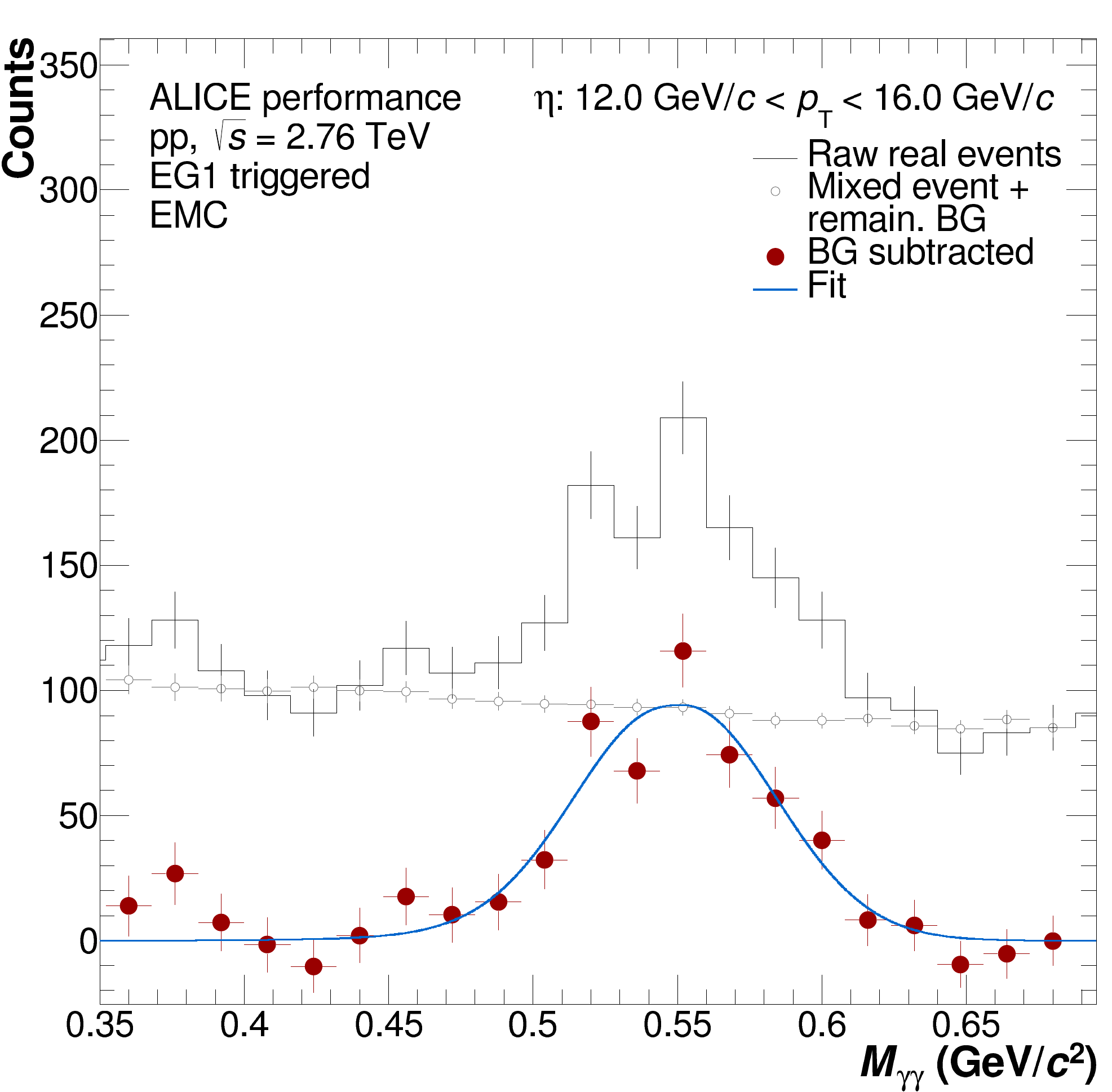}\\
    \includegraphics[width=0.45\textwidth]{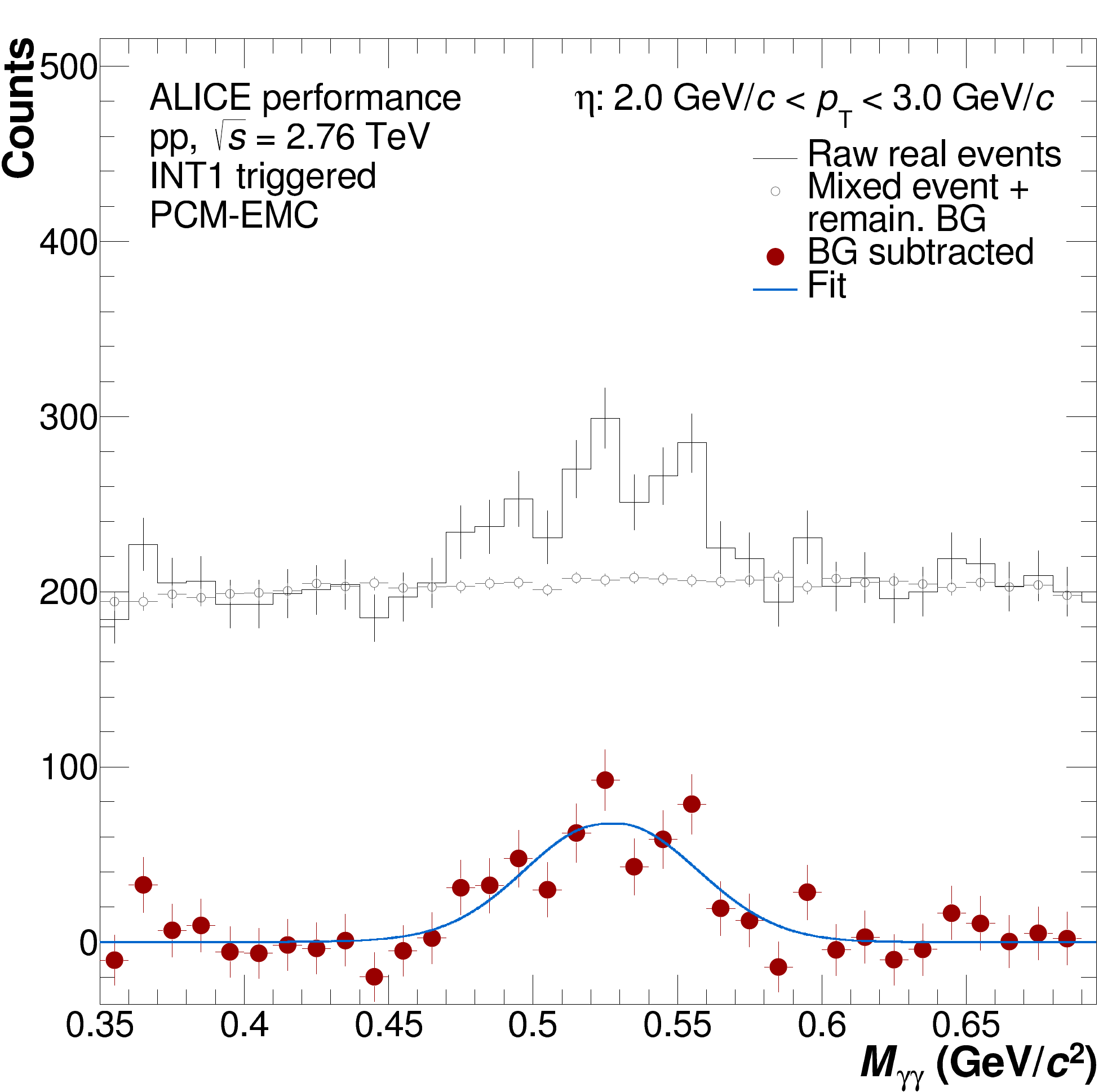}\hspace{0.3cm}
    \includegraphics[width=0.45\textwidth]{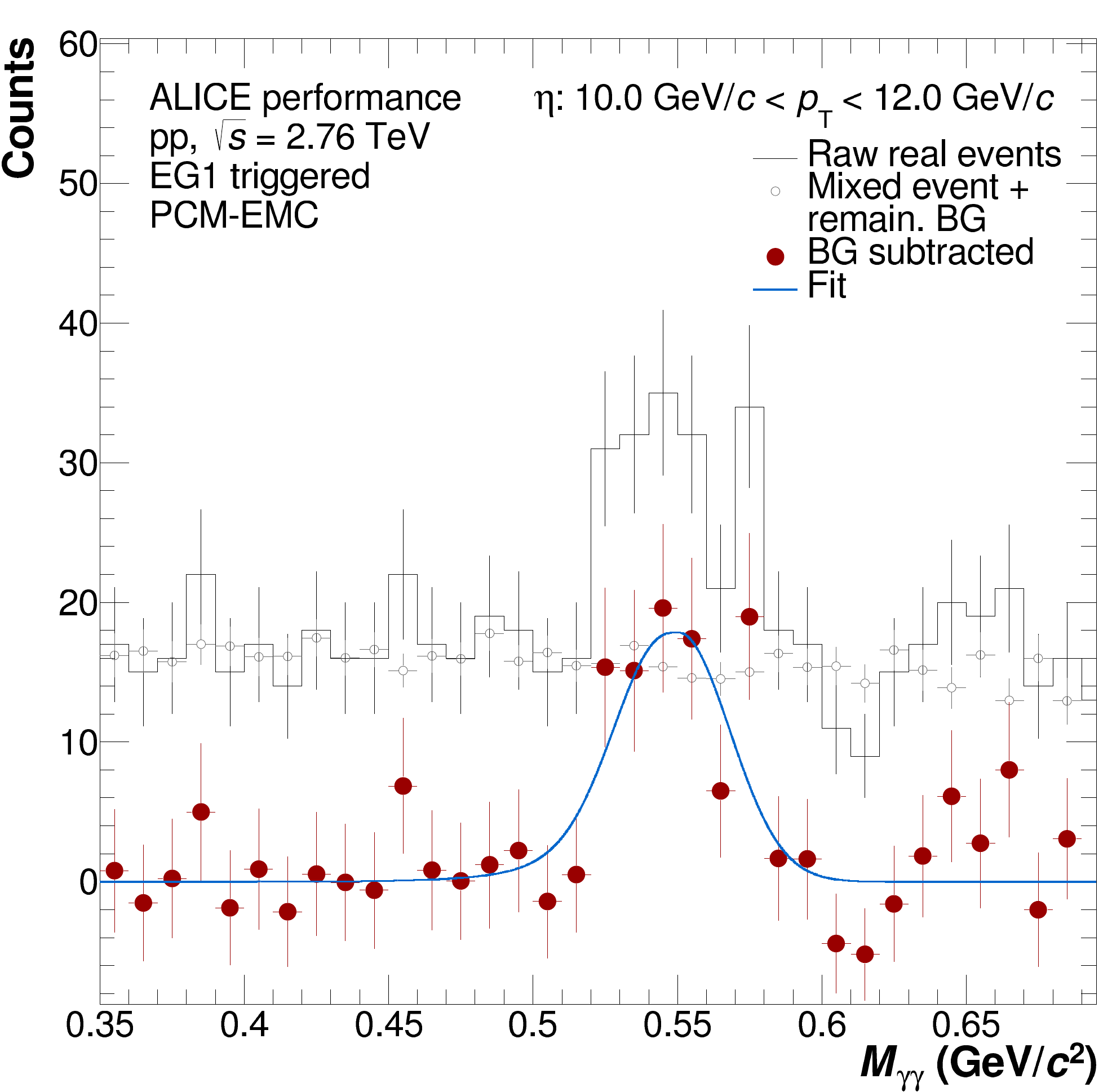}
    \caption{Invariant mass distributions in the $\eta$ peak region for INT1~(left panels) and EG1~(right panels) triggers and \EMC~(top panels) and \PCMEMC~(bottom panels) methods.} 
    \label{fig:etamasspeaks}
  \end{figure}
\fi
\ifmasspartsub
  \begin{figure}[t!]
    \includegraphics[width=0.45\textwidth]{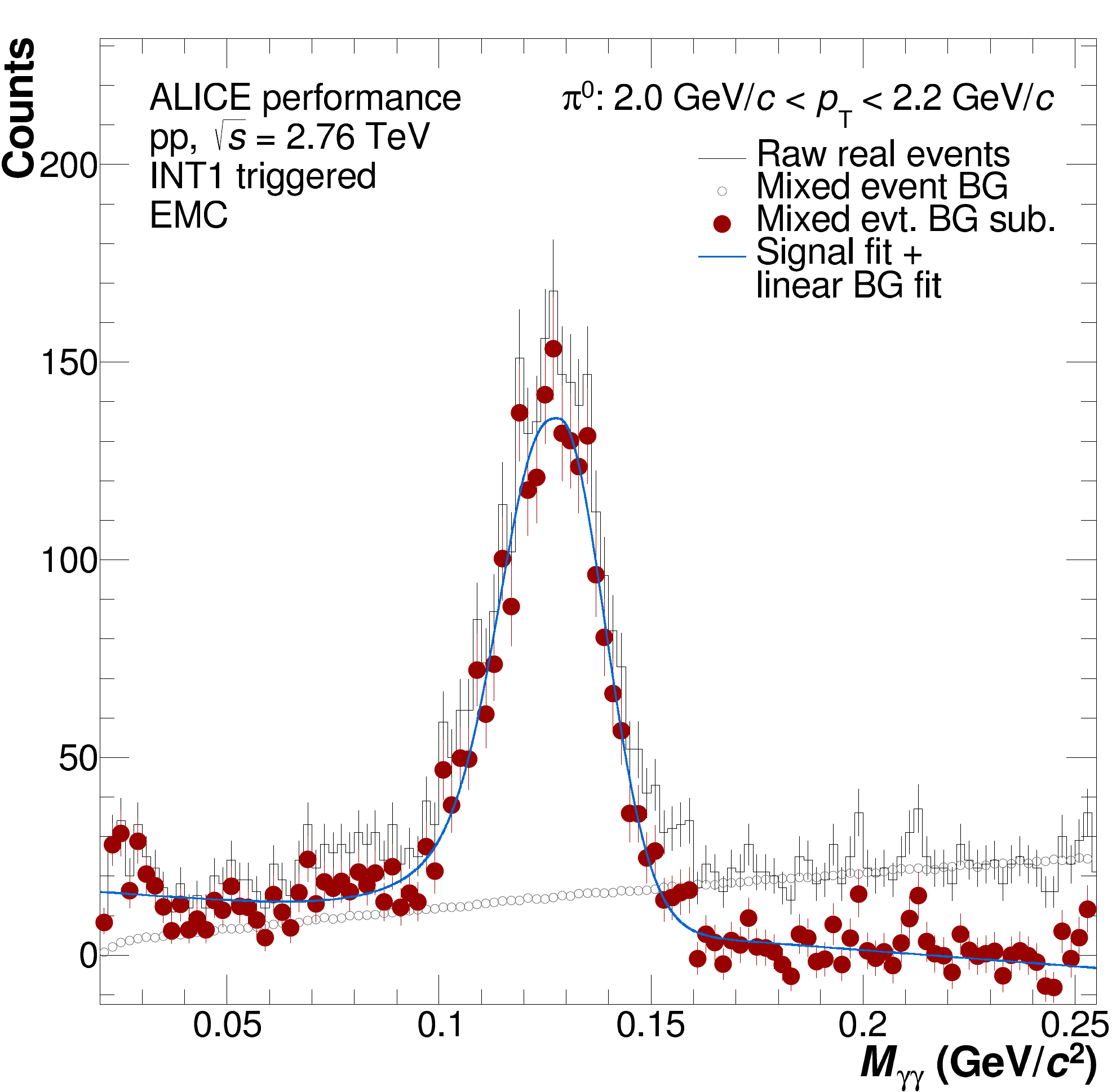}\hspace{0.3cm}
    \includegraphics[width=0.45\textwidth]{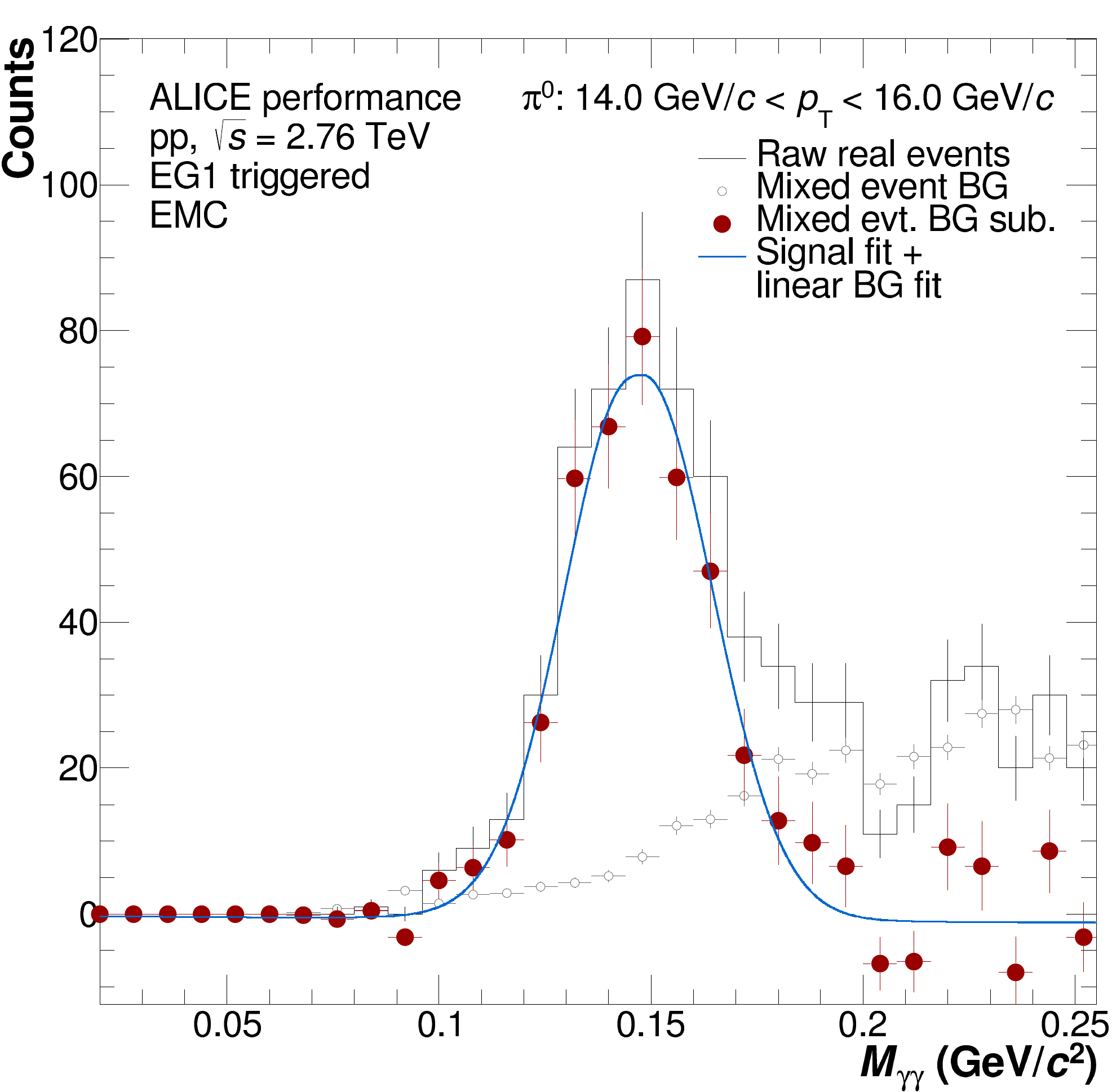}\\
    \includegraphics[width=0.45\textwidth]{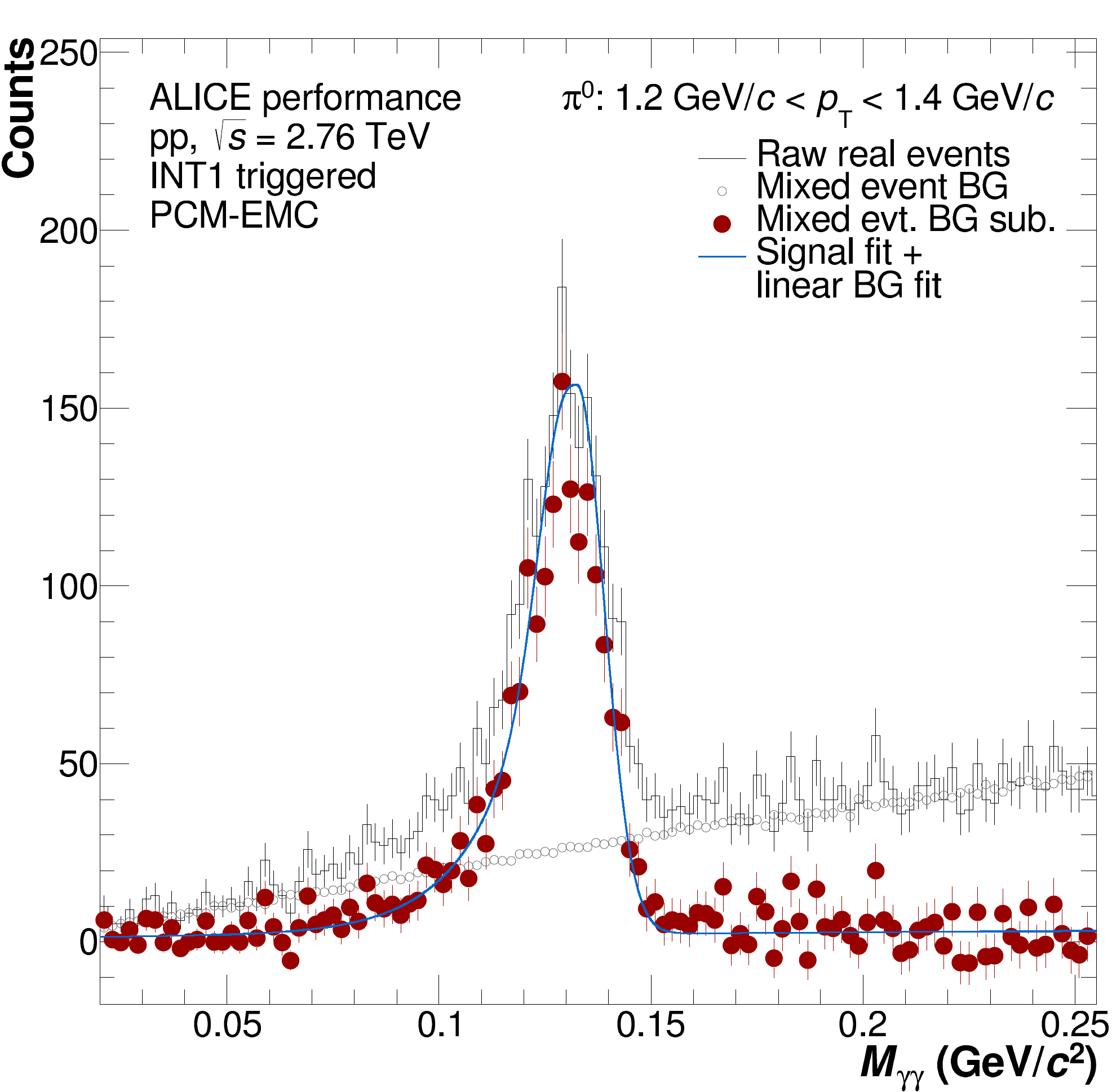}\hspace{0.3cm}
    \includegraphics[width=0.45\textwidth]{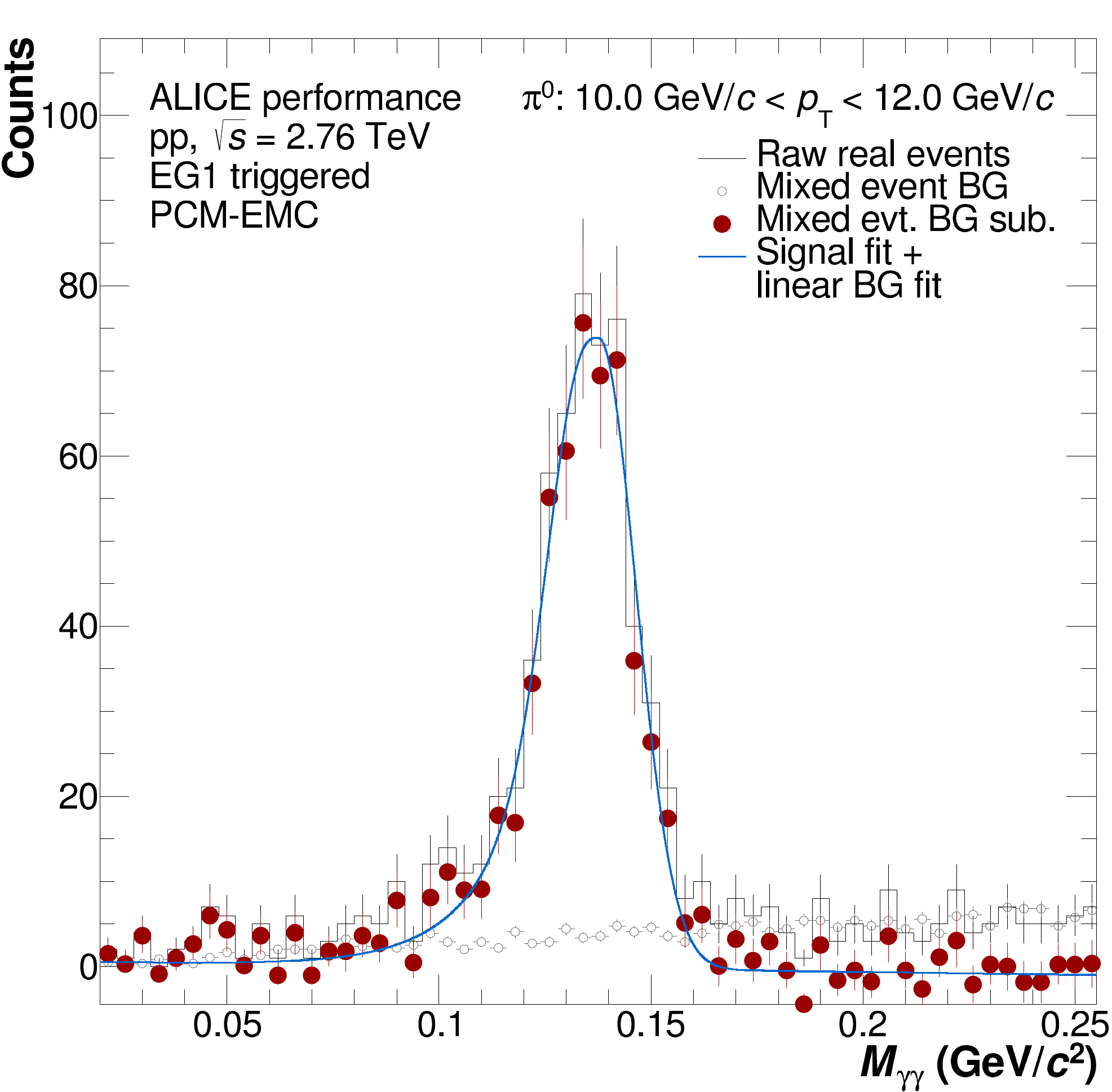}\\
    \caption{Invariant mass distributions in the $\pz$ peak region for INT1~(left panels) and EG1~(right panels) triggers and \EMC~(top panels) and \PCMEMC~(bottom panels) methods.} 
    \label{fig:pi0masspeaks}
  \end{figure}
  \begin{figure}[t!]
    \includegraphics[width=0.45\textwidth]{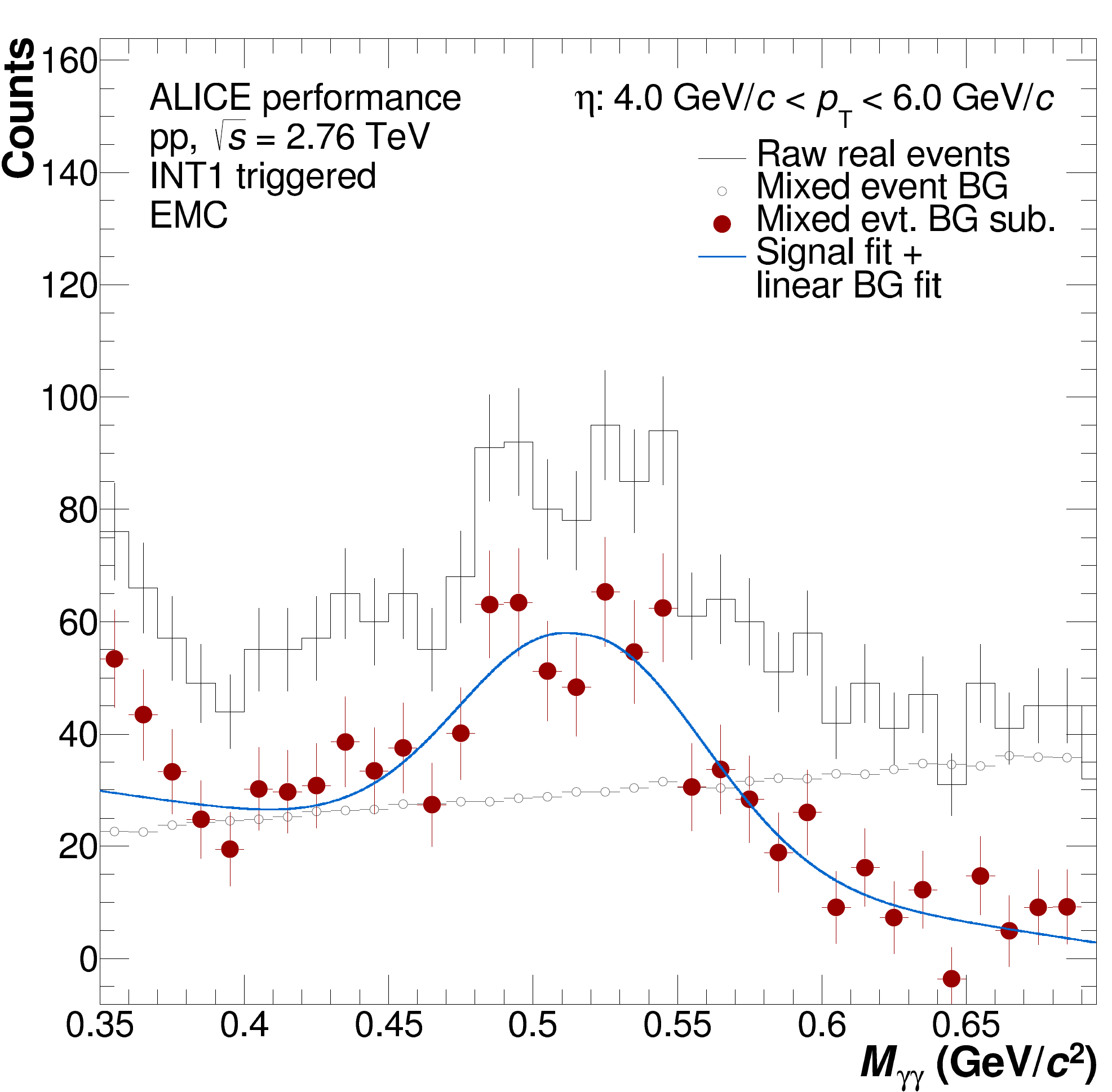}\hspace{0.3cm}
    \includegraphics[width=0.45\textwidth]{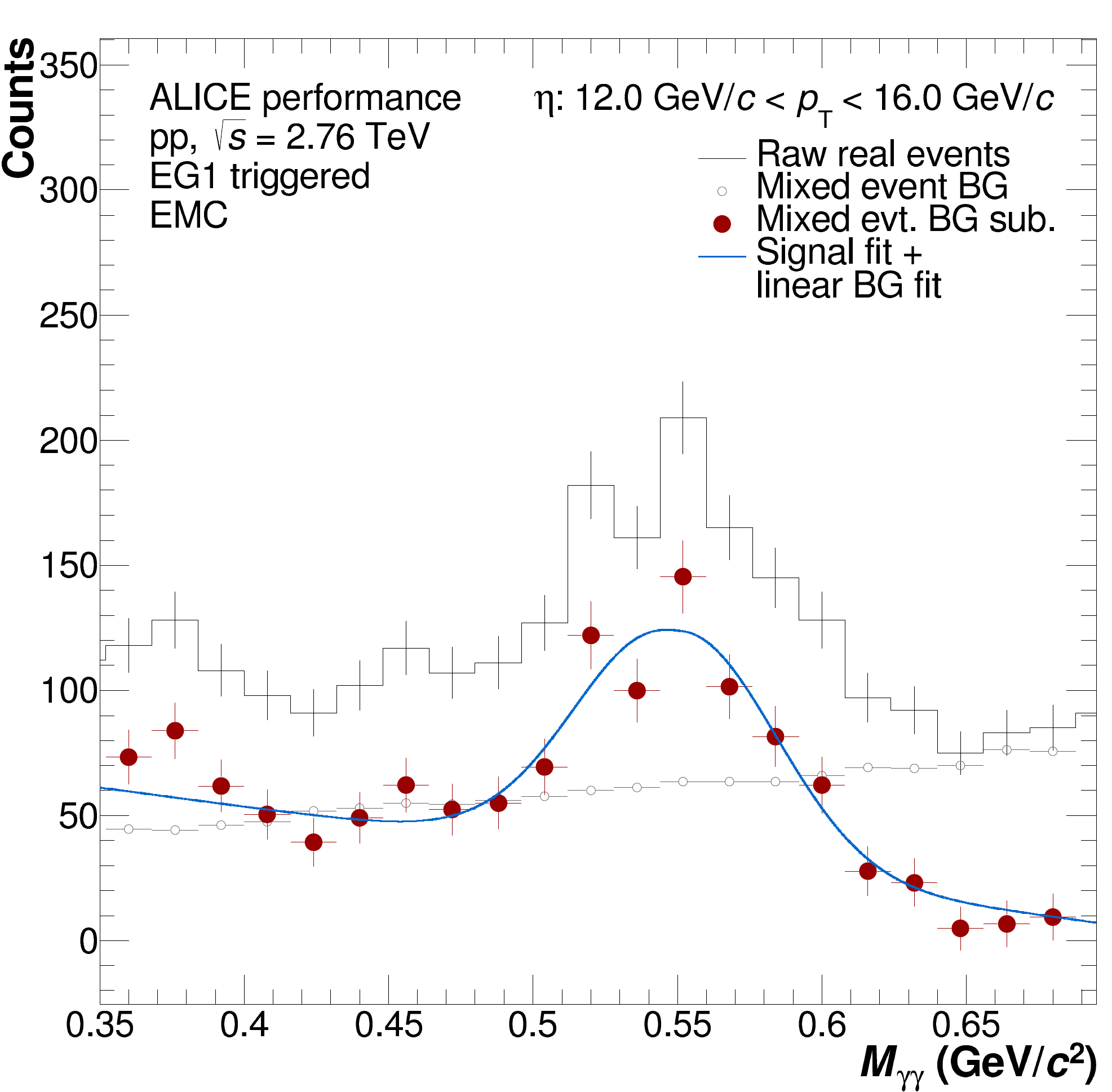}\\
    \includegraphics[width=0.45\textwidth]{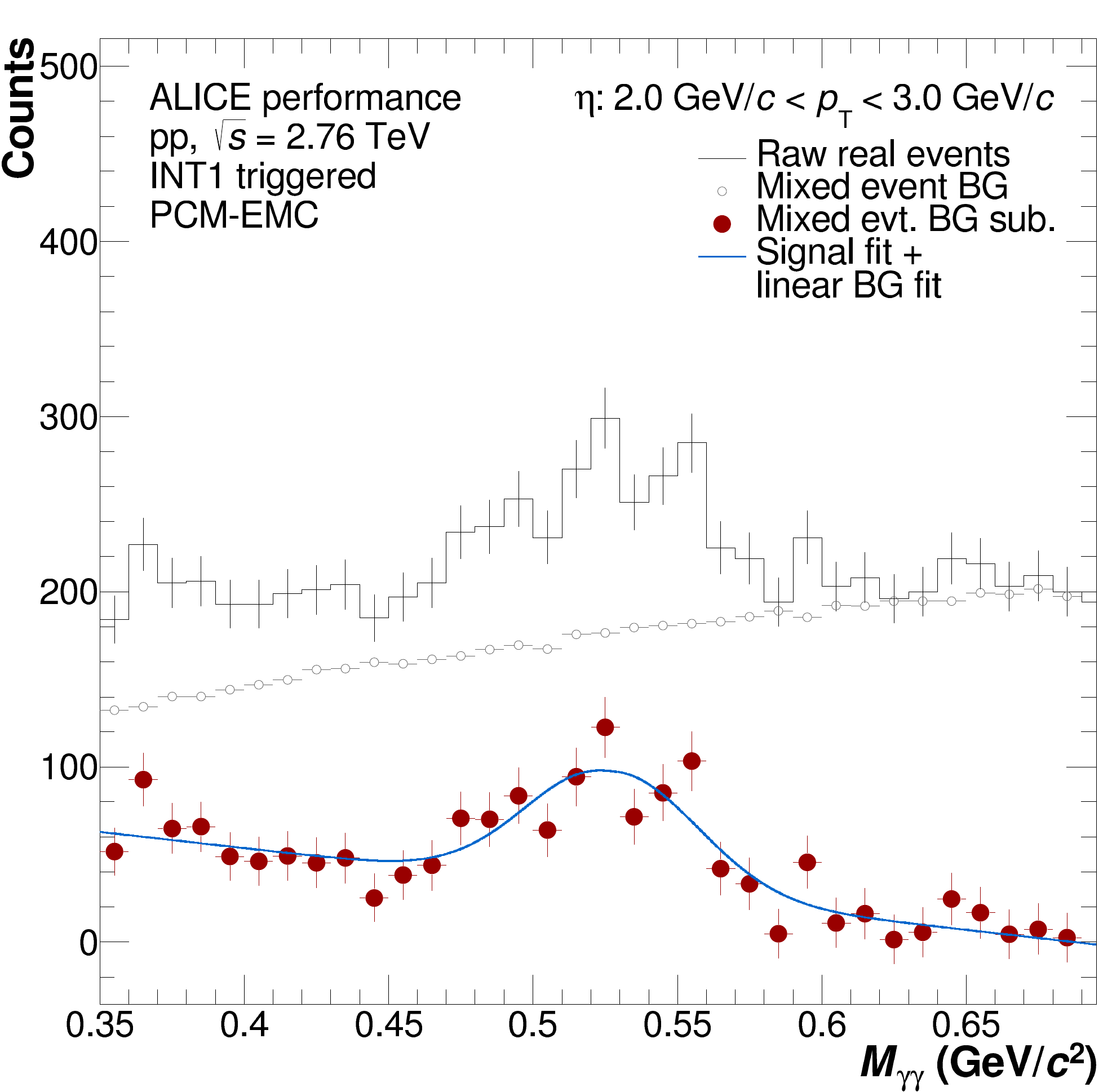}\hspace{0.3cm}
    \includegraphics[width=0.45\textwidth]{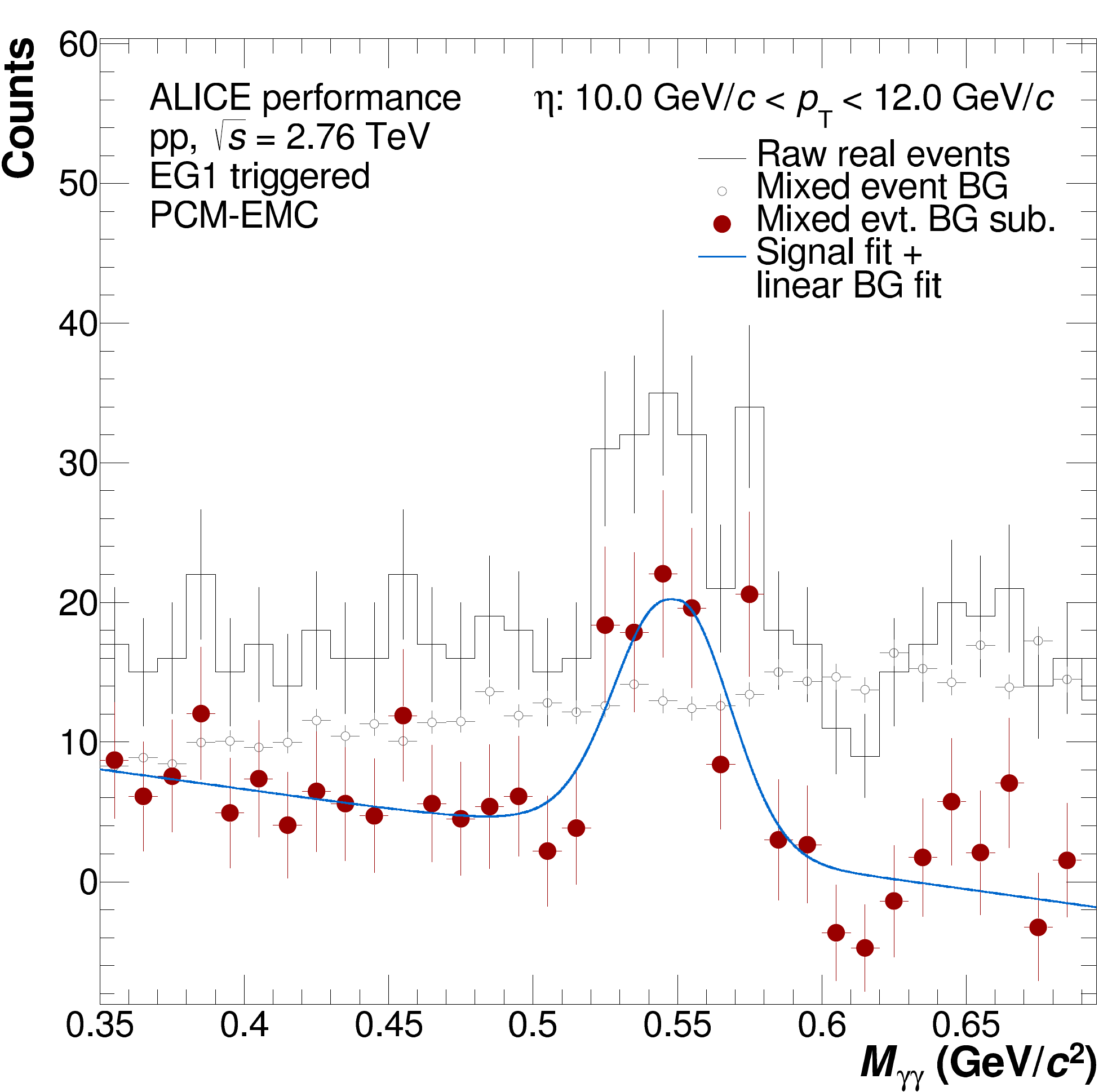}
    \caption{Invariant mass distributions in the $\eta$ peak region for INT1~(left panels) and EG1~(right panels) triggers and \EMC~(top panels) and \PCMEMC~(bottom panels) methods.} 
    \label{fig:etamasspeaks}
  \end{figure}    
\fi
\ifmasssepbg
  \begin{figure}[t!]
    \includegraphics[width=0.45\textwidth]{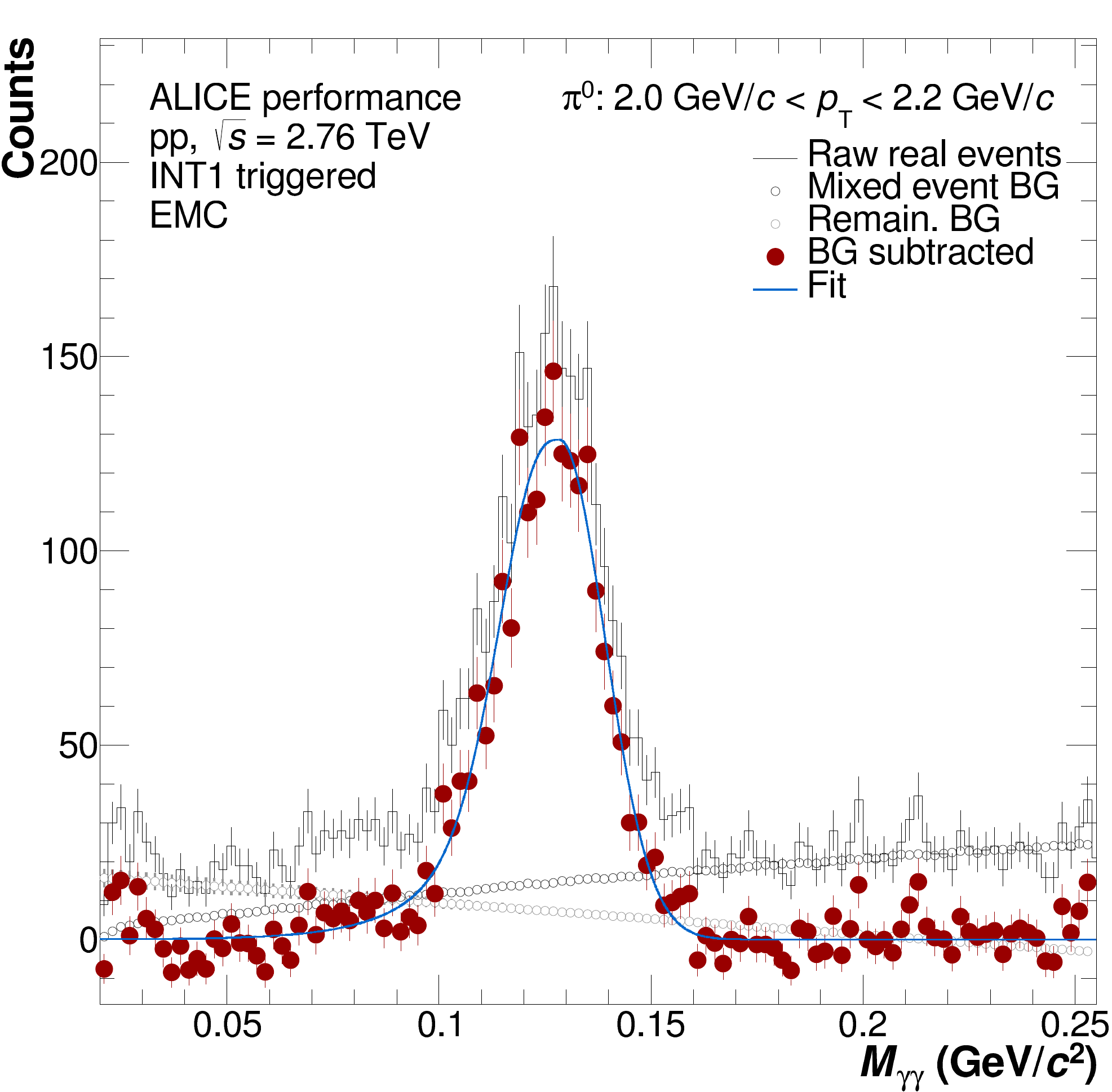}\hspace{0.3cm}
    \includegraphics[width=0.45\textwidth]{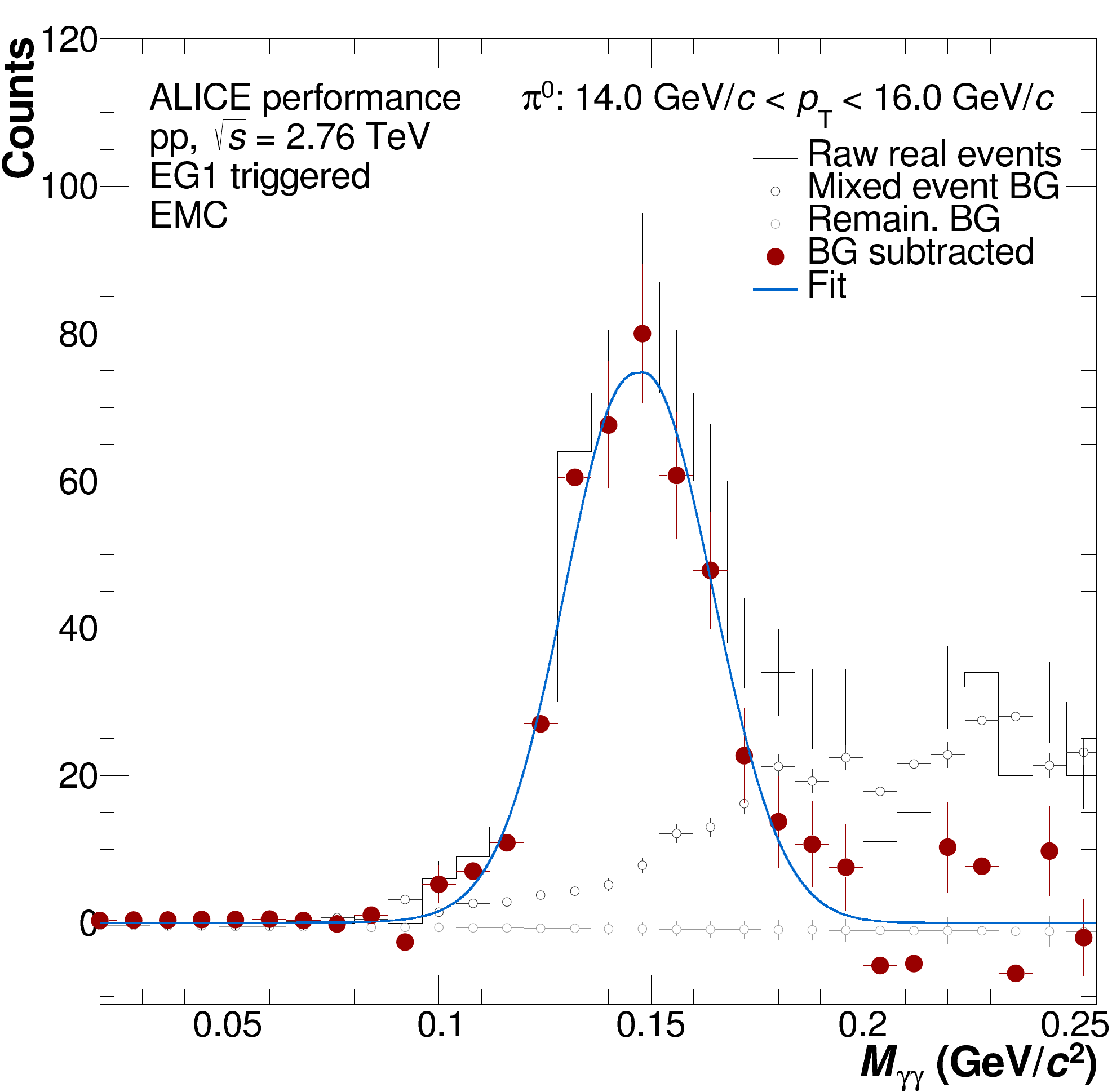}\\
    \includegraphics[width=0.45\textwidth]{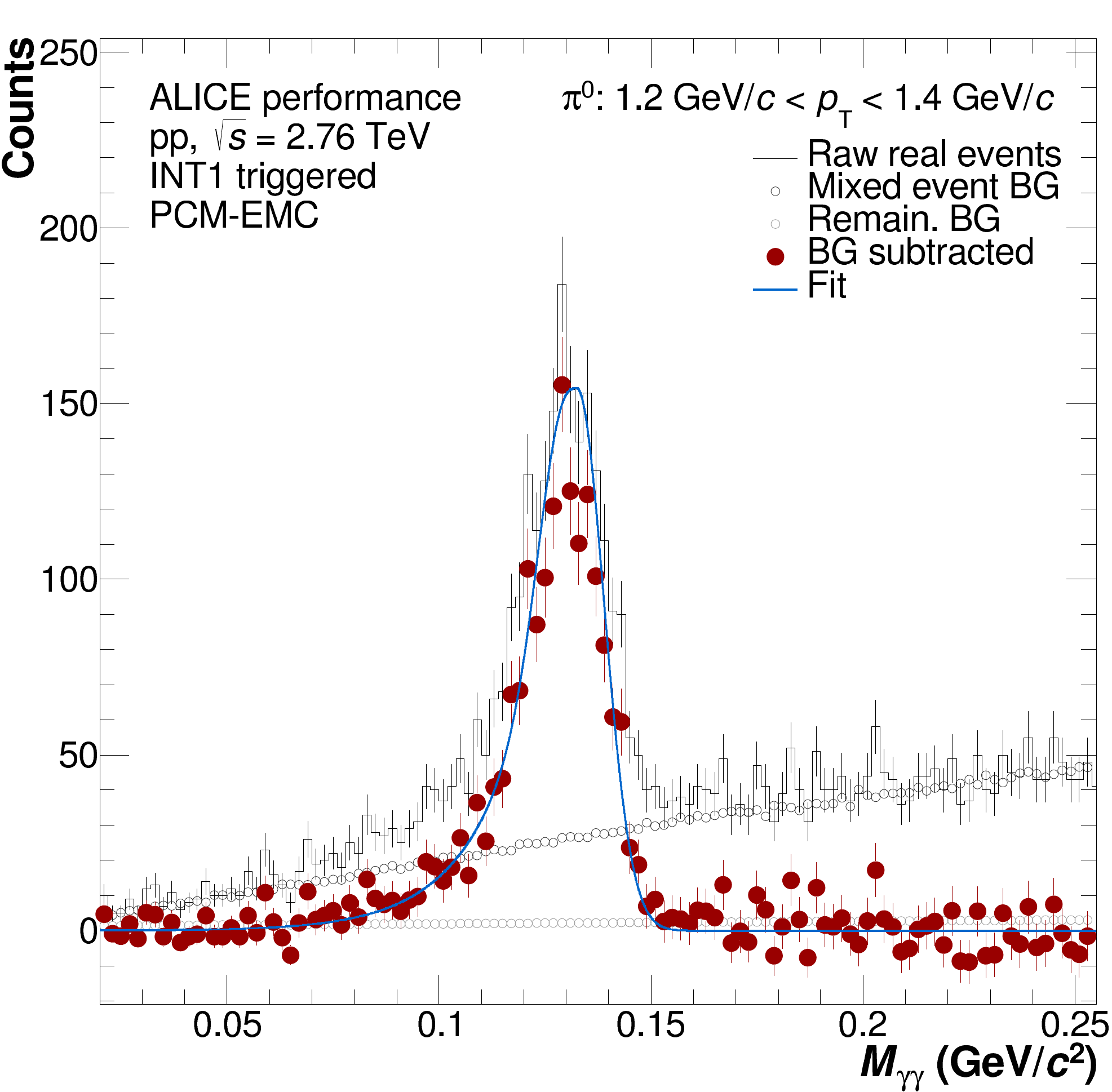}\hspace{0.3cm}
    \includegraphics[width=0.45\textwidth]{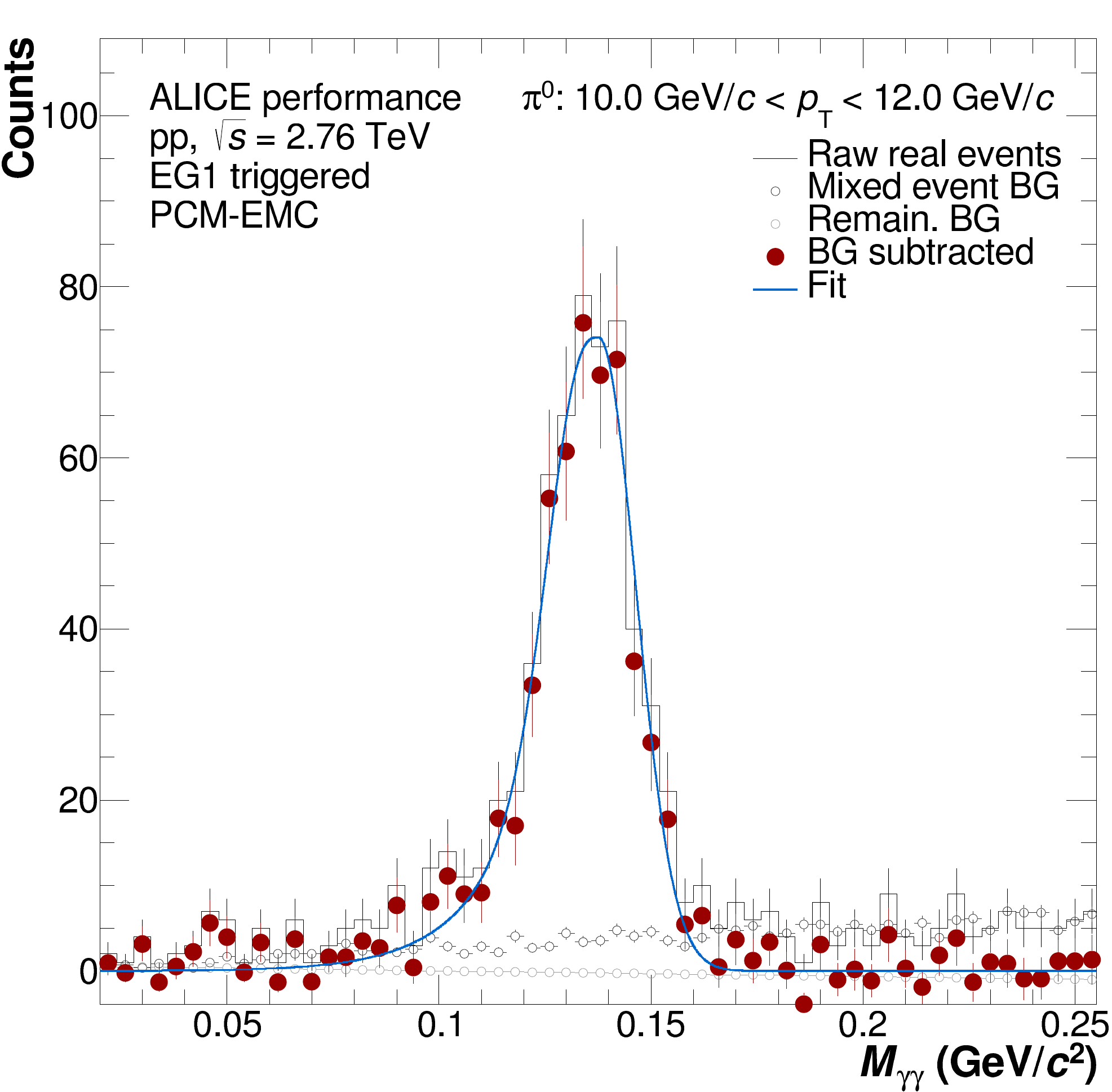}\\
    \caption{Invariant mass distributions in the $\pz$ peak region for INT1~(left panels) and EG1~(right panels) triggers and \EMC~(top panels) and \PCMEMC~(bottom panels) methods.} 
    \label{fig:pi0masspeaks}
  \end{figure}
  \begin{figure}[t!]
    \includegraphics[width=0.45\textwidth]{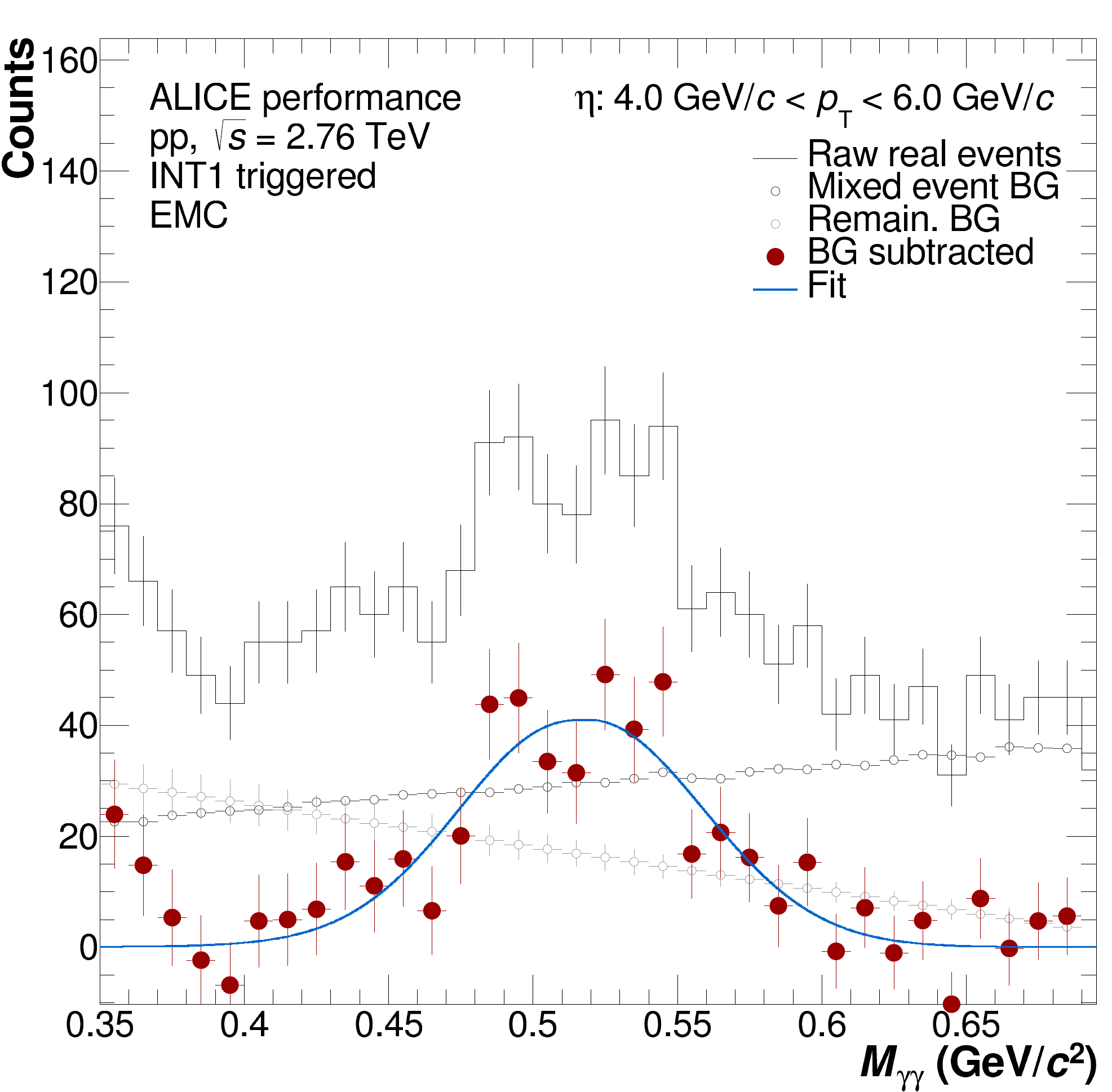}\hspace{0.3cm}
    \includegraphics[width=0.45\textwidth]{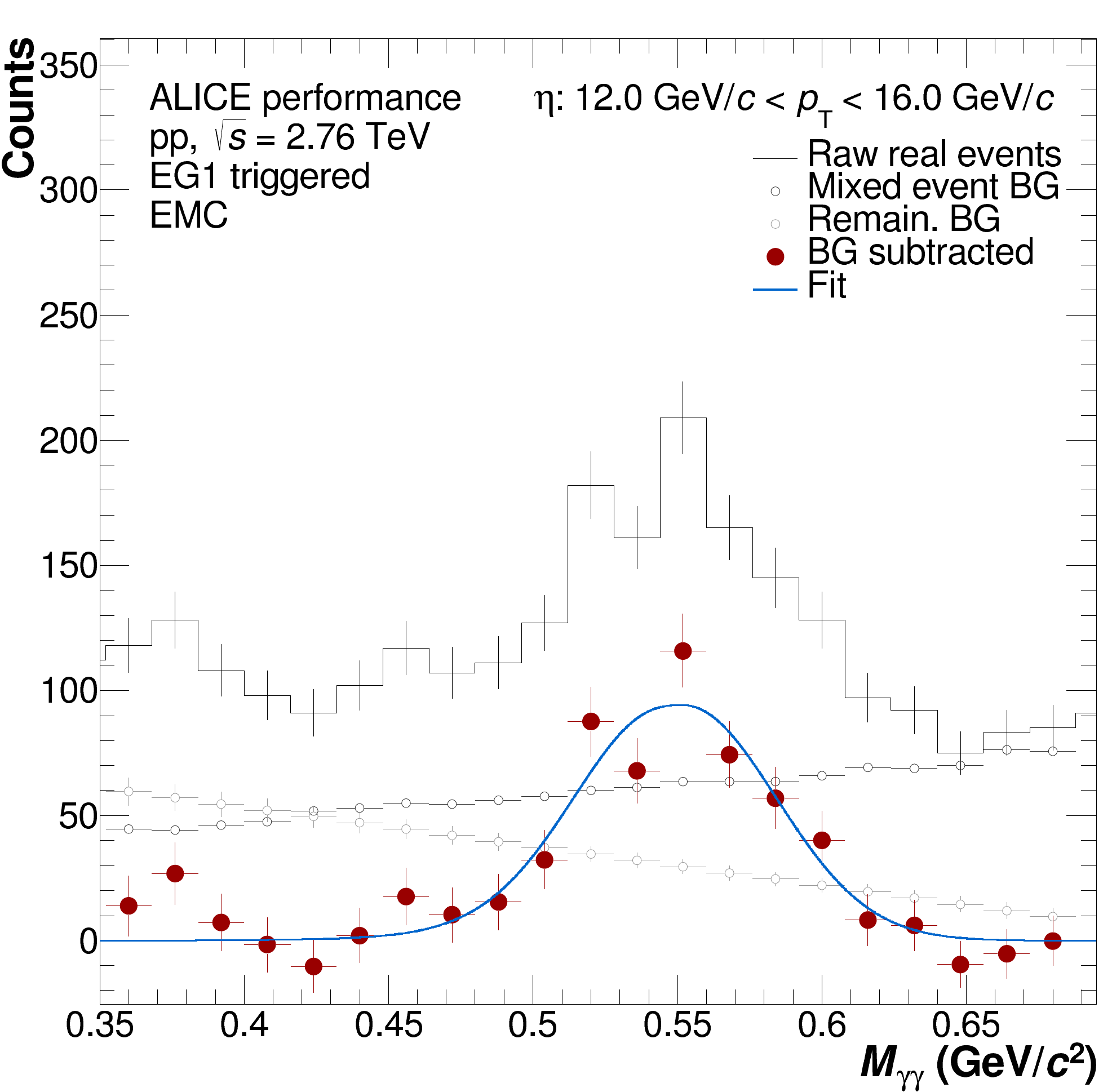}\\
    \includegraphics[width=0.45\textwidth]{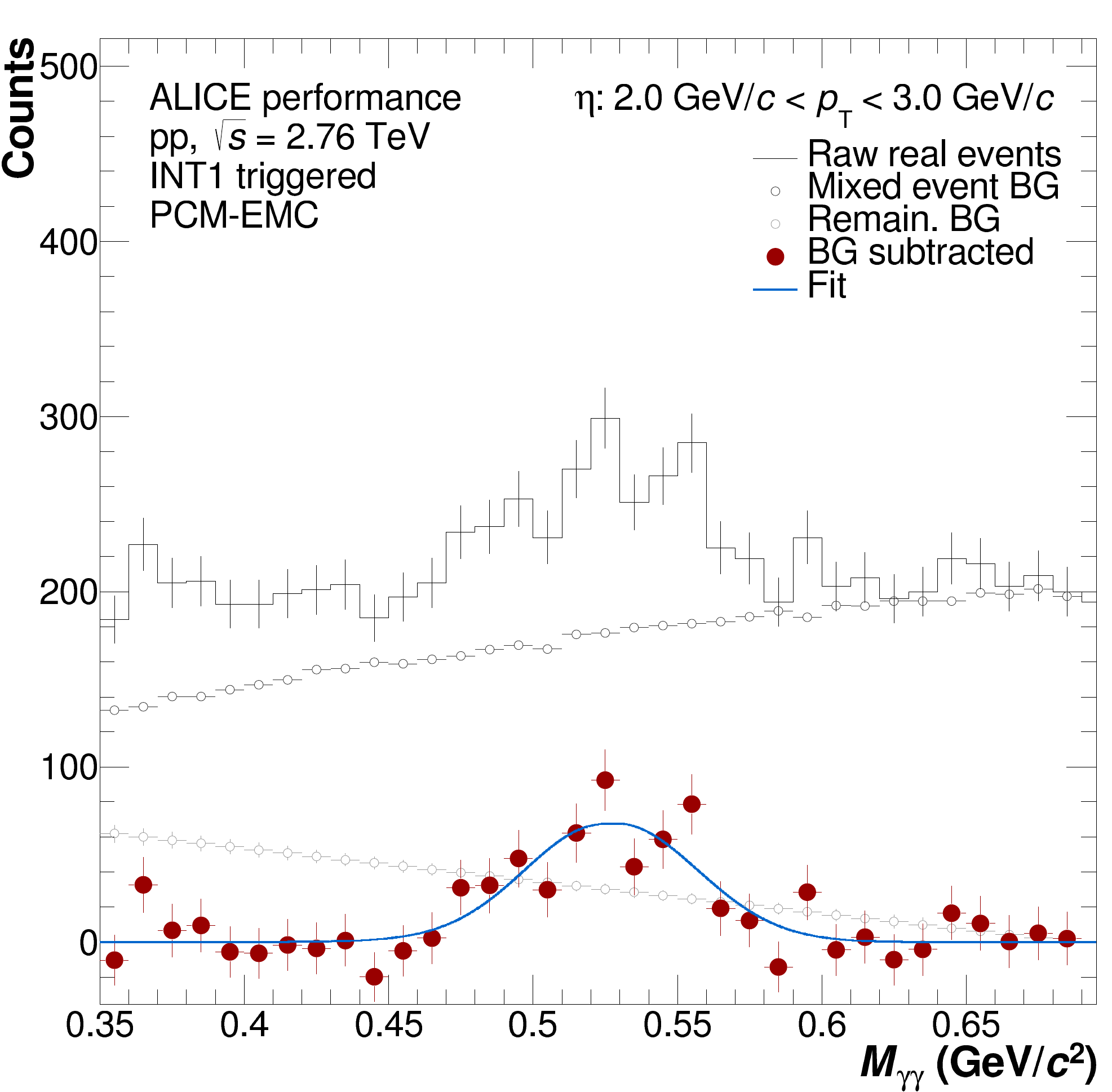}\hspace{0.3cm}
    \includegraphics[width=0.45\textwidth]{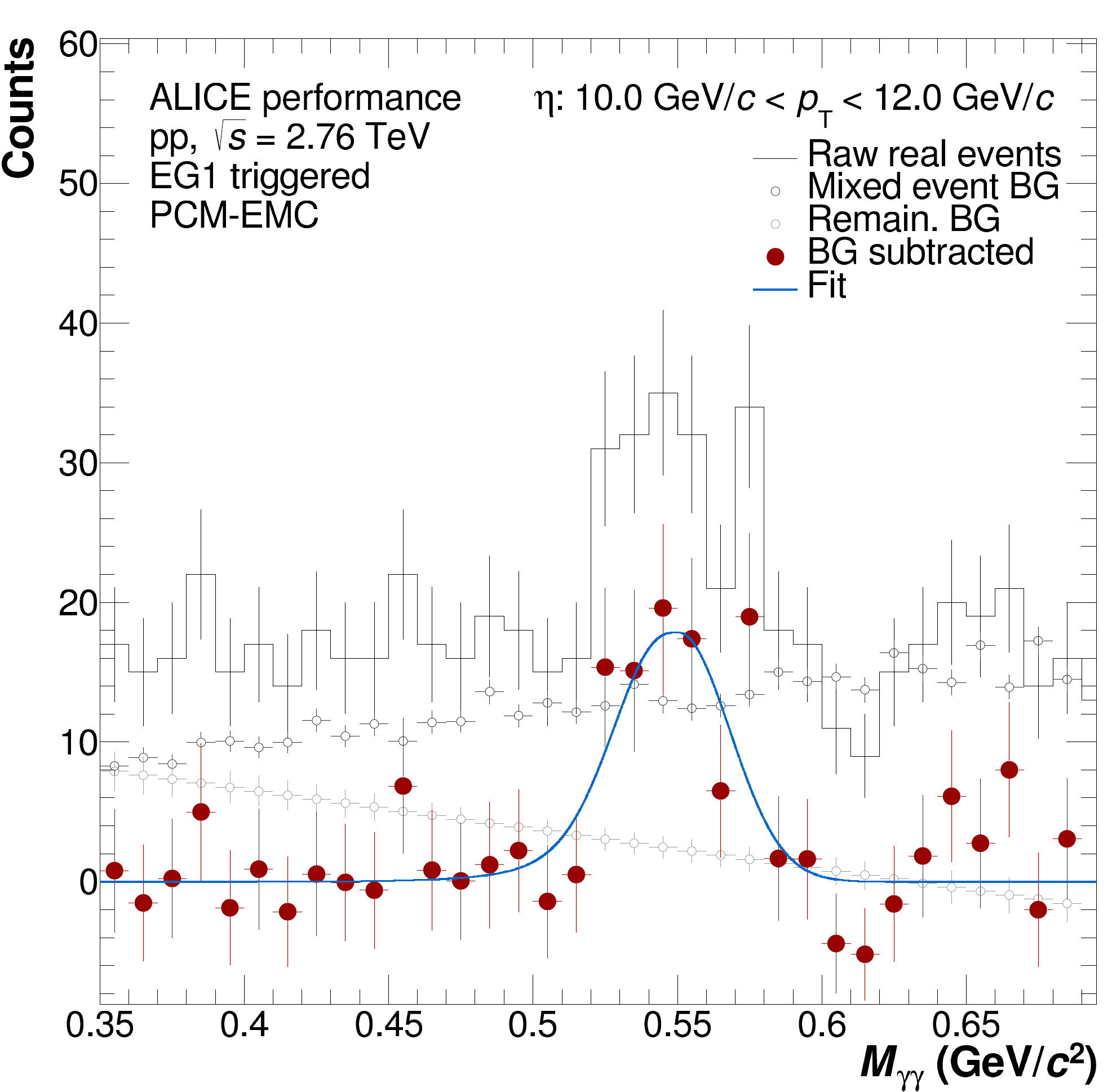}
    \caption{Invariant mass distributions in the $\eta$ peak region for INT1~(left panels) and EG1~(right panels) triggers and \EMC~(top panels) and \PCMEMC~(bottom panels) methods.} 
    \label{fig:etamasspeaks}
  \end{figure}    
\fi
Example invariant mass distributions obtained by correlating photons reconstructed with EMCal or by one photon from \PCM\ and one from EMCal are shown in \Fig{fig:pi0masspeaks} for neutral pions and \Fig{fig:etamasspeaks} for $\eta$ mesons.
The combinatorial background was calculated using the mixed event technique~\cite{Kopylov:1974th} using event pools\com{ with a depth of $80$ photons for various event classes} binned by primary vertex position, multiplicity and transverse momentum.
The mixed-event background has been normalized to the right side of the $\pz (\eta)$ peak. 
Additionally, a residual correlated background estimated using a linear fit was subtracted. 
Only pairs with a minimum opening angle of $0.02$~($0.005$)~mrad for \EMC~(\PCM\ and \PCMEMC) methods were considered for signal and background construction.
Finally, pairs are restricted to rapidity of $|y| < 0.8$.

\begin{figure}[t!]
  \centering
  \includegraphics[width=0.48\textwidth]{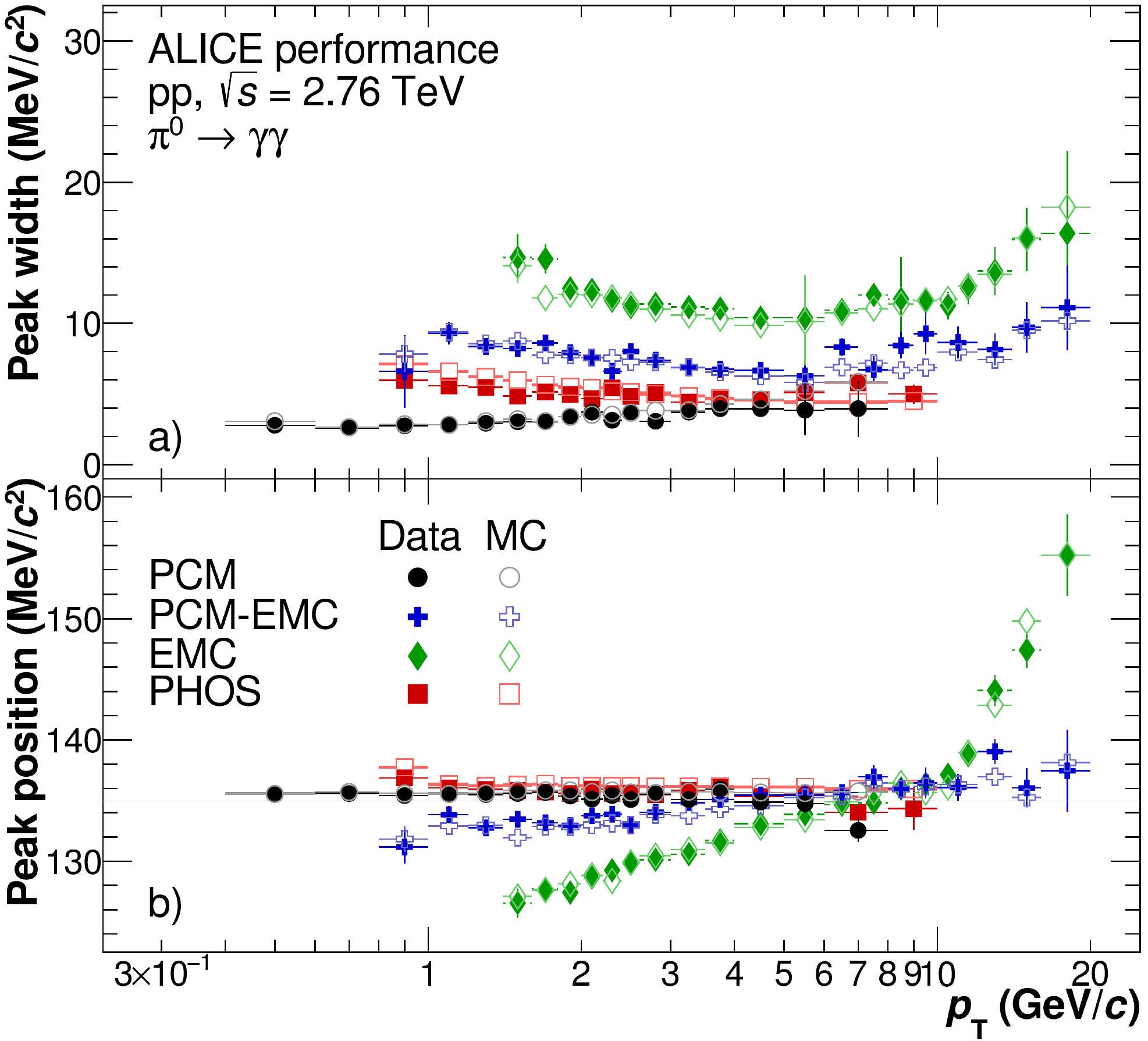}\hspace{0.3cm}
  \includegraphics[width=0.48\textwidth]{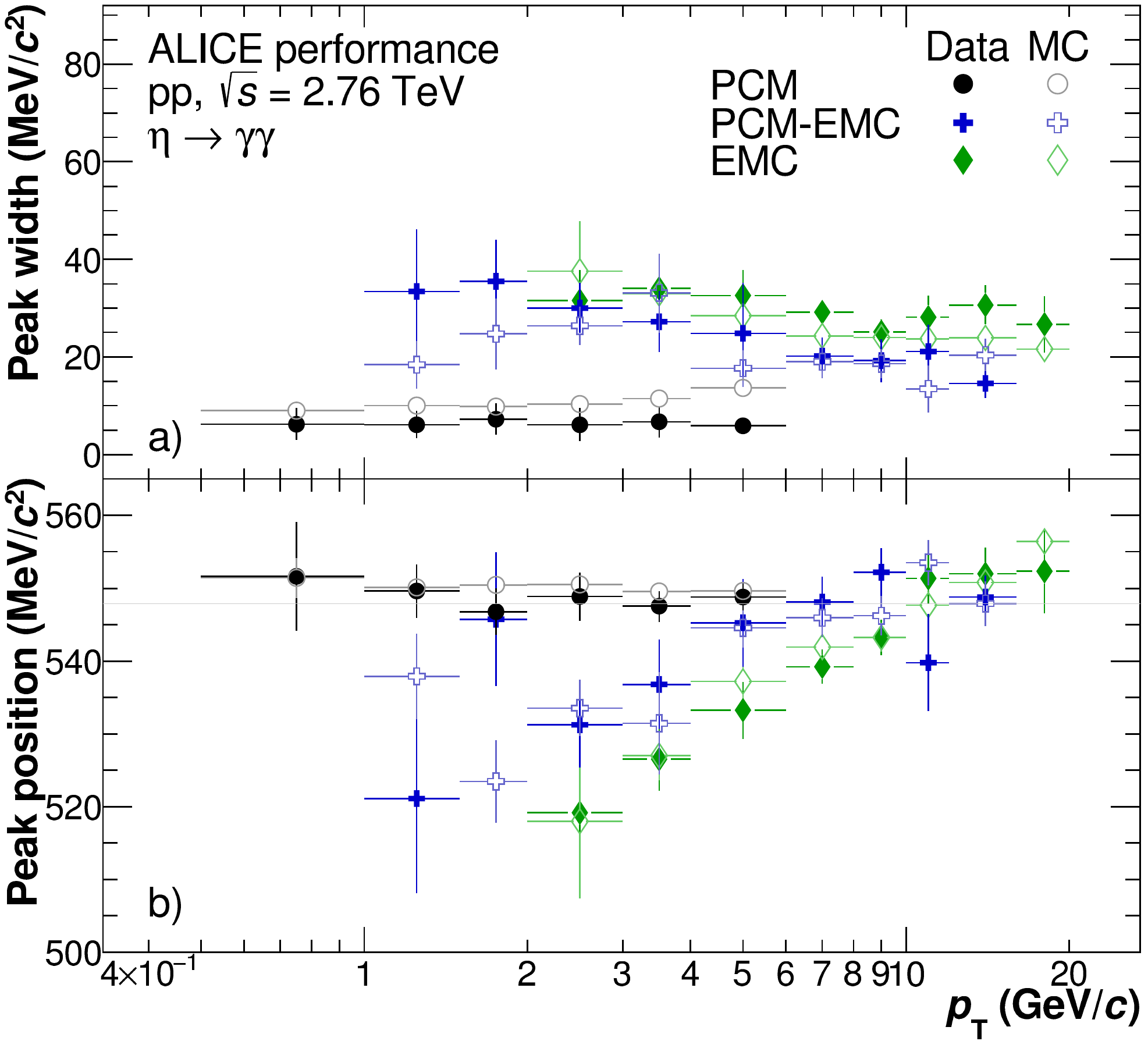}
  \caption{Neutral pion (left panels) and $\eta$ meson (right panels) mass position~(bottom panels) and width~(top panels) for the \PCM, \PCMEMC\ and \EMC\ methods. The performance of \PHOS\ for $\pz$ is taken from \Ref{Abelev:2014ypa}.
           Data are displayed as closed symbols, simulations as open symbols.} 
  \label{fig:massposition}
\end{figure}

A Gaussian with an exponential tail on the left side\com{~\cite{Koch:2011}} was fitted to the subtracted invariant mass distributions, in order to determine the mass position and width of the peak.
The results of the fits for the mass position and widths of neutral pions and $\eta$ mesons are shown in \Fig{fig:massposition}.
The performance of \PHOS\ from \Ref{Abelev:2014ypa} in the case of $\pz$ is added for completeness.
For all systems, the data for both $\pz$ and $\eta$ are reproduced by the MC simulations to a precision on average better than 0.3\% for the mass position. 
For \EMC, the $\pt$-dependence of the mass position is especially pronounced, due to non-linearity effects for low $\pt$ clusters, shower merging and shower overlaps, and decay asymmetry enhanced by the employed triggers at high $\pt$.
The widths of the meson peaks are similarly well described, with the expected ordering for the various methods.
In particular, the peak widths of the \PCMEMC\ fits are between the standalone measurements of \PCM\ and \EMC\ and are comparable to the PHOS measurement above 7 \gevc. 
This illustrates that the inclusion of one photon from \PCM\ significantly improves the resolution of the neutral meson measurements. 

The neutral meson raw yield was extracted by integrating the background-subtracted invariant mass distributions around the measured peak mass.
The integration windows for the different reconstruction techniques were adjusted based on the average width of the meson peaks and their signal shape:
($M_{\pz}-0.035$, $M_{\pz}+0.010$), ($M_\eta-0.047$, $M_\eta+0.023$) for \PCM,
($M_{\pz}-0.032$, $M_\pz+0.022$), ($M_\eta-0.060$, $M_\eta+0.055$) for \PCMEMC, and
($M_{\pz}-0.05$, $M_\pz+0.04$), ($M_\eta-0.080$, $M_\eta+0.08$) for \EMC.
For both mesons, an asymmetric range around the measured mass position was used to account for the low mass tail originating not only from the bremsstrahlung energy loss of conversion electrons and positrons, but also from additional missing energy in the EMCal due to the partial reconstruction of the photon. 

\begin{figure}[t!]
 \centering
 \includegraphics[width=0.48\textwidth]{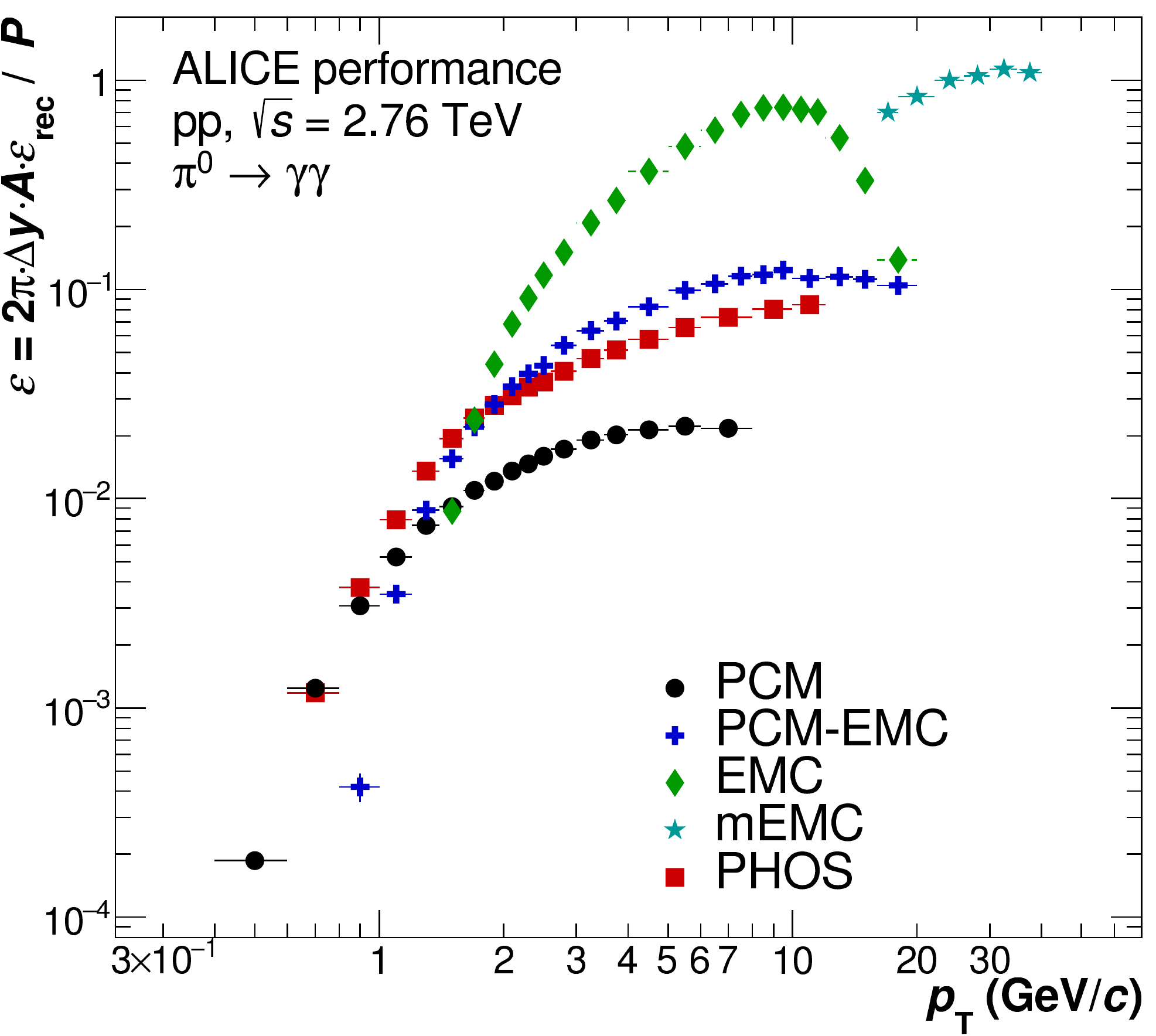}\hspace{0.3cm}
 \includegraphics[width=0.48\textwidth]{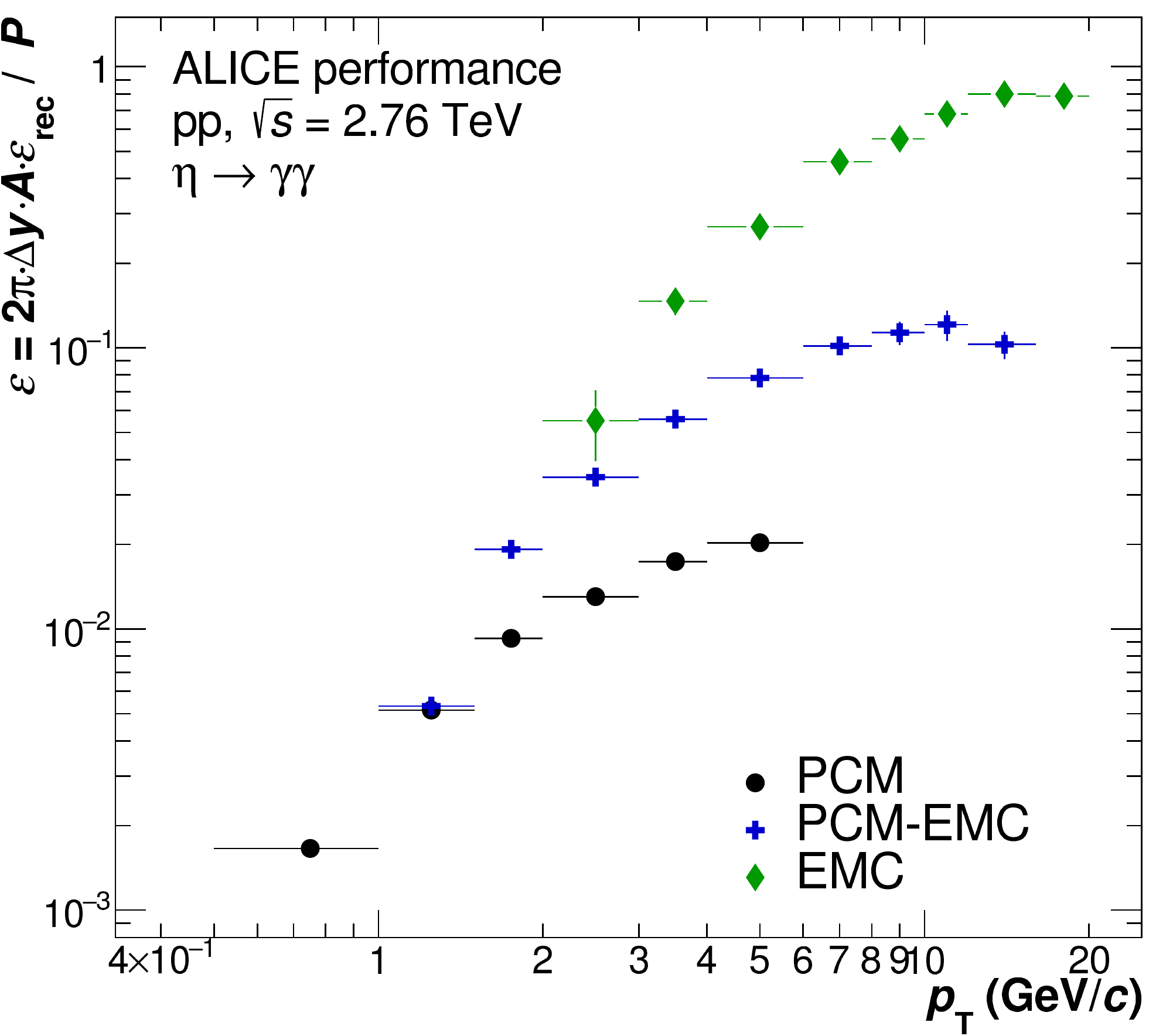}
 \caption{Normalized efficiency for different methods of neutral pion (left panel) and $\eta$ meson~(right panel) reconstruction methods.
          The values for \PHOS\ are taken from \cite{Abelev:2014ypa}.} 
 \label{fig:effacc}
\end{figure}

The corrections for the geometric acceptance and reconstruction efficiency for the different mesons were calculated using MC simulations\com{ with a full ALICE GEANT setup} as mentioned in \Sect{sec:datasamples}. 
The acceptance for the EMCal reconstruction techniques was calculated as the fraction of $\pi^0$~($\eta$), whose decay photons point to the EMCal surface ($|\eta| < 0.67,~\unit[1.40]{rad}< \varphi < \unit[3.15]{rad}$), compared to the $\pi^0$~($\eta$) generated with $|y| < 0.8$.
In the case of \PCMEMC, only one photon was required to point to the EMCal surface, while the other was required to be within the acceptance of the TPC~($|\eta| < 0.9,~\unit[0]{rad}< \varphi < \unit[2\pi]{rad}$). 
The output from the full event MC simulations was reconstructed and analyzed in the same way as the data.
The reconstruction efficiency was calculated as the fraction of reconstructed mesons\com{ passing all selection criteria and performing the yield extraction} compared to the mesons whose decay photons passed the acceptance criteria.
The normalized efficiency $\varepsilon$~(see \Eq{eq:effi}) as a function of meson $\pt$ is shown in \Fig{fig:effacc} for the various methods.
For \EMC, $\varepsilon$ rises at low $\pt$ and reaches its maximum at about $0.8$ at $10$~\gevc. 
Subsequently, $\varepsilon$ drops due to the merging of the two clusters\com{ that cannot be split by the clustering algorithm anymore}, and is already a factor of $5$ smaller at about $15$~\gevc.
In the case of the $\eta$, the efficiency at $15$ \gevc\ is not yet affected by the cluster merging due to its higher mass.
The efficiency for \PCMEMC\ is approximately a factor $10$ smaller than for \EMC\ for both mesons due to the conversion probability of about $0.09$ in the respective pseudorapidity window.
For the $\pz$, it is similar to that of PHOS. 
The small decrease at higher $\pt$ for the \PCMEMC\ results from shower overlaps of the \EMC\ photon with one of the conversion legs, and thus a stronger rejection of the EMCal photons due to track matching.
Relative to \PCMEMC, $\varepsilon$ for \PCM\ is suppressed by the conversion probability affecting both decay photons.

The correction for secondaries from hadronic interactions depends on $\pt$ for the \EMC-related methods.
It ranges from $1.2$\% at the lowest $\pt$ to $0.1$\%~($0.4$\%) above $3$~\gevc\ for the \PCMEMC\ (\EMC) method. 
For \PCM, the correction amounts to less than $0.2$\% independent of $\pt$. 
However, the contribution of the neutral pions from \kzs\ is strongly $\pt$ dependent due to the tight selection criteria forcing the photons to point to the primary vertex.
The correction drops quickly from about $8$\% to less than $1$\% at $4$~\gevc.
For the \PCMEMC\ and \EMC, the corresponding correction amounts to $0.9$\% and $1.6$\%, respectively, independent of $\pt$ in the measured $\pt$ range. 
Contributions from other weak decays are below 0.1\% and thus neglected for all reconstruction techniques.

\subsection{Single cluster analysis}
\label{sec:singlePi0}
At high $\pt$ the showers induced by the two decay photons from a neutral pion merge into a single EMCal cluster, and therefore are unidentifiable in an invariant mass analysis.
Hence, for $\pzs$ above 15 GeV/$c$ we use a different approach, namely to reconstruct and identify $\pzs$ based only on single clusters, exploiting that clusters at high $\pt$ mostly originate from merged $\pz$ decay photons.

\begin{figure}[t!]
 \centering 
 \includegraphics[width=0.48\textwidth]{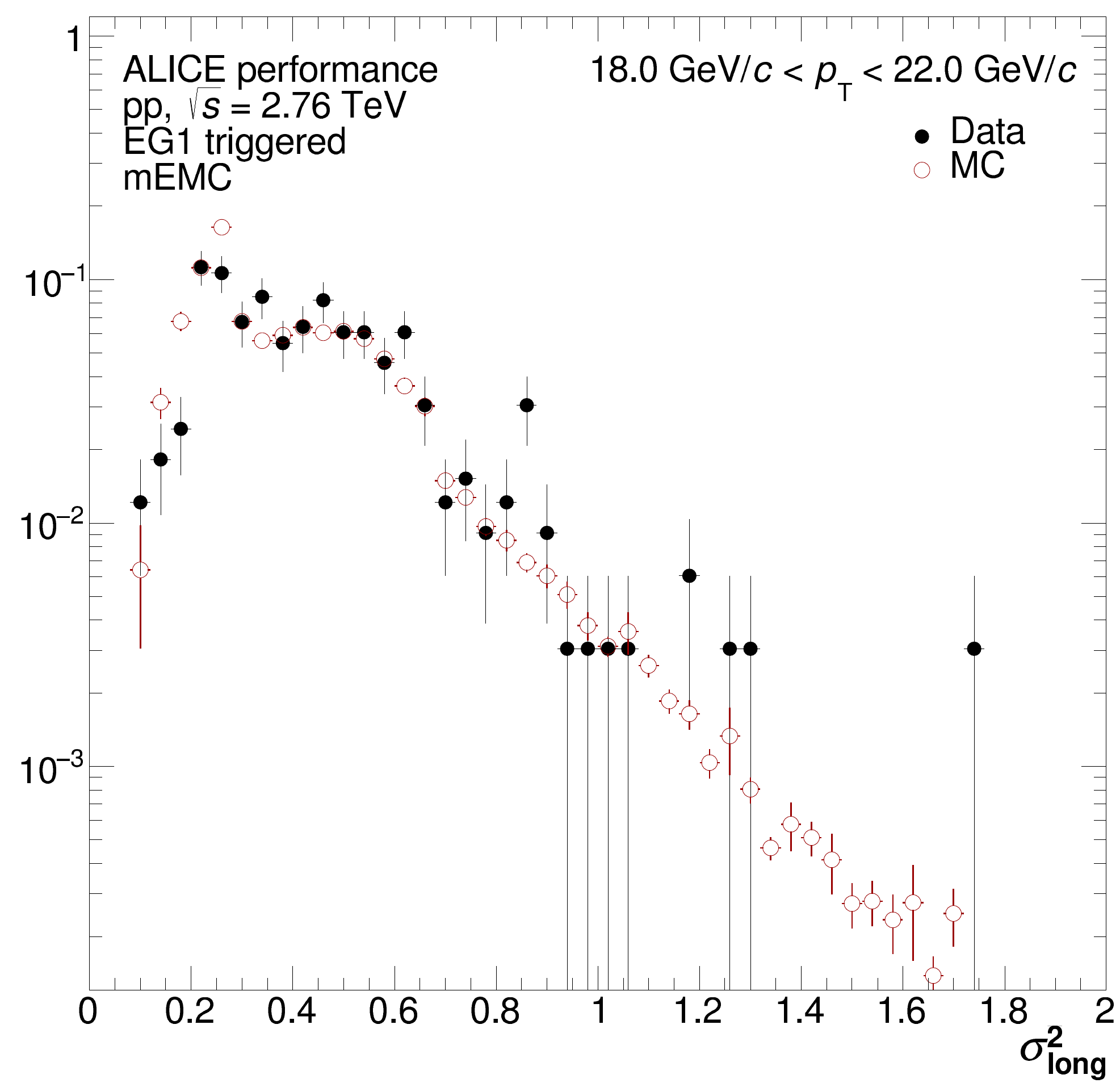}\hspace{0.3cm}
 \includegraphics[width=0.48\textwidth]{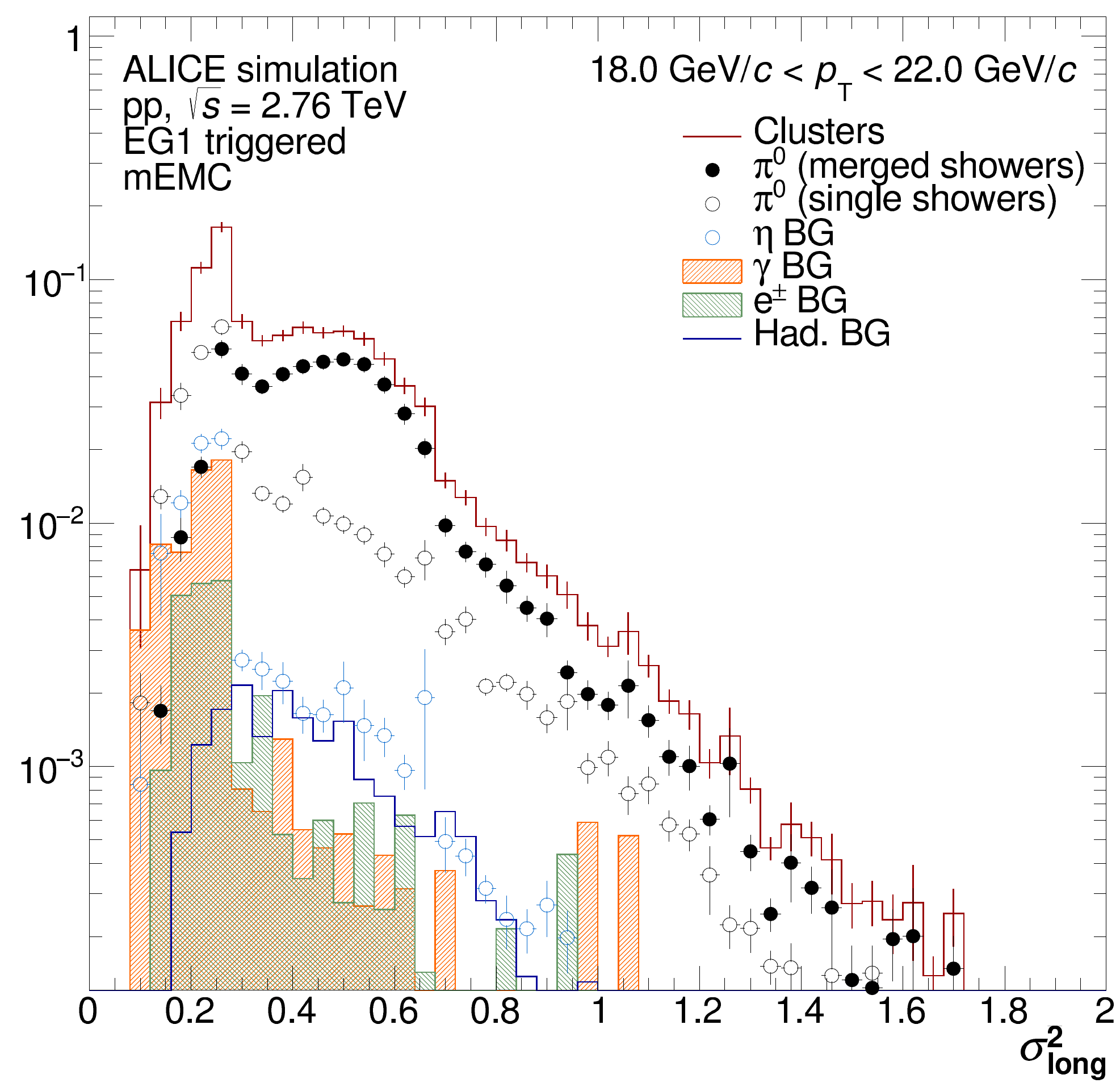}
 \caption{Shower shape ($\lzt$) distributions for $\pz$ candidates with $18<p_{\rm T}<22$~\gevc\ compared in data and MC (left panel), and corresponding signal and background contributions in MC (right panel).}
 \label{fig:ShowerShape}
\end{figure}

Merged clusters from $\pz$ decays tend to be more elongated than clusters from photons and electrons, and their deformation is reflected by the shower shape $\lzt$, defined in \Eq{eq:lambda02}.
The shower shape distributions are shown for data and MC in \Fig{fig:ShowerShape} for $\pz$ candidates, i.e.\ clusters fulfilling the selection criteria listed in \Tab{tab:emccuts} except $\lzt$.
The $\lzt$ distribution is found to be fairly well described by the MC, in particular for $\lzt>0.3$.
For $\lzt>0.3$, the dominant contribution to $\pz$ candidates is from merged $\pz$ showers, while for $\lzt<0.3$ clusters dominate where only the energy of one decay photon contributed.
The most significant background is from decay photons of the $\eta$ meson and direct photons, located mainly at $\lzt<0.3$.
Hence, for the \mEMC\ measurement, $\pz$ candidates are simply required to have $\lzt>0.27$ in order to discriminate from $\eta$ decay and direct photons. 
Only candidates with a rapidity of $|y| < 0.6$ are considered. 

The corrections for the geometric acceptance, reconstruction efficiency, and purity were calculated using MC simulations\com{ with a full ALICE GEANT setup} as described in \Sect{sec:datasamples}.
The resulting efficiency is shown in \Fig{fig:effacc} compared to the other neutral pion reconstruction techniques. 
At high $\pt$, \mEMC\ clearly has an advantage due to its larger coverage compared to PHOS, and the exploitation of merging of the $\pz$ decay photons in the EMCal. 

The $\pz$ reconstruction efficiency was calculated by comparing the reconstructed with generator-level $\pt$ distributions within a rapidity of $|y| < 0.6$. 
By comparing measured and generated $\pt$ of the neutral pion, the $\pt$ resolution correction is included in the inefficiency correction.
The resolution is significantly different for candidate clusters containing all or only parts of the decay products, i.e.\ single photons or conversions.
If all $\pz$ decay products contribute to the cluster, the mean momentum difference between reconstructed and generated $\pt$ is smaller than $2\%$ with an RMS of $16$--$25$\% above $20$~\gevc. 
Otherwise, the mean momentum difference can reach up to $30$\% depending on the fraction of decay particles which could be reconstructed and whether they converted in the detector material.

The purity represents the fraction of reconstructed clusters that pass all the selections and are from a $\pz$ decay. 
For $\pt>16$ GeV/$c$, it is almost constant at around $90$\% with variations of $1$--$2$\%.
As can be seen in \Fig{fig:ShowerShape}, the largest contamination in the considered $\lzt$ window originates from the $\eta$ meson decay~($\approx 5\%$ after fine-tuning the $\etatopi$ ratio to the measured value), closely followed by the hadronic background consisting mainly of charged pions~($\approx 2\%$) and \kzl~($\approx 1.8\%$).
The contamination from $\eta$ mesons rises by about $2$\% towards higher momenta, while the contamination from the other two sources decrease by about $0.5$\%.
Fragmentation photons contribute to the background about $1.2$\%.
Their contribution was additionally scaled up by up to a factor $2$, given by the ratio of fragmentation photons to direct photons according to NLO pQCD calculations~\cite{Gordon:1993qc,Vogelsang:1997cq}, to account for direct photons which are not included in generator.
Lastly, prompt electrons contribute to the contamination about $0.7$\%.

The correction for secondary pions from \kzs\ decays amounts to approximately $5$\%, as their reconstruction efficiency is very similar to that of primary $\pzs$, albeit with worse resolution. 
In addition, corrections for $\pzs$ from weak decays from \kzl\ and $\Lambda$~(together only about 0.3\%) and from secondary hadronic interactions~($2.2$\%) were applied.

\section{Systematic uncertainties}
\label{sec:corrsys}
The sources of systematic uncertainties associated with the various measurement techniques and their magnitude in different $\pt$ ranges, chosen to reflect the strengths of the various methods, are given in \Tab{tab:SysPi0} for the $\pz$ meson, in \Tab{tab:SysEta} for $\eta$ meson and in \Tab{tab:SysEtaToPi0} for the $\etatopi$ ratio.
Since the measurements obtained with \PCMEMC, \EMC\ and \mEMC\ are a combination of multiple triggers, the systematic uncertainties associated with each method reflect the contribution of different triggered data samples weighted by their statistical uncertainties. 
The uncertainties for the $\etatopi$ were evaluated directly on the ratio in order to cancel correlated uncertainties between the $\pz$ and $\eta$ measurements.
In the following, we first describe the uncertainties on photon candidates reconstructed with \EMC\ and \PCM, then those on the meson level, and finally those related to the overall normalization, in the same order as given in the tables.

\textbf{EMCal clustering: }
The uncertainty on clustering quantifies the mismatch in the description of the clusterization process between data and simulation. 
It incorporates the uncertainties arising from the variation of the minimum energy and time on cluster and cell level, the minimum number of cells per cluster as well as the variation of the $\lzt$ selection on the clusters. 
For \mEMC, varying the selection on $\lzt$ is especially important since it quantifies the uncertainty of how well the $\lzt$ distributions of the background are described in the simulation, and was varied from $0.27$ to $0.25$ and $0.3$.
The corresponding uncertainties range between $2.1$\% and $6.2$\% depending on $\pt$ and method.

\textbf{EMCal cluster energy calibration: }
To estimate the uncertainty of the cluster energy calibration, the remaining relative difference between data and simulation in the mass position of the neutral pion was used.
On average, the difference is $0.3$\%, which leads to an uncertainty on the spectra of about $2$\% taking into account that they approximately fall with $\pt^{-6}$.
In addition, the correction of the simulations for relative energy scale and residual misalignment, described in \Section{sec:diclusterPi0}, was varied by changing the underlying parametrization of the mass position correction with $\pt$.
We chose only correction factors where the measured neutral pion mass position could be reproduced by the simulation to better than $1.5$\% over all $\pt$.
The overall resulting uncertainties range between $2.0$\% and $5.5$\% depending on $\pt$ and method.
For the $\eta$ meson~(\etatopi\ ratio), the uncertainties are approximately a factor $1.5$~($2$) larger at similar $\pt$ due to lower photon energies entering at the same meson $\pt$.

\textbf{Track matching to cluster: }
The uncertainty introduced by the imperfection of the cluster-track matching procedure was studied by repeating the measurements with different track-matching parameters. 
The criteria were varied from tight selections, which removed only centrally matched clusters, to rather loose selections allowing a distance of $2$--$3$ cells depending on $\varphi$ and $\eta$.
At low $\pt$ the uncertainties on the $\pz$ measurement are below $2$\%, while with increasing $\pt$ higher track densities due to the jettier environment become more important and lead to uncertainties of about $7$\%. 
In the case of the $\eta$, the uncertainties are generally larger, between $4.9$\% and $8.9$\%,  due to the worse signal-to-background ratio.
For the $\etatopi$ ratio, the uncertainty of the $\eta$ alone is used, since part of the uncertainty is expected to cancel.

\begin{table}[t!]
 \small
 \centering
 \begin{tabular}{l |ccc |ccc |ccc |c}
  \hline
  $\pt$ interval (\gevc)    & \multicolumn{3}{c|}{1.4--1.6}     & \multicolumn{3}{c|}{3.0--3.5}          & \multicolumn{3}{c|}{16--20}           & \multicolumn{1}{c}{30--35}  \\
  Method                    & \PCM\  & \PE\   & \EMC\           & \PCM\   & \PE\    & \EMC\              & \PE\     & \EMC\     & \mEMC\         & \mEMC\     \\ \hline
  EMCal clustering          & -      & 2.4\%  & 4.9\%           & -       & 2.1\%   & 2.3\%              & 6.2\%    & 4.4\%     & 4.6\%          & 5.9\%      \\ 
  EMCal energy calib.       & -      & 2.0\%  & 4.9\%           & -       & 2.1\%   & 2.5\%              & 5.4\%    & 5.5\%     & 4.2\%          & 4.8\%      \\ 
  Track matching            & -      & 0.9\%  & 1.8\%           & -       & 1.4\%   & 1.7\%              & 6.9\%    & 6.7\%     & 5.4\%          & 6.1\%      \\ 
  Secondary track reco.     & 1.6\%  & 1.1\%  & -               & 0.9\%   & 0.8\%   & -                  & 5.7\%    & -         & -              & -          \\ 
  Electron PID              & 1.3\%  & 0.7\%  & -               & 1.5\%   & 0.6\%   &  -                 & 12.7\%   & -         & -              & -          \\ 
  \PCM\ photon PID          & 1.7\%  & 1.4\%  & -               & 2.3\%   & 1.1\%   & -                  & 13.4\%   & -         & -              & -          \\ 
  Signal extraction         & 1.9\%  & 1.5\%  & 2.4\%           & 4.0\%   & 1.9\%   & 1.5\%              & 3.4\%    & 14.1\%    & -              & -          \\ 
  Efficiency                & -      & 2.0\%  & 2.0\%           & -       & 3.6\%   & 2.5\%              & 2.1\%    & 2.1\%     & 8.4\%          & 7.1\%      \\ 
  Secondary correction      & -      & -      & -               & -       & -       & -                  & -        & -         & 1.8\%          & 1.8\%      \\ 
  Inner material            & 9.0\%  & 4.5\%  & -               & 9.0\%   & 4.5\%   & -                  & 4.5\%    & -         & -              & -          \\ 
  Outer material            & -      & 4.2\%  & 4.2\%           & -       & 4.2\%   & 4.2\%              & 4.2\%    & 4.2\%     & 4.2\%          & 4.2\%      \\ 
  Trigger norm.+pileup      & 0.8\%  & -      & -               & 0.4\%   & 1.1\%   & 0.5\%              & 7.5\%    & 5.5\%     & 8.0\%          & 8.8\%      \\ \hline
  Tot.\ sys.\ uncertainty   & 9.6\%  & 7.6\%  & 8.9\%           & 10.3\%  & 8.3\%   & 6.5\%              & 24.5\%   & 18.6\%    & 14.9\%         & 15.6\%     \\ \hline\hline
  Stat.\ uncertainty        & 2.8\%  & 2.0\%  & 6.5\%           & 5.1\%   & 3.3\%   & 2.8\%              & 14.8\%   & 15.6\%    & 5.7\%          & 11.3\%     \\
  \hline
 \end{tabular}
  \caption{Systematic uncertainty for various sources and methods assigned to the $\pz$ measurement at different $\pt$ intervals. For comparison, the total systematic and the statistical uncertainties are also given. \PE\ stands for \PCMEMC.}
  \label{tab:SysPi0}
\end{table}

\textbf{Secondary track reconstruction: }
The uncertainty on the secondary track reconstruction quantifies the uncertainty related to secondary track finding used in \PCM.
It is estimated by variation of the TPC found-over-findable cluster selection and the minimum $\pt$ cut as well as reducing the acceptance for the conversion photons in \PhiConv\ requiring them to approximately point towards the EMCal direction. 
The uncertainty depends on the precision of the relative alignment and track matching efficiency between TPC and ITS in different sectors of the TPC, and hence can vary for different data taking periods and trigger conditions.
For the EMCal triggers, for instance, the conversion photons are mainly sampled in the region directly in front of the EMCal, where the ITS had larger inefficiencies than in other areas.
The uncertainties range from $0.8$\% to $5.7$\%.

\textbf{Electron PID: }
Systematic uncertainty on the electron identification for the \PCM\ photon reconstruction was estimated by varying the TPC $\dEdx$-based electron inclusion as well as the pion rejection selections.
The corresponding uncertainties are small at low $\pt$~($\approx 1$\%), where there is good separation between electrons and pions, but reach up to $12.7$\% at high $\pt$, where electrons and pions can not be efficiently separated any longer.

\textbf{PCM photon PID: }
The uncertainty assigned to the \PCM\ photon reconstruction combines the contributions from varying the criteria for the photon quality and Armenteros-Podolanski selections. 
The uncertainties are slightly larger than those on the electron PID, with similar $\pt$ dependence, since both the electron and the photon PID selections attempt to reduce the contamination which increases with increasing $\pt$.
For the \etatopi\ ratio, it is one of the dominant uncertainties, in particular at high $\pt$, as only a small fraction cancels in the ratio due to the different decay kinematics of the two mesons.

\begin{table}[t!]
 \centering
 \small
 \begin{tabular}{l |cc |ccc |cc}
  \hline
  $\pt$ interval (\gevc)      & \multicolumn{2}{c|}{1--1.5}     & \multicolumn{3}{c|}{3--4}             & \multicolumn{2}{c}{10--12}  \\
  Method                      & \PCM\   & \PCMEMC\              & \PCM\   & \PCMEMC\  & \EMC\           & \PCMEMC\  & \EMC\       \\ \hline
  EMCal clustering            & -       & 3.1\%                 & -       & 3.1\%     & 2.7\%           & 3.6\%     & 3.1\%       \\
  EMCal energy calib.         & -       & 3.0\%                 & -       & 3.2\%     & 4.5\%           & 5.0\%     & 6.8\%       \\
  Track matching              & -       & 8.9\%                 & -       & 4.9\%     & 5.7\%           & 6.6\%     & 8.8\%       \\
  Secondary track reco.       & 3.7\%   & 3.3\%                 & 1.6\%   & 3.3\%     & -               & 4.1\%     & -           \\
  Electron PID                & 2.1\%   & 2.5\%                 & 2.4\%   & 2.2\%     &  -              & 5.2\%     & -           \\
  \PCM\ photon PID            & 3.9\%   & 7.7\%                 & 3.9\%   & 7.3\%     & -               & 11.2\%    & -           \\
  Signal extraction           & 6.0\%   & 16.4\%                & 6.0\%   & 8.1\%     & 9.3\%           & 11.8\%    & 3.5\%       \\
  Efficiency                  & -       & 5.0\%                 & -       & 5.0\%     & 5.7\%           & 5.8\%     & 5.3\%       \\
  Inner material              & 9.0\%   & 4.5\%                 & 9.0\%   & 4.5\%     & -               & 4.5\%     & -           \\ 
  Outer material              & -       & 4.2\%                 & -       & 4.2\%     & 4.2\%           & 4.2\%     & 4.2\%       \\ 
  Trigger norm.+pileup        & 1.8\%   & -                     & 1.9\%   & -         & 2.8\%           & 7.0\%     & 7.2\%       \\ \hline
  Tot.\ sys.\ uncertainty     & 12.3\%  & 22.5\%                & 11.9\%  & 15.5\%    & 14.3\%          & 22.6\%    & 15.5\%      \\ \hline\hline
  Stat.\ uncertainty          & 20.4\%  & 43.4\%                & 17.2\%  & 16.7\%    & 10.8\%          & 21.3\%    & 8.9\%       \\ \hline
 \end{tabular}
  \caption{Systematic uncertainty for various sources and methods assigned to the $\eta$ measurement at different $\pt$ intervals. For comparison, the total systematic and the statistical uncertainties are also given.}
  \label{tab:SysEta}
\end{table}
\begin{table}[t]
 \centering
 \small
 \begin{tabular}{l |cc |ccc |cc}
  \hline
  $\pt$ interval (\gevc)      & \multicolumn{2}{c|}{1--1.5}     & \multicolumn{3}{c|}{3--4}   & \multicolumn{2}{c}{10--12}   \\
  Method                      & \PCM\   & \PCMEMC\              & \PCM\   & \PCMEMC\  & \EMC\         & \PCMEMC\  & \EMC\    \\ \hline
  EMCal clustering            & -       & 4.1\%                 & -       & 4.2\%     & 2.4\%         & 6.0\%     & 2.8\%    \\
  EMCal energy calib.         & -       & 4.1\%                 & -       & 4.3\%     & 4.6\%         & 6.6\%     & 7.6\%    \\
  Track matching              & -       & 8.9\%                 & -       & 4.9\%     & 5.7\%         & 6.6\%     & 9.0\%    \\
  Secondary track reco.       & 3.7\%   & 4.5\%                 & 1.6\%   & 4.2\%     & -             & 8.1\%     & -        \\
  Electron PID                & 2.1\%   & 3.3\%                 & 2.4\%   & 3.2\%     & -             & 7.0\%     & -        \\
  \PCM\ photon PID            & 3.9\%   & 7.7\%                 & 4.0\%   & 6.5\%     & -             & 12.7\%    & -        \\
  Signal extraction           & 6.1\%   & 16.6\%                & 7.0\%   & 9.1\%     & 9.3\%         & 10.5\%    & 8.5\%    \\
  Efficiency                  & -       & 5.4\%                 & -       & 5.4\%     & 3.8\%         & 7.0\%     & 4.3\%    \\ \hline
  Tot.\ sys.\ uncertainty     & 8.4\%   & 22.5\%                & 8.5\%   & 15.6\%    & 12.6\%        & 23.8\%    & 15.4\%   \\ \hline\hline
  Stat. uncertainty           & 20.4\%  & 44.1\%                & 17.7\%  & 17.9\%    & 10.9\%        & 22.1\%    & 8.8\%    \\ \hline
 \end{tabular}
  \caption{Systematic uncertainty for various sources and methods assigned to the $\etatopi$ measurement at different $\pt$ intervals.  For comparison, the total systematic and the statistical uncertainties are also given.}
  \label{tab:SysEtaToPi0}
\end{table}

\textbf{Signal extraction: }
The uncertainties arising from the signal extraction for the invariant mass analyses were estimated by varying the integration window, the background normalization region as well as the minimum opening angle, and requiring a mild asymmetry of the decay photons. 
For the neutral pion, the signal extraction uncertainty for \PCM\ ranges from $1.9$\% at low $\pt$ to $4.0$\% at higher $\pt$, due to the good momentum resolution of the tracks. 
For \PCMEMC, the equivalent uncertainty ranges from $1.5$\% to $3.4$\% at low and high $\pt$, respectively, while for \EMC\ it ranges from $2.4$\% at low to $1.5$\% at intermediate and $14.1$\% at high $\pt$. 
Above $10$~\gevc\ the signal extraction uncertainty for the \EMC\ arises from the merging of the two photon clusters, and the exact dependence of the corresponding description in the simulation. 
For the $\eta$ meson the signal extraction uncertainty generally is larger since the signal-to-background ratio is smaller, particularly at low $\pt$.
For \PCM\ the uncertainty is $6.0$\%, for \PCMEMC\ it ranges from $16.4$\% to $8.1$\% to $11.8$\% and for \EMC\ from $9.3$\% to $3.5$\%~\gevc\ at low, intermediate and high $\pt$, respectively.
Unlike in the case of the $\pz$, the uncertainty for \EMC\ decreases with increasing $\pt$ since the merging of the clusters for the $\eta$ meson only sets in at much higher $\pt$~(around $35$~\gevc).
For the \etatopi\ ratio, the signal extraction uncertainties of the $\pz$ and $\eta$ mesons contribute independently. 

\textbf{Efficiency: }
The uncertainties on the efficiency were estimated using different MC generators to vary the input spectrum for the efficiency calculation, to quantify effects affecting the $\pt$ resolution. 
Also, the uncertainties on the modeling of the efficiency bias in the simulation were included.
For \PCMEMC\ and \EMC\ the uncertainties range from $2.0$\% to $3.6$\% depending on $\pt$ for the $\pz$, while they are between $5$\% and $5.8$\% for the $\eta$ meson.
For the \etatopi\ measurement, the uncertainties were added quadratically, without including the trigger-related uncertainties, which largely cancel.
In the case of \mEMC, the uncertainty on the $\pt$ resolution is particularly important, since it strongly depends on whether the neutral pion could be reconstructed with all decay particles contributing to the single cluster or just some of them.
To estimate the uncertainty due to a possible imperfection of the MC simulation in the contribution of the various possibilities, the fractions of the respective reconstruction possibilities were varied by 20\% each, leading to an uncertainty on the efficiency of $8.4$\% at mid ($17$~\gevc) and $7.1$\% at high $\pt$~($32.5$~\gevc).

\textbf{Secondary correction: }
The correction for secondary $\pz$ was estimated applying the efficiency and acceptance from the full ALICE GEANT3 simulation to a fast MC simulation of the decay kinematics based on the parametrized K$^0_S$ (K$^0_L$) and $\Lambda$ spectra \cite{Abelev:2014laa}. 
The corresponding uncertainty was obtained by varying the kaon and $\Lambda$ yield within their measured uncertainties.
Since the correction due to the secondaries is only $1$--$2$\%, for all but the \mEMC\ reconstruction technique, even a variation of $15$\% on the input yields leads to a negligible contribution compared to other uncertainties. 
For \mEMC, where the correction is about $5$\%, an uncertainty of $\approx 0.5\%$ was obtained.
In addition, $\approx 1.5$\% were added to the uncertainty to account for the limited precision in the shape and size of the correction factors of the full simulations for the pions from K$^0_S$, K$^0_L$ and $\Lambda$, which was estimated by varying the parametrization underlying the efficiencies for secondary $\pz$.

\textbf{Inner material: }
The uncertainty related to the knowledge of the inner~(radius $<\unit[180]{cm}$) material budget reflects the uncertainty of the conversion probability of photons, and hence dominantly affects the \PCM\ measurements.
It was estimated to be $4.5\%$ independent of $\pt$ based on detailed comparison between simulation and data for pp collisions at $\sqrt{s} = \unit[7]{TeV}$~\cite{Abelev:2012cn}.
Thus, it affects the \PCM\ meson measurements with $9$\%, while it only contributes 4.5\% to \PCMEMC.
In \etatopi, the uncertainty cancels as both mesons are affected in the same way.

\textbf{Outer material: }
For the reconstructed photons in the EMCal, a possible mismatch between the material present in reality and assumed in the simulation in front of the EMCal may cause an error in the absorption rate or the production of secondary pions. 
In most cases, however, the photon simply converts and at least one of its daughter electrons can be reconstructed in the EMCal so that the $\pz$ likely will be reconstructed as well, although with degraded $\pt$ resolution. 
The probability to still reconstruct the neutral meson increases with increasing conversion radius, i.e.\ the closer the conversion happens to the surface of the EMCal. 
Most of the material is located at most $1.5$~m away from the EMCal, namely the TPC outer wall, the Transition Radiation Detector~(TRD) and the Time-Of-Flight~(TOF) detector plus their support structures. 
The TRD was only fully installed in the LHC shutdown period after 2013.
For the 2011 and 2013 data there were regions in $\varphi$ without TRD modules in front of the EMCal.
Hence, the net-effect of the material in front of the EMCal could be studied by comparing fully corrected $\pz$ yields for different $\varphi$ regions with or without the TRD in front of the EMCal.
From the observed difference\com{ between the yields with and without TRD modules in front of the EMCal} measured using the \EMC\ and \PCMEMC\ measurements, an uncertainty on the neutral meson yields of $4.2$\% independent of $\pt$ was derived, and assigned to all measurements involving the EMCal. 
For \etatopi\ the uncertainty is assumed to cancel as both mesons should be affected in a similar way.

\textbf{Trigger normalization and pileup: }
The uncertainties for the trigger normalization were calculated by varying the range for the fit of the plateau region~(see \Fig{fig:trignorm}) for the different trigger combinations, leading to the respective rejection factors with their uncertainties given in \Tab{tab:EventStat}.
Since the final spectra for each measurement technique using the EMCal are composed of several triggers, the contributions of the respective trigger rejection uncertainties enter the final measurement with different magnitudes depending on $\pt$. 
The uncertainties range between $0.5$\% and $8.8$\%.
For \etatopi\ the uncertainties cancel as the ratio was measured per trigger and reconstruction method and combined afterwards.
For \PCM\ only minimum bias triggers were used, and hence no uncertainty due to the trigger rejection was assigned. 
However, an uncertainty of $0.8$\% to $0.4$\% was taken into account for the out-of-bunch pileup subtraction described in \cite{Abelev:2014ypa}.
The pileup uncertainty is about $1.8$\% for the $\eta$ meson. 
It largely cancels in the \etatopi\ ratio, however, and the remaining error can be neglected compared to other error sources.

\begin{table}[t]
  \small
 \begin{tabular}{l lllll}
  \textbf{Meson}      & $\mathbf{A_{e}}$ (pb GeV$^{-2} c^3$)  & $\mathbf{T_{e}}$ (GeV$/c$)    & $\mathbf{A}$ (pb GeV$^{-2} c^3$)  & $\mathbf{T}$  (GeV$/c$)     & $\mathbf{n_{\rm br}}$        \\ \hline
  $\pz$               & $(0.79 \pm 0.35) \cdot 10^{9}$        & $0.566 \pm 0.035$             & $(74.3 \pm 12.9) \cdot 10^9$      & $0.441 \pm 0.021$           & $ 3.083 \pm 0.027$  \\
  $\eta$              & $(18.5 \pm 22.1) \cdot 10^{9}$        & $0.149 \pm 0.070$             & $(1.4 \pm 1.0) \cdot 10^9$        & $0.852 \pm 0.136$           & $ 3.318 \pm 0.122$  \\ \hline
 \end{tabular}
  \caption{Parameters of the two-component model, \Eq{eq:mfit}~\cite{Bylinkin:2014fta,Bylinkin:2014qea}, which are used to parametrize the neutral pion and $\eta$ meson spectra, respectively, for the comparisons to models and among the different methods.}
  \label{tab:TCMfits}
\end{table}

\section{Results}
\label{sec:results}
Since the meson measurements with \PHOS, \PCM, \EMC, \PCMEMC\ and \mEMC\ \com{are statistically independent and }have partly uncorrelated systematic uncertainties, their combination will increase the precision of the respective cross section measurements. 
The BLUE method~\cite{Lyons:1988rp,Valassi:2003mu,Valassi:2013bga} was used to calculate the combined spectra of the $\pz$ and $\eta$ mesons as well as the \etatopi\ ratio. 
For the combination of the spectra, the full correlation matrix was taken into account by estimating the correlated and uncorrelated part of the systematics for all pairs of measurements versus $\pt$. 
Correlations are most apparent between the three EMC related measurements (\EMC, \PCMEMC and  \mEMC), as well as for the \PCMEMC\ and \PCM\ results. 
At high $\pt$, for instance, the uncertainties are dominated by the uncertainty on $\Rt$ which is largely common between the EMCal triggered analyses.
Uncertainties between \PHOS, \PCM, and \EMC~(\mEMC) are uncorrelated.
The combined spectra were fitted with a two-component model~(TCM)
\begin{equation}
E \frac{\mbox{d}^3 \sigma}{\mbox{d}p^3} = A_{\rm e} \exp{\frac{(M-\sqrt{\pt^2 + M^2})}{T_{\rm e}}} + {A}{\left( 1+ \frac{\pt^2}{n_{\rm br}T^2}\right)^{-n_{\rm br}}}
\label{eq:mfit}
\end{equation}
introduced by Bylinkin and Rostovtsev~\cite{Bylinkin:2014fta,Bylinkin:2014qea}, which serves as convenient parametrization of the data without aiming for a physics interpretation.
The parameters for the $\pz$ and $\eta$ fits are given in \Tab{tab:TCMfits} for $\chi^2/n_{\rm dof}$ values of better than $0.5$ taking statistical and systematic uncertainties in quadrature.
Unlike for Tsallis~\cite{Tsallis:1987eu} and power-law distributions, which at high and low $\pt$, respectively, systematically deviate from the data, the TCM parameterization describes the data over the full measured range to better than 10\%.

\begin{figure}[t!]
  \centering
  \includegraphics[width=0.49\textwidth]{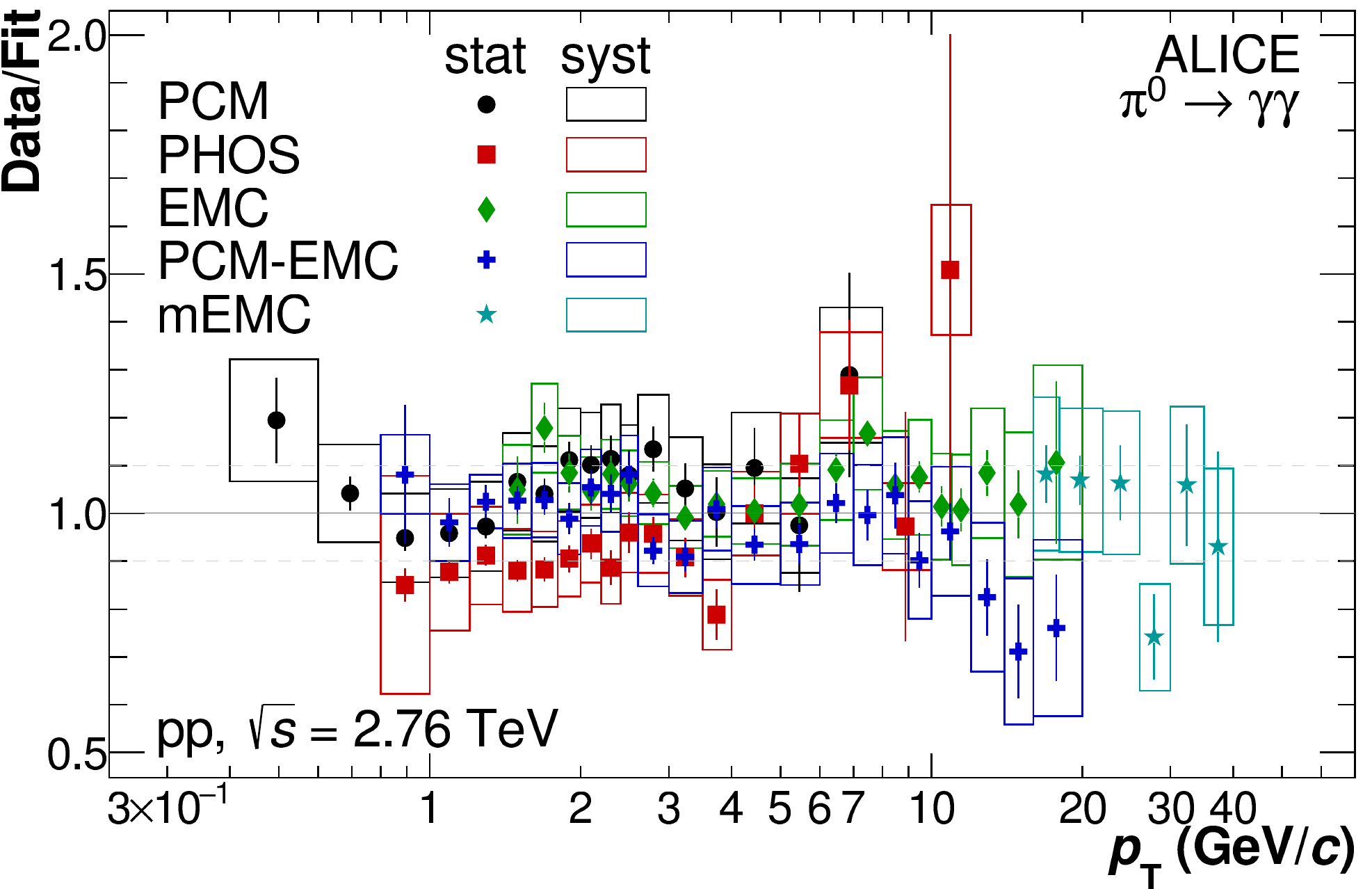}\hspace{0.1cm}
  \includegraphics[width=0.49\textwidth]{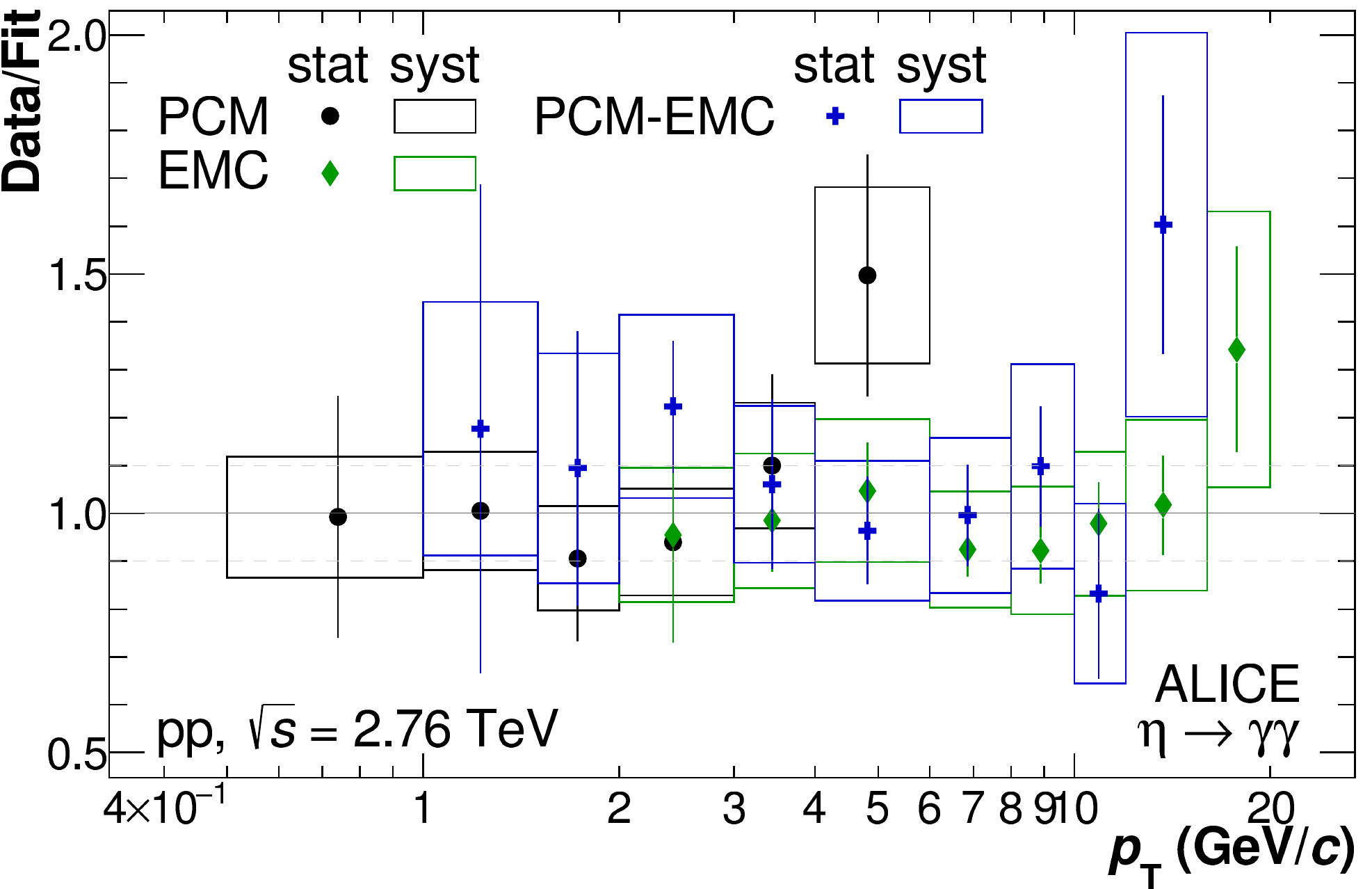}
  \caption{Comparison of the individual measurements in their respective measured transverse momentum ranges relative to the two-component model fits~\cite{Bylinkin:2014fta,Bylinkin:2014qea} of the final spectra.
           The final spectra are obtained by combining the individual measurements in the overlapping \pt\ regions with the highest granularity using the full correlation matrix as defined in the BLUE-algorithm~\cite{Lyons:1988rp,Valassi:2003mu,Valassi:2013bga}.} 
  \label{fig:comptechniques}
\end{figure}

\Figure{fig:comptechniques} shows a comparison of the individual measurements in their respective measured $\pt$ ranges summarized in \Tab{tab:ptrange} to the two-component model fits for the $\pz$ and $\eta$ mesons.
As already mentioned above, the $\pz$ spectrum in \pp\ collisions at $\sqrt{s} = 2.76$ TeV has been measured by ALICE using the \PHOS\ and \PCM~\cite{Abelev:2014ypa}. 
The new results obtained with the different \EMC\ measurements and with the hybrid \PCMEMC\ method are consistent with these earlier results, and the combination with the former measurements improves the precision of the data.
The figure also demonstrates an approximately fourfold extension of the $\pt$ reach of the measurement by using the EMCal.
The $\eta$ measurement, which is the first such measurement at $\sqrt{s} = 2.76$ TeV, spans from $0.6$~\gevc\ to $20$~\gevc.
There is good agreement within the statistical uncertainties among the different detection techniques.
Above $\pt>4$~\gevc, the result is dominated by the EMCal measurements. 

\begin{table}[t!]
\centering  
 \begin{tabular}{lccc}
          Method              & $\pz$         & $\eta$            & $\etatopi$      \\\hline
          \PCM                & 0.4--8.0      & 0.5--6.0          & 0.5--6.0        \\
          \PHOS               & 0.8--12.0     & n/a               & n/a             \\
          \EMC                & 1.4--20.0     & 2.0--20.0         & 2.0--20.0       \\
          \PCMEMC             & 0.8--20.0     & 1.0--16.0         & 1.0--16.0       \\
          \mEMC               & 16.0--40.0    & n/a               & n/a             \\\hline
 \end{tabular}
 \caption{Summary of the \pt\ reach (in GeV/$c$) of the various reconstruction methods for $\pz$, $\eta$ and $\etatopi$.}
 \label{tab:ptrange}
\end{table} 

\begin{figure}[t!]
  \centering
  \includegraphics[width=0.49\textwidth]{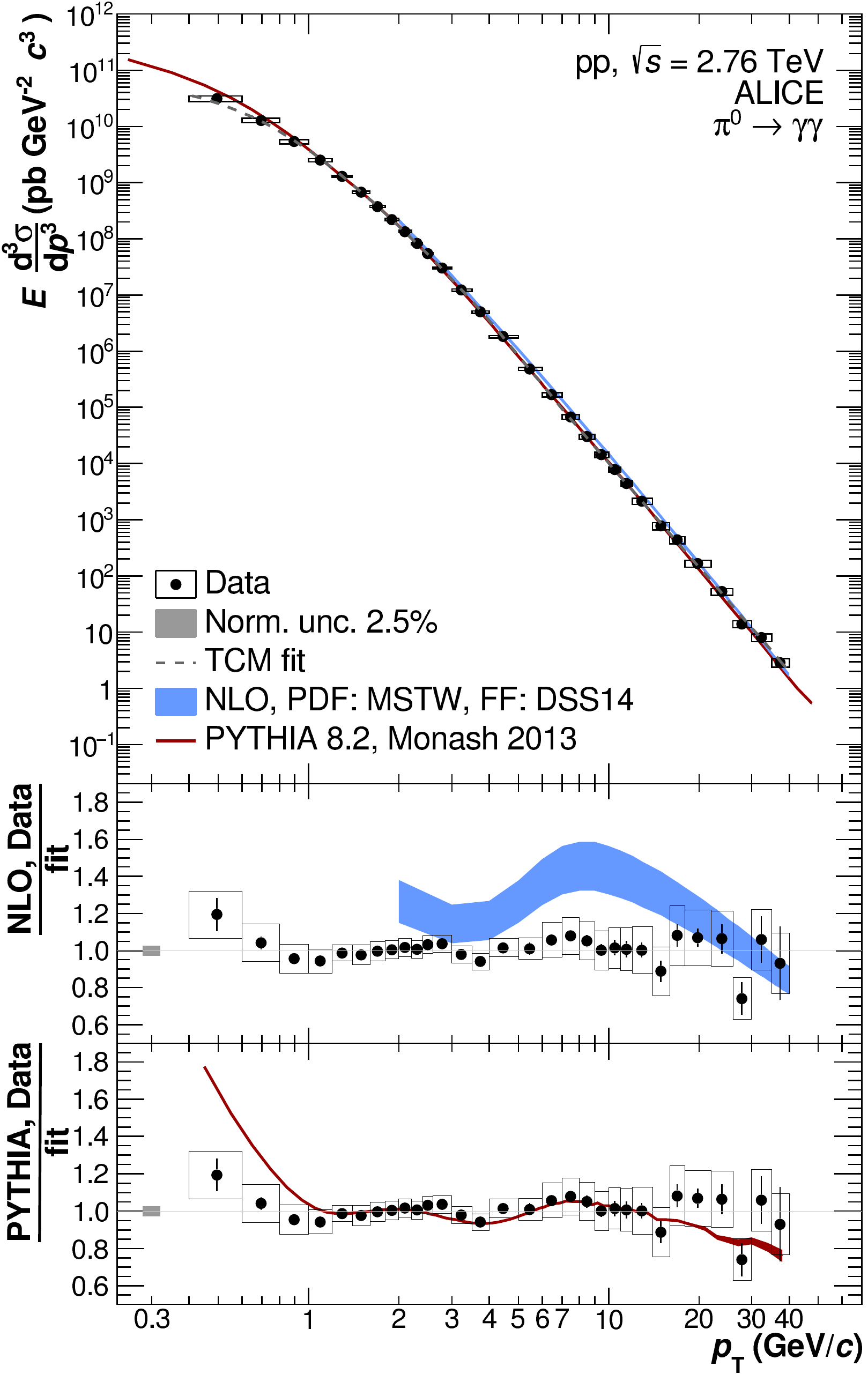}\hspace{0.1cm}
  \includegraphics[width=0.49\textwidth]{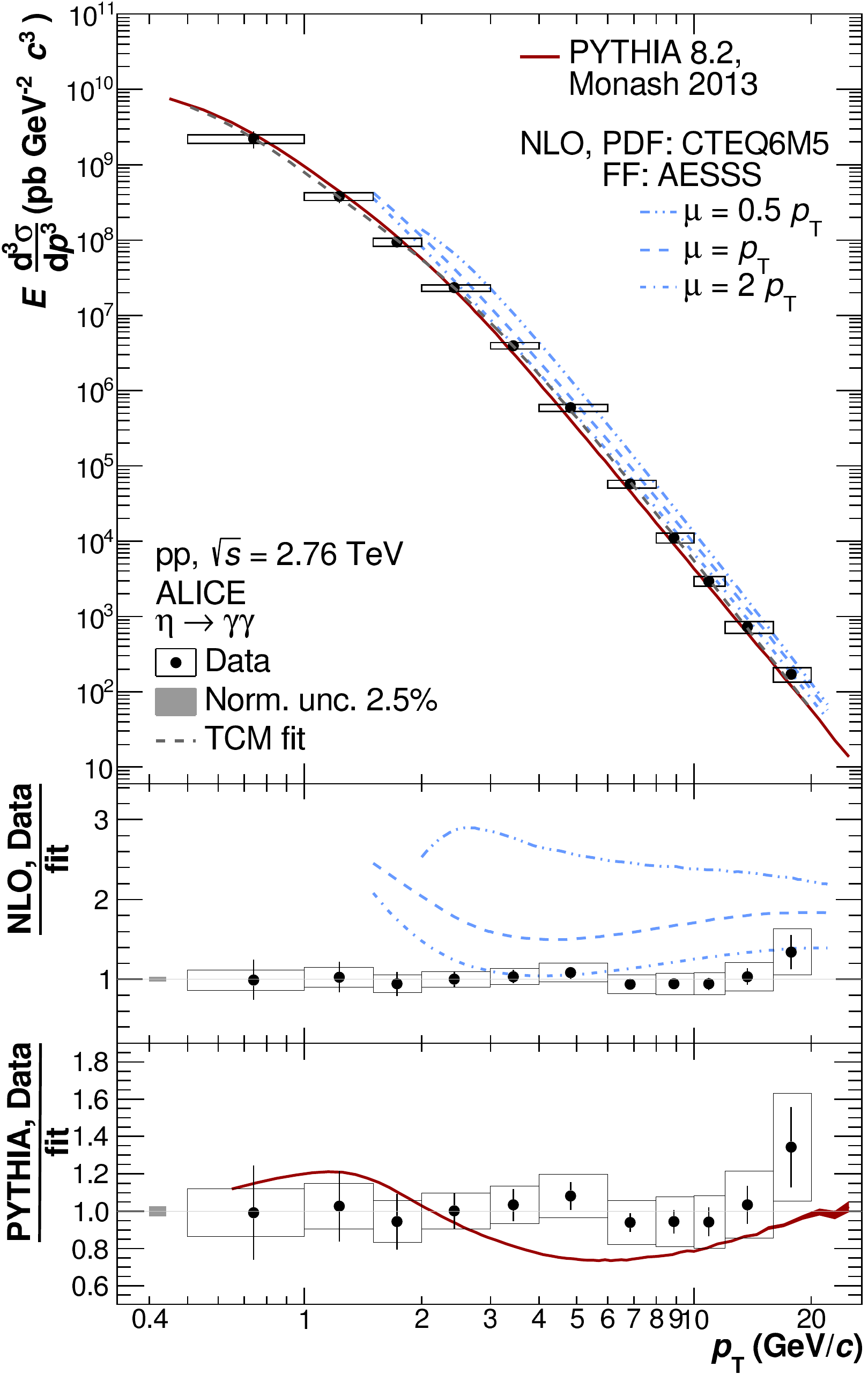}
  \caption{Invariant differential cross section of the $\pz$ (left, top panel) and $\eta$ meson (right, top panel) for \pp\ collisions at $\sqrt{s} = 2.76$ TeV. 
           The data are compared to \Pythia\ 8.2~\cite{Sjostrand:2014zea} generator-level simulations using the Monash 2013 tune as well as recent NLO pQCD calculations~\cite{deFlorian:2014xna,Aidala:2010bn}. 
           The ratios of the data and the calculations to the respective two-component model fits~\cite{Bylinkin:2014fta,Bylinkin:2014qea} to the data are shown in the lower panels. 
           The horizontal error bars denote statistical, the boxes systematic uncertainties.} 
  \label{fig:spectrum}
\end{figure}

\begin{figure}[t!]
  \centering
  \includegraphics[width=0.7\textwidth]{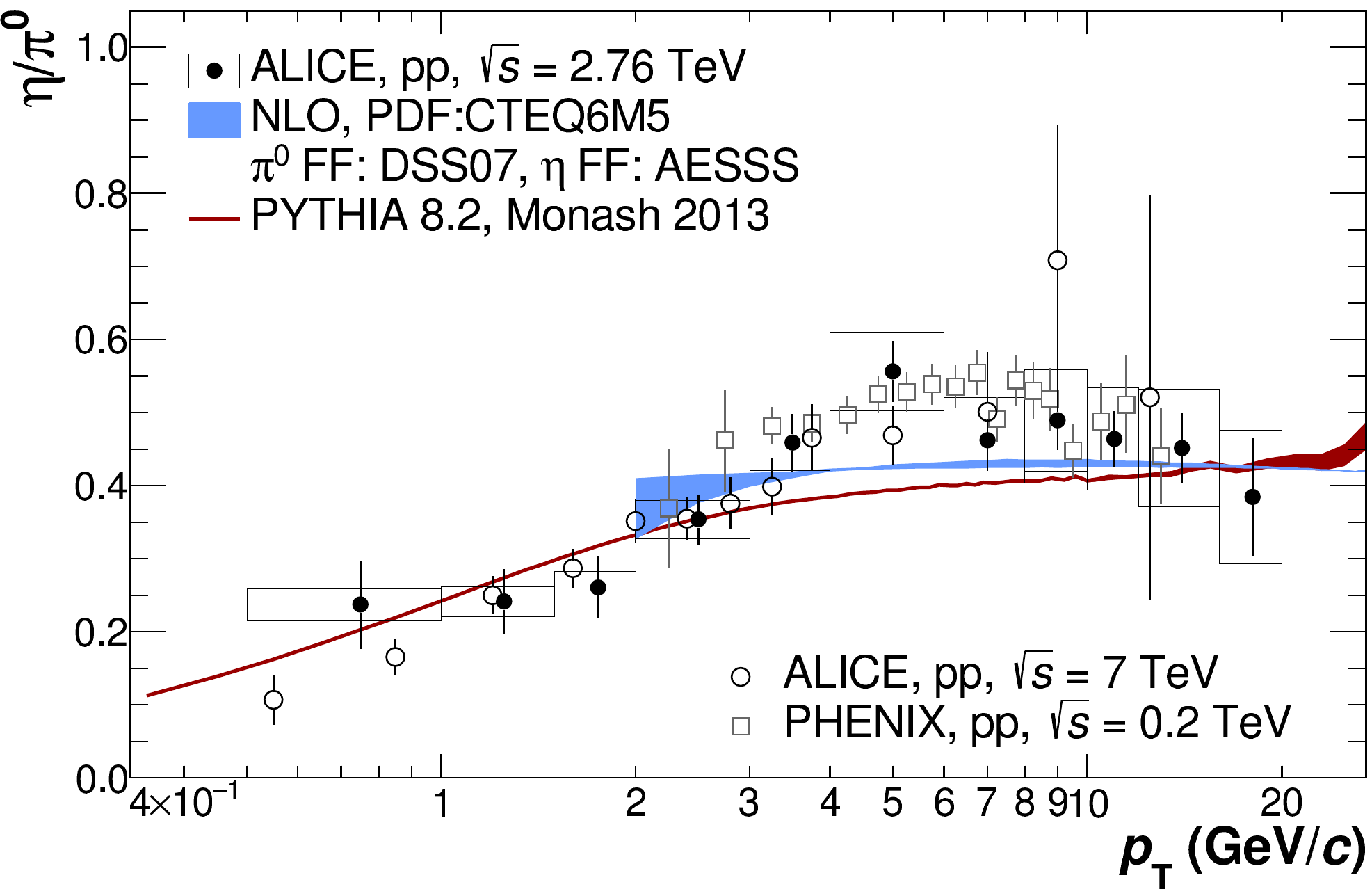}
  \caption{Measured $\eta/\pi^0$ ratio in \pp\ collisions at $\sqrt{s} = 2.76$ TeV compared to NLO pQCD calculations~\cite{deFlorian:2007aj,Aidala:2010bn} and \Pythia\ 8.2 ~\cite{Sjostrand:2007gs} generator-level simulations using the Monash 2013 tune.
           The horizontal error bars denote statistical, the boxes systematic uncertainties.
           The data at $\sqrt{s}=0.2$~TeV~\cite{Adare:2010cy} and $\sqrt{s}=7$~TeV~\cite{Abelev:2012cn} are shown with statistical and systematic uncertainties added in quadrature.}
  \label{fig:etatopi}
\end{figure}

\Figure{fig:spectrum} shows the combined $\pz$ and $\eta$ cross sections in \pp\ collisions at $\sqrt{s} = 2.76$~TeV, and \Fig{fig:etatopi} the corresponding \etatopi\ ratio. 
As mentioned earlier, the data were parameterized with a two-component model of Bylinkin and Rostovtsev~\cite{Bylinkin:2014qea}~(see \Tab{tab:TCMfits}) and compared to recent NLO pQCD calculations~\cite{deFlorian:2014xna,Aidala:2010bn}, and \Pythia\ 8.2~\cite{Sjostrand:2014zea} generator-level simulations using the widely-used Monash 2013 tune~\cite{Skands:2014pea}.
A large fraction of hadrons at low $\pt$ is produced in pp collisions via soft parton interactions and from resonance decays, which cannot be well described within the framework of pQCD, but are taken into account in the event-generator approach. 
For the $\pz$, the pQCD calculation~\cite{deFlorian:2014xna}, which uses the DSS14 fragmentation functions seems to have a different shape than the data.
It overpredicts the data by about $30$\% at intermediate $\pt$ ($5$ \gevc $< \pt < 16$ \gevc), while it agrees with the data at higher $\pt$.
The \Pythia\ 8.2 calculation describes the data well, except below $1$~\gevc, where it overpredicts the data by up to $30$\%.
For $\pt$ above $15$~\gevc\ \Pythia\ has a tendency to underpredict the data by about $10$\%; however this slight difference is covered by the uncertainties of the measurement.
For the $\eta$ meson, the data and the NLO pQCD calculation~\cite{Aidala:2010bn}, which uses the AESSS fragmentation functions, agree within the uncertainties for $\mu = 2\pt$ for factorization and fragmentation scale, while for $\mu = 0.5\pt$ the calculation overpredicts the data by up to a factor of $2$--$3$, leaving room for future improvements in the understanding of the strange versus non-strange quark fragmentation functions.
The \Pythia\ 8.2 simulation with the Monash 2013 tune performs slightly worse for the $\eta$ than for the $\pz$, in particular for $\pt>3$~\gevc\ where it underpredicts the data by about $20$--$30$\%.
In the \etatopi\ ratio, parts of the systematic uncertainties cancel not only for the data but also for the NLO pQCD calculation.
Thus, even the predictions using older fragmentation functions for the $\pz$~\cite{deFlorian:2007aj} and the $\eta$~\cite{Aidala:2010bn}, which can not reproduce the individual spectra~\cite{Abelev:2014ypa}, are in good agreement for the \etatopi\ measurement.
\Pythia\ 8.2 using the Monash 2013 tune can reproduce the $\pt$ dependence of the ratio; however it underpredicts the ratio by about $20$--$30$\% above $3$~\gevc, albeit still in agreement with the data to within $1$--$2\sigma$.
The measured \etatopi\ ratio is found to agree with previous measurements in \pp\ collisions at $\sqrt{s} = 0.2$ TeV~\cite{Adare:2010cy} and $\sqrt{s} = 7$ TeV~\cite{Abelev:2012cn} suggesting that $\etatopi$ is collision-energy independent. 
Above $4$~\gevc, both mesons exhibit a similar power-law behavior with $n_{\pz} = 6.29 \pm 0.02^{\text{stat}} \pm 0.04^{\text{sys}}$ and $n_{\eta} = 6.38 \pm 0.09^{\text{stat}} \pm 0.15^{\text{sys}}$ with $\chi^2/n_{\rm dof}$ of below $1.8$.
This is also reflected in the \etatopi\ ratio, which above $4$~\gevc\ reaches a value of $0.48 \pm 0.02^{\text{stat}} \pm 0.04^{\text{sys}}$. 

\section{Summary}
\label{sec:summary}
The invariant differential cross sections for inclusive \pz and $\eta$ production at midrapidity in pp collisions at $\sqrt{s}=2.76$~TeV were measured over a large range in transverse momentum of $0.4<\pt<40$~GeV/$c$ and $0.6<\pt<20$~GeV/$c$, respectively.
To achieve these measurements, for the $\pz$~($\eta$) five~(three) different reconstruction techniques and multiple higher-level triggers involving the EMCal in ALICE were exploited. 
In particular, a new single-cluster, shower-shape based method was developed to identify high-$\pt$ neutral pions whose decay photons overlap in the EMCal.
Above $4$~\gevc, both the $\pz$ and $\eta$ cross sections are found to exhibit a similar power-law behavior with an exponent of about $6.3$.
The data were compared to state-of-the-art NLO pQCD calculations which are found to reproduce the neutral pion cross section within $30\%$, while the deviations for the $\eta$ meson are significantly larger.
Calculations using \Pythia\ 8.2 at generator-level with the Monash 2013 tune turn out to be consistent with the $\pz$ measurement, except below $1$~\gevc, where the calculation overpredicts the data by up to $50$\%.
For the $\eta$, the agreement is slightly worse than for the $\pz$, in particular for $\pt>3$~\gevc\ where the calculation underpredicts the data by about $20$--$30$\%.
The \etatopi\ ratio, which was found to be described by the calculations to within $1$--$2\sigma$, is $0.48\pm 0.02^{\text{stat}} \pm 0.04^{\text{sys}}$ above $4$~\gevc, consistent with previous measurements.
The new data provide significant constraints for future calculations of hadron spectra over a large range in $\pt$.

\newenvironment{acknowledgement}{\relax}{\relax}
\begin{acknowledgement}
\section*{Acknowledgments}
We thank Werner Vogelsang and Marco Stratmann for providing us with their calculations.

\ifdraft
\else

The ALICE Collaboration would like to thank all its engineers and technicians for their invaluable contributions to the construction of the experiment and the CERN accelerator teams for the outstanding performance of the LHC complex.
The ALICE Collaboration gratefully acknowledges the resources and support provided by all Grid centres and the Worldwide LHC Computing Grid (WLCG) collaboration.
The ALICE Collaboration acknowledges the following funding agencies for their support in building and running the ALICE detector:
A. I. Alikhanyan National Science Laboratory (Yerevan Physics Institute) Foundation (ANSL), State Committee of Science and World Federation of Scientists (WFS), Armenia;
Austrian Academy of Sciences and Nationalstiftung f\"{u}r Forschung, Technologie und Entwicklung, Austria;
Ministry of Communications and High Technologies, National Nuclear Research Center, Azerbaijan;
Conselho Nacional de Desenvolvimento Cient\'{\i}fico e Tecnol\'{o}gico (CNPq), Universidade Federal do Rio Grande do Sul (UFRGS), Financiadora de Estudos e Projetos (Finep) and Funda\c{c}\~{a}o de Amparo \`{a} Pesquisa do Estado de S\~{a}o Paulo (FAPESP), Brazil;
Ministry of Science \& Technology of China (MSTC), National Natural Science Foundation of China (NSFC) and Ministry of Education of China (MOEC) , China;
Ministry of Science, Education and Sport and Croatian Science Foundation, Croatia;
Ministry of Education, Youth and Sports of the Czech Republic, Czech Republic;
The Danish Council for Independent Research | Natural Sciences, the Carlsberg Foundation and Danish National Research Foundation (DNRF), Denmark;
Helsinki Institute of Physics (HIP), Finland;
Commissariat \`{a} l'Energie Atomique (CEA) and Institut National de Physique Nucl\'{e}aire et de Physique des Particules (IN2P3) and Centre National de la Recherche Scientifique (CNRS), France;
Bundesministerium f\"{u}r Bildung, Wissenschaft, Forschung und Technologie (BMBF) and GSI Helmholtzzentrum f\"{u}r Schwerionenforschung GmbH, Germany;
Ministry of Education, Research and Religious Affairs, Greece;
National Research, Development and Innovation Office, Hungary;
Department of Atomic Energy Government of India (DAE) and Council of Scientific and Industrial Research (CSIR), New Delhi, India;
Indonesian Institute of Science, Indonesia;
Centro Fermi - Museo Storico della Fisica e Centro Studi e Ricerche Enrico Fermi and Istituto Nazionale di Fisica Nucleare (INFN), Italy;
Institute for Innovative Science and Technology , Nagasaki Institute of Applied Science (IIST), Japan Society for the Promotion of Science (JSPS) KAKENHI and Japanese Ministry of Education, Culture, Sports, Science and Technology (MEXT), Japan;
Consejo Nacional de Ciencia (CONACYT) y Tecnolog\'{i}a, through Fondo de Cooperaci\'{o}n Internacional en Ciencia y Tecnolog\'{i}a (FONCICYT) and Direcci\'{o}n General de Asuntos del Personal Academico (DGAPA), Mexico;
Nationaal instituut voor subatomaire fysica (Nikhef), Netherlands;
The Research Council of Norway, Norway;
Commission on Science and Technology for Sustainable Development in the South (COMSATS), Pakistan;
Pontificia Universidad Cat\'{o}lica del Per\'{u}, Peru;
Ministry of Science and Higher Education and National Science Centre, Poland;
Korea Institute of Science and Technology Information and National Research Foundation of Korea (NRF), Republic of Korea;
Ministry of Education and Scientific Research, Institute of Atomic Physics and Romanian National Agency for Science, Technology and Innovation, Romania;
Joint Institute for Nuclear Research (JINR), Ministry of Education and Science of the Russian Federation and National Research Centre Kurchatov Institute, Russia;
Ministry of Education, Science, Research and Sport of the Slovak Republic, Slovakia;
National Research Foundation of South Africa, South Africa;
Centro de Aplicaciones Tecnol\'{o}gicas y Desarrollo Nuclear (CEADEN), Cubaenerg\'{\i}a, Cuba, Ministerio de Ciencia e Innovacion and Centro de Investigaciones Energ\'{e}ticas, Medioambientales y Tecnol\'{o}gicas (CIEMAT), Spain;
Swedish Research Council (VR) and Knut \& Alice Wallenberg Foundation (KAW), Sweden;
European Organization for Nuclear Research, Switzerland;
National Science and Technology Development Agency (NSDTA), Suranaree University of Technology (SUT) and Office of the Higher Education Commission under NRU project of Thailand, Thailand;
Turkish Atomic Energy Agency (TAEK), Turkey;
National Academy of  Sciences of Ukraine, Ukraine;
Science and Technology Facilities Council (STFC), United Kingdom;
National Science Foundation of the United States of America (NSF) and United States Department of Energy, Office of Nuclear Physics (DOE NP), United States of America.
\fi
\end{acknowledgement}

\bibliographystyle{utphys}
\bibliography{biblio}{}
\newpage
\appendix
\section{The ALICE Collaboration}
\label{app:collab}
\ifdraft
\else



\begingroup
\small
\begin{flushleft}
S.~Acharya$^\textrm{\scriptsize 139}$,
D.~Adamov\'{a}$^\textrm{\scriptsize 87}$,
M.M.~Aggarwal$^\textrm{\scriptsize 91}$,
G.~Aglieri Rinella$^\textrm{\scriptsize 34}$,
M.~Agnello$^\textrm{\scriptsize 30}$,
N.~Agrawal$^\textrm{\scriptsize 47}$,
Z.~Ahammed$^\textrm{\scriptsize 139}$,
N.~Ahmad$^\textrm{\scriptsize 17}$,
S.U.~Ahn$^\textrm{\scriptsize 69}$,
S.~Aiola$^\textrm{\scriptsize 143}$
% ...
\renewcommand\labelenumi{\textsuperscript{\theenumi}~}

\section*{Affiliation notes}
\renewcommand\theenumi{\roman{enumi}}
\begin{Authlist}
\item \Adef{0}Deceased
\item \Adef{idp1757952}{Also at: Dipartimento DET del Politecnico di Torino, Turin, Italy}
\item \Adef{idp1777344}{Also at: Georgia State University, Atlanta, Georgia, United States}
\item \Adef{idp3983968}{Also at: M.V. Lomonosov Moscow State University, D.V. Skobeltsyn Institute of Nuclear, Physics, Moscow, Russia}
\item \Adef{idp4320208}{Also at: Department of Applied Physics, Aligarh Muslim University, Aligarh, India}
\end{Authlist}

\section*{Collaboration Institutes}
\renewcommand\theenumi{\arabic{enumi}~}

$^{1}$A.I. Alikhanyan National Science Laboratory (Yerevan Physics Institute) Foundation, Yerevan, Armenia
\\
$^{2}$Benem\'{e}rita Universidad Aut\'{o}noma de Puebla, Puebla, Mexico
\\
$^{3}$Bogolyubov Institute for Theoretical Physics, Kiev, Ukraine
\\
$^{4}$Bose Institute, Department of Physics 
and Centre for Astroparticle Physics and Space Science (CAPSS), Kolkata, India
\\
$^{5}$Budker Institute for Nuclear Physics, Novosibirsk, Russia
\\
$^{6}$California Polytechnic State University, San Luis Obispo, California, United States
\\
$^{7}$Central China Normal University, Wuhan, China
\\
$^{8}$Centre de Calcul de l'IN2P3, Villeurbanne, Lyon, France
\\
$^{9}$Centro de Aplicaciones Tecnol\'{o}gicas y Desarrollo Nuclear (CEADEN), Havana, Cuba
\\
$^{10}$Centro de Investigaciones Energ\'{e}ticas Medioambientales y Tecnol\'{o}gicas (CIEMAT), Madrid, Spain
\\
$^{11}$Centro de Investigaci\'{o}n y de Estudios Avanzados (CINVESTAV), Mexico City and M\'{e}rida, Mexico
\\
$^{12}$Centro Fermi - Museo Storico della Fisica e Centro Studi e Ricerche ``Enrico Fermi', Rome, Italy
\\
$^{13}$Chicago State University, Chicago, Illinois, United States
\\
$^{14}$China Institute of Atomic Energy, Beijing, China
\\
$^{15}$COMSATS Institute of Information Technology (CIIT), Islamabad, Pakistan
\\
$^{16}$Departamento de F\'{\i}sica de Part\'{\i}culas and IGFAE, Universidad de Santiago de Compostela, Santiago de Compostela, Spain
\\
$^{17}$Department of Physics, Aligarh Muslim University, Aligarh, India
\\
$^{18}$Department of Physics, Ohio State University, Columbus, Ohio, United States
\\
$^{19}$Department of Physics, Sejong University, Seoul, South Korea
\\
$^{20}$Department of Physics, University of Oslo, Oslo, Norway
\\
$^{21}$Department of Physics and Technology, University of Bergen, Bergen, Norway
\\
$^{22}$Dipartimento di Fisica dell'Universit\`{a} 'La Sapienza'
and Sezione INFN, Rome, Italy
\\
$^{23}$Dipartimento di Fisica dell'Universit\`{a}
and Sezione INFN, Cagliari, Italy
\\
$^{24}$Dipartimento di Fisica dell'Universit\`{a}
and Sezione INFN, Trieste, Italy
\\
$^{25}$Dipartimento di Fisica dell'Universit\`{a}
and Sezione INFN, Turin, Italy
\\
$^{26}$Dipartimento di Fisica e Astronomia dell'Universit\`{a}
and Sezione INFN, Bologna, Italy
\\
$^{27}$Dipartimento di Fisica e Astronomia dell'Universit\`{a}
and Sezione INFN, Catania, Italy
\\
$^{28}$Dipartimento di Fisica e Astronomia dell'Universit\`{a}
and Sezione INFN, Padova, Italy
\\
$^{29}$Dipartimento di Fisica `E.R.~Caianiello' dell'Universit\`{a}
and Gruppo Collegato INFN, Salerno, Italy
\\
$^{30}$Dipartimento DISAT del Politecnico and Sezione INFN, Turin, Italy
\\
$^{31}$Dipartimento di Scienze e Innovazione Tecnologica dell'Universit\`{a} del Piemonte Orientale and INFN Sezione di Torino, Alessandria, Italy
\\
$^{32}$Dipartimento Interateneo di Fisica `M.~Merlin'
and Sezione INFN, Bari, Italy
\\
$^{33}$Division of Experimental High Energy Physics, University of Lund, Lund, Sweden
\\
$^{34}$European Organization for Nuclear Research (CERN), Geneva, Switzerland
\\
$^{35}$Excellence Cluster Universe, Technische Universit\"{a}t M\"{u}nchen, Munich, Germany
\\
$^{36}$Faculty of Engineering, Bergen University College, Bergen, Norway
\\
$^{37}$Faculty of Mathematics, Physics and Informatics, Comenius University, Bratislava, Slovakia
\\
$^{38}$Faculty of Nuclear Sciences and Physical Engineering, Czech Technical University in Prague, Prague, Czech Republic
\\
$^{39}$Faculty of Science, P.J.~\v{S}af\'{a}rik University, Ko\v{s}ice, Slovakia
\\
$^{40}$Faculty of Technology, Buskerud and Vestfold University College, Tonsberg, Norway
\\
$^{41}$Frankfurt Institute for Advanced Studies, Johann Wolfgang Goethe-Universit\"{a}t Frankfurt, Frankfurt, Germany
\\
$^{42}$Gangneung-Wonju National University, Gangneung, South Korea
\\
$^{43}$Gauhati University, Department of Physics, Guwahati, India
\\
$^{44}$Helmholtz-Institut f\"{u}r Strahlen- und Kernphysik, Rheinische Friedrich-Wilhelms-Universit\"{a}t Bonn, Bonn, Germany
\\
$^{45}$Helsinki Institute of Physics (HIP), Helsinki, Finland
\\
$^{46}$Hiroshima University, Hiroshima, Japan
\\
$^{47}$Indian Institute of Technology Bombay (IIT), Mumbai, India
\\
$^{48}$Indian Institute of Technology Indore, Indore, India
\\
$^{49}$Indonesian Institute of Sciences, Jakarta, Indonesia
\\
$^{50}$Inha University, Incheon, South Korea
\\
$^{51}$Institut de Physique Nucl\'eaire d'Orsay (IPNO), Universit\'e Paris-Sud, CNRS-IN2P3, Orsay, France
\\
$^{52}$Institute for Nuclear Research, Academy of Sciences, Moscow, Russia
\\
$^{53}$Institute for Subatomic Physics of Utrecht University, Utrecht, Netherlands
\\
$^{54}$Institute for Theoretical and Experimental Physics, Moscow, Russia
\\
$^{55}$Institute of Experimental Physics, Slovak Academy of Sciences, Ko\v{s}ice, Slovakia
\\
$^{56}$Institute of Physics, Academy of Sciences of the Czech Republic, Prague, Czech Republic
\\
$^{57}$Institute of Physics, Bhubaneswar, India
\\
$^{58}$Institute of Space Science (ISS), Bucharest, Romania
\\
$^{59}$Institut f\"{u}r Informatik, Johann Wolfgang Goethe-Universit\"{a}t Frankfurt, Frankfurt, Germany
\\
$^{60}$Institut f\"{u}r Kernphysik, Johann Wolfgang Goethe-Universit\"{a}t Frankfurt, Frankfurt, Germany
\\
$^{61}$Institut f\"{u}r Kernphysik, Westf\"{a}lische Wilhelms-Universit\"{a}t M\"{u}nster, M\"{u}nster, Germany
\\
$^{62}$Instituto de Ciencias Nucleares, Universidad Nacional Aut\'{o}noma de M\'{e}xico, Mexico City, Mexico
\\
$^{63}$Instituto de F\'{i}sica, Universidade Federal do Rio Grande do Sul (UFRGS), Porto Alegre, Brazil
\\
$^{64}$Instituto de F\'{\i}sica, Universidad Nacional Aut\'{o}noma de M\'{e}xico, Mexico City, Mexico
\\
$^{65}$IRFU, CEA, Universit\'{e} Paris-Saclay, F-91191 Gif-sur-Yvette, France, Saclay, France
\\
$^{66}$iThemba LABS, National Research Foundation, Somerset West, South Africa
\\
$^{67}$Joint Institute for Nuclear Research (JINR), Dubna, Russia
\\
$^{68}$Konkuk University, Seoul, South Korea
\\
$^{69}$Korea Institute of Science and Technology Information, Daejeon, South Korea
\\
$^{70}$KTO Karatay University, Konya, Turkey
\\
$^{71}$Laboratoire de Physique Corpusculaire (LPC), Clermont Universit\'{e}, Universit\'{e} Blaise Pascal, CNRS--IN2P3, Clermont-Ferrand, France
\\
$^{72}$Laboratoire de Physique Subatomique et de Cosmologie, Universit\'{e} Grenoble-Alpes, CNRS-IN2P3, Grenoble, France
\\
$^{73}$Laboratori Nazionali di Frascati, INFN, Frascati, Italy
\\
$^{74}$Laboratori Nazionali di Legnaro, INFN, Legnaro, Italy
\\
$^{75}$Lawrence Berkeley National Laboratory, Berkeley, California, United States
\\
$^{76}$Moscow Engineering Physics Institute, Moscow, Russia
\\
$^{77}$Nagasaki Institute of Applied Science, Nagasaki, Japan
\\
$^{78}$National and Kapodistrian University of Athens, Physics Department, Athens, Greece, Athens, Greece
\\
$^{79}$National Centre for Nuclear Studies, Warsaw, Poland
\\
$^{80}$National Institute for Physics and Nuclear Engineering, Bucharest, Romania
\\
$^{81}$National Institute of Science Education and Research, Bhubaneswar, India
\\
$^{82}$National Nuclear Research Center, Baku, Azerbaijan
\\
$^{83}$National Research Centre Kurchatov Institute, Moscow, Russia
\\
$^{84}$Niels Bohr Institute, University of Copenhagen, Copenhagen, Denmark
\\
$^{85}$Nikhef, Nationaal instituut voor subatomaire fysica, Amsterdam, Netherlands
\\
$^{86}$Nuclear Physics Group, STFC Daresbury Laboratory, Daresbury, United Kingdom
\\
$^{87}$Nuclear Physics Institute, Academy of Sciences of the Czech Republic, \v{R}e\v{z} u Prahy, Czech Republic
\\
$^{88}$Oak Ridge National Laboratory, Oak Ridge, Tennessee, United States
\\
$^{89}$Petersburg Nuclear Physics Institute, Gatchina, Russia
\\
$^{90}$Physics Department, Creighton University, Omaha, Nebraska, United States
\\
$^{91}$Physics Department, Panjab University, Chandigarh, India
\\
$^{92}$Physics Department, University of Cape Town, Cape Town, South Africa
\\
$^{93}$Physics Department, University of Jammu, Jammu, India
\\
$^{94}$Physics Department, University of Rajasthan, Jaipur, India
\\
$^{95}$Physikalisches Institut, Eberhard Karls Universit\"{a}t T\"{u}bingen, T\"{u}bingen, Germany
\\
$^{96}$Physikalisches Institut, Ruprecht-Karls-Universit\"{a}t Heidelberg, Heidelberg, Germany
\\
$^{97}$Physik Department, Technische Universit\"{a}t M\"{u}nchen, Munich, Germany
\\
$^{98}$Purdue University, West Lafayette, Indiana, United States
\\
$^{99}$Pusan National University, Pusan, South Korea
\\
$^{100}$Research Division and ExtreMe Matter Institute EMMI, GSI Helmholtzzentrum f\"ur Schwerionenforschung GmbH, Darmstadt, Germany
\\
$^{101}$Rudjer Bo\v{s}kovi\'{c} Institute, Zagreb, Croatia
\\
$^{102}$Russian Federal Nuclear Center (VNIIEF), Sarov, Russia
\\
$^{103}$Saha Institute of Nuclear Physics, Kolkata, India
\\
$^{104}$School of Physics and Astronomy, University of Birmingham, Birmingham, United Kingdom
\\
$^{105}$Secci\'{o}n F\'{\i}sica, Departamento de Ciencias, Pontificia Universidad Cat\'{o}lica del Per\'{u}, Lima, Peru
\\
$^{106}$Sezione INFN, Bari, Italy
\\
$^{107}$Sezione INFN, Bologna, Italy
\\
$^{108}$Sezione INFN, Cagliari, Italy
\\
$^{109}$Sezione INFN, Catania, Italy
\\
$^{110}$Sezione INFN, Padova, Italy
\\
$^{111}$Sezione INFN, Rome, Italy
\\
$^{112}$Sezione INFN, Trieste, Italy
\\
$^{113}$Sezione INFN, Turin, Italy
\\
$^{114}$SSC IHEP of NRC Kurchatov institute, Protvino, Russia
\\
$^{115}$Stefan Meyer Institut f\"{u}r Subatomare Physik (SMI), Vienna, Austria
\\
$^{116}$SUBATECH, IMT Atlantique, Universit\'{e} de Nantes, CNRS-IN2P3, Nantes, France
\\
$^{117}$Suranaree University of Technology, Nakhon Ratchasima, Thailand
\\
$^{118}$Technical University of Ko\v{s}ice, Ko\v{s}ice, Slovakia
\\
$^{119}$Technical University of Split FESB, Split, Croatia
\\
$^{120}$The Henryk Niewodniczanski Institute of Nuclear Physics, Polish Academy of Sciences, Cracow, Poland
\\
$^{121}$The University of Texas at Austin, Physics Department, Austin, Texas, United States
\\
$^{122}$Universidad Aut\'{o}noma de Sinaloa, Culiac\'{a}n, Mexico
\\
$^{123}$Universidade de S\~{a}o Paulo (USP), S\~{a}o Paulo, Brazil
\\
$^{124}$Universidade Estadual de Campinas (UNICAMP), Campinas, Brazil
\\
$^{125}$Universidade Federal do ABC, Santo Andre, Brazil
\\
$^{126}$University of Houston, Houston, Texas, United States
\\
$^{127}$University of Jyv\"{a}skyl\"{a}, Jyv\"{a}skyl\"{a}, Finland
\\
$^{128}$University of Liverpool, Liverpool, United Kingdom
\\
$^{129}$University of Tennessee, Knoxville, Tennessee, United States
\\
$^{130}$University of the Witwatersrand, Johannesburg, South Africa
\\
$^{131}$University of Tokyo, Tokyo, Japan
\\
$^{132}$University of Tsukuba, Tsukuba, Japan
\\
$^{133}$University of Zagreb, Zagreb, Croatia
\\
$^{134}$Universit\'{e} de Lyon, Universit\'{e} Lyon 1, CNRS/IN2P3, IPN-Lyon, Villeurbanne, Lyon, France
\\
$^{135}$Universit\'{e} de Strasbourg, CNRS, IPHC UMR 7178, F-67000 Strasbourg, France, Strasbourg, France
\\
$^{136}$Universit\`{a} degli Studi di Pavia, Pavia, Italy
\\
$^{137}$Universit\`{a} di Brescia, Brescia, Italy
\\
$^{138}$V.~Fock Institute for Physics, St. Petersburg State University, St. Petersburg, Russia
\\
$^{139}$Variable Energy Cyclotron Centre, Kolkata, India
\\
$^{140}$Warsaw University of Technology, Warsaw, Poland
\\
$^{141}$Wayne State University, Detroit, Michigan, United States
\\
$^{142}$Wigner Research Centre for Physics, Hungarian Academy of Sciences, Budapest, Hungary
\\
$^{143}$Yale University, New Haven, Connecticut, United States
\\
$^{144}$Yonsei University, Seoul, South Korea
\\
$^{145}$Zentrum f\"{u}r Technologietransfer und Telekommunikation (ZTT), Fachhochschule Worms, Worms, Germany
\endgroup

\fi
\end{document}